\documentclass[preprint]{aastex631}

\newcommand{\sqcm}{${\rm cm^{-2}}$}

\newcommand{\mum}{${\rm \mu m}$}
  
\newcommand{\waven}{${\rm cm^{-1}}$}

\newcommand{\ocnm}{${\rm OCN^{-}}$}
\newcommand{\ammonia}{${\rm NH_{3}}$}
\newcommand{\water}{${\rm H_{2}O}$}
\newcommand{\methanol}{${\rm CH_{3}OH}$}
\newcommand{\methane}{${\rm CH_{4}}$}
\newcommand{\cotwo}{${\rm CO_{2}}$}
\newcommand{\av}{${\it A_{\rm V}}$}

\shorttitle{Ices toward Massive YSOs: I. OCS, CO, OCN$^-$, and
  CH$_3$OH}

\shortauthors{Boogert, Brewer, Brittain, and Emerson}

\received{8/26/2022}
\accepted{10/17/2022 for publication in the Astrophysical Journal}
\published{12/09/2022 in ApJ 941, 32}

\begin{document}

\title{Survey of Ices toward Massive Young Stellar Objects: I. OCS,
  CO, OCN$^-$, and CH$_3$OH}

\author[0000-0001-9344-0096]{A. C. A. Boogert}

\affiliation{Institute for Astronomy, University of Hawai'i at
  Manoa, 2680 Woodlawn Drive, Honolulu, HI 96822, USA;
  aboogert@hawaii.edu}

\altaffiliation{Staff Astronomer at the Infrared Telescope Facility,
  which is operated by the University of Hawaii under contract
  80HQTR19D0030 with the National Aeronautics and Space
  Administration.}

\author[0000-0001-7081-9415]{K. Brewer}

\affiliation{Institute for Astronomy, University of Hawai'i at
  Manoa, 2680 Woodlawn Drive, Honolulu, HI 96822, USA}

\author[0000-0003-1720-1899]{A. Brittain}

\affiliation{Institute for Astronomy, University of Hawai'i at
  Manoa, 2680 Woodlawn Drive, Honolulu, HI 96822, USA}

\author[0000-0002-1141-751X]{K. S. Emerson}

\affiliation{Institute for Astronomy, University of Hawai'i, 640
  N. Aohoku Place, Hilo, HI 96720, USA}

\begin{abstract}
An important tracer of the origin and evolution of cometary ices is
the comparison with ices found in dense clouds and towards Young
Stellar Objects (YSOs).  We present a survey of ices in the 2-5
\mum\ spectra of 23 massive YSOs, taken with the NASA InfraRed
Telescope Facility SpeX spectrometer. The 4.90 \mum\ absorption band
of OCS ice is detected in 20 sight-lines, more than five times the
previously known detections.  The absorption profile shows little
variation and is consistent with OCS embedded in CH$_3$OH-rich ices,
and proton-irradiated H$_2$S or SO$_2$-containing ices.  The OCS
column densities correlate well with those of CH$_3$OH and OCN$^-$,
but not with H$_2$O and apolar CO ice.  This association of OCS with
CH$_3$OH and OCN$^-$ firmly establishes their formation location deep
inside dense clouds or protostellar envelopes.  The median composition
of this ice phase towards massive YSOs, as a percentage of H$_2$O, is
CO:CH$_3$OH:OCN$^-$:OCS=24:20:1.53:0.15. CS, due to its low abundance,
is likely not the main precursor to OCS.  Sulfurization of CO is
likely needed, although the source of this sulfur is not well
constrained.  Compared to massive YSOs, low mass YSOs and dense clouds
have similar or somewhat lower CO and CH$_3$OH ice abundances, but
less OCN$^-$ and more apolar CO, while OCS awaits detection. Comets
tend to be under-abundant in carbon-bearing species, but this does not
appear to be the case for OCS, perhaps signalling OCS production in
protoplanetary disks.
\end{abstract}


\section{Introduction}~\label{sec:intro}

Ices are the dominant phase of molecules such as \water, CO, \cotwo,
and \methanol\ in the densest regions of molecular clouds (e.g.,
\citealt{chiar95, whittet98, boogert11, boogert15}). Through the
process of star and protoplanetary disk formation, occuring in these
same regions, these ices may be the building blocks of comets (e.g.,
\citealt{mumma11}), which deliver them to Earth, affecting the
composition of the Earth's oceans and perhaps even the formation of
life. The interstellar ices may be incorporated in comets in pristine
(dense cloud) form, or after physical and chemical modification
(``processing'') along the way. They may also be destroyed and the
interstellar heritage is lost (e.g., \citealt{visser09,
  pontoppidan14}).

The relation between dense cloud ices and solar system materials is a
key topic in astronomy.  Observational constraints are derived from
the comparison of the ice composition and structure towards quiescent
dense clouds, Young Stellar Objects (YSOs), and comets.  It is well
established that simple, hydrogenated molecules such as \water\ and
\methane\ form efficiently on the surfaces of cold dust grains in
dense molecular clouds (e.g., \citealt{tielens82}), yielding
\water-rich icy grain mantles. CO$_2$ is the second most abundant
species in these early ices, as it forms from the reaction between CO
and OH (e.g., \citealt{ioppolo11}).  Pure CO will, due to its
volatility, freeze out deeper into clouds (\av$>$3-9 mag). Laboratory
experiments and simulations (e.g., \citealt{watanabe02, cuppen09})
have proven that this CO-rich ``apolar'' ice layer hydrogenates to
H$_2$CO and CH$_3$OH ices. And indeed, CH$_3$OH is very abundant in
some dense cloud environments (10-30\% relative to H$_2$O;
\citealt{boogert11, chiar11, chu20, goto21}). These CO-rich ices may
well be the sites of Complex Organic Molecules (COMs) formation in
dense clouds, in particular due to hydrogen abstraction reactions,
creating radicals (e.g., HCO) and subsequently COMs such as
glycolaldehyde \citep{fedoseev15}.

Upon incorporation into YSO envelopes and disks, elevated dust
temperatures are expected to modify the ice composition and
structure. Indeed, H$_2$O ice crystallization is regularly observed by
the profile of the 3.0 \mum\ absorption band (e.g.,
\citealt{brooke99, terada12}). The increased mobility of radicals at
higher temperature is expected to be a route to COM formation
(e.g., \citealt{herbst09}). Sublimation of the ices creates so called
Hot Cores, which exhibit a rich warm gas phase chemistry
\citep{charnley92}.  It has also been noted that CO and CO$_2$ seem
systematically under-abundant by a factor of a few towards massive
YSOs (MYSOs) compared to low mass YSOs \citep{pontoppidan08, oberg11},
possibly due to the warmer envelopes of the former.  Interestingly,
cometary CO, CO$_2$ and CH$_3$OH ice abundances are also relatively
low. Does this relate to the origin of the Sun in a cluster near a
massive star, in agreement with Sun formation theories
\citep{adams10}?

The minor ice species \ocnm\ and OCS offer important insights into the
evolution of interstellar ices. The 4.62 \mum\ band from OCN$^-$ was
detected in both low and high mass YSO envelopes
\citep{vanbroekhuizen05}. While an association with CO--rich ices is
expected, with HNCO a likely precursor, it is notably absent in the
spectra of dense clouds \citep{whittet01}, also in lines of sight with
abundant CH$_3$OH ice, indicative of a rich CO chemistry
\citep{chu20}. Perhaps an elevated dust temperature is needed for
\ocnm\ formation. While acid-base reactions of \ammonia\ with HNCO,
forming the NH$_4^+$OCN$^-$ salt, proceed at temperatures of 10 K
\citep{raunier04}, enhancement is expected at higher temperatures,
when the mobility of species in the ice increases. The detection of
abundant salts in the comet 67P/Churyumov-Gerasimenko links the
cometary ices with those in the protostellar envelopes
\citep{altwegg20}.

An absorption feature at 4.90 \mum\ was securely detected in three
sight-lines, all of which tracing MYSO envelopes, and was attributed
to the C-O stretch mode of OCS ice \citep{palumbo95, palumbo97}.  A
possible identification with the 2$\nu _2$ overtone mode of CH$_3$OH
\citep{grim91} was dismissed because compared to the strength of the
3.53 \mum\ C-H stretch mode, the observed 4.90 \mum\ band is a factor
of $\sim$5 too strong, and in pure CH$_3$OH ices and mixtures with
H$_2$O, the laboratory band is too broad. \citet{palumbo97} did
conclude from the absorption band profile, however, that OCS is likely
mixed with CH$_3$OH ices. This indicates that, like OCN$^-$, OCS is a
product of the CO-rich phase of the ice mantles.

The uncertain location of sulfur in dense clouds is the topic of many
studies (see \citealt{laas19} and references therein).  The models of
\citet{laas19} locate much of the sulfur in organics, explaining the
unexpectedly low abundance of H$_2$S ice \citep{smith91}.  Although it
seems that OCS consumes just a fraction of the cosmic sulfur budget,
further investigations are needed to constrain the location and
chemistry of sulfur.  No other sulfur-bearing ices are presently
available for this purpose. SO$_2$ was tentatively detected at 7.58
\mum\ in one sight-line \citep{boogert97}, and, interestingly, was
also found to be located in CH$_3$OH-rich ices.

Comet observations, most strikingly the in situ measurements of comet
67P/Churyumov-Gerasimenko by the Rosetta mission, have shown great
molecular complexity, including glycine, carbon chains, salts, and a
plethora of sulfur-bearing molecules (e.g., \citealt{mumma11,
  altwegg19}). What is the origin of this complexity? Molecular
complexity in dense clouds would affect all stars and planets formed
within them, but an origin in protoplanetary disks depends on local
factors (e.g., stellar radiation field, disk size and structure).  The
rich collection of sulfur-bearing species in comet 67P makes sulfur an
attractive tracer of the history of cometary ices
\citep{drozdovskaya19}.

A tentative detection of the 4.90 \mum\ band towards the edge-on
protoplanetary disk IRC L1041-2 by the AKARI mission was presented by
\citet{aikawa12}. It would correspond to an OCS abundance ten times
that towards MYSOs \citep{palumbo97}. The detection of OCS towards the
envelopes and disks surrounding low mass YSOs and background stars
tracing dense clouds is awaiting observations with the James Webb
Space Telescope (JWST).

While low mass YSOs and dense cloud background stars seem more
appropriate to study the history of cometary ices, the study of ices
towards MYSOs has proven to be insightful as well. MYSOs evolve over
shorter time scales, and likely experience different radiation and
heating effects. In addition, most stars, including the Sun, are
formed in clusters \citep{adams10}, where feedback as well as material
exchange \citep{levison10} could affect the protoplanetary
environment.

The MYSO sample that has been studied over the past four decades
consists of only $\sim$15 objects (e.g., \citealt{, gibb04,
  oberg11}). We have embarked on an IRTF/SpeX spectroscopic survey in
the 2-5.3 \mum\ wavelength range, with the goal to better sample ices
towards MYSOs. We will quadruple the MYSO sample by selecting targets
from the Red MSX Survey (RMS; \citealt{lumsden13},
\citealt{pomohaci17}).  JWST studies of this sample will be limited
due to detector saturation. The NIRSpec Integral Field Unit mode
saturates in the $M$-band at $\sim$6.6 magnitudes, and at weaker
magnitudes for the Micro Shutter Array and Fixed Slit
modes\footnote{\url{https://www.stsci.edu/jwst/science-planning/proposal-planning-toolbox/sensitivity-and-saturation-limits}}.
Also, the NIRCam Wide Field Slitless Spectroscopy saturates at $M\leq
6.1$ mag.  Even with NIRCam, all but a few of the targets in our
sample would saturate the detector. These SpeX data will enable a
statistical study of the ice abundances, relating them to existing and
future samples of dense clouds and low mass YSOs.

This paper is structured as follows. In \S\ref{sec:sou}, the selection
of MYSO targets from the Red MSX Survey is described. The IRTF/SpeX
observations and data reduction process are discussed in
\S\ref{sec:obs}. The continuum is determined in \S\ref{sec:cont}, the
removal of contaminating gas phase CO lines is addressed in
\S\ref{sec:gas}, and the ice band absorption profiles are fitted and
column densities are derived in \S\ref{sec:prof}. Laboratory
spectroscopy of solid state OCS, taken from the literature, is
summarized in \S\ref{sec:labocs}. In \S\ref{sec:correl} we search for
correlations between the column densities of H$_2$O, OCN$^-$, OCS, CO,
and CH$_3$OH. The ice abundances are compared across MYSOs, low mass
YSOs. dense cloud background stars, and comets in \S\ref{sec:histo}.
In the discussion, we address the chemical origin of OCS and OCN$^-$
(\S\ref{sec:dis} and \S\ref{sec:dis_ocn}), and the origin of cometary
ices (\S\ref{sec:dis_comet}). The conclusions and future work are
summarized in \S\ref{sec:concl}.

\begin{deluxetable}{lllllllp{3cm}}
\tabletypesize{\scriptsize}
\tablecolumns{5}
\tablewidth{0pc}
\tablecaption{Source Sample~\label{t:sample}}
\tablehead{
\multicolumn{2}{c}{Source\tablenotemark{a}}& \colhead{K\tablenotemark{a}}   & \colhead{J-K\tablenotemark{a}} & \colhead{L\tablenotemark{a}} & \colhead{Date\tablenotemark{b}} & \colhead{Standard\tablenotemark{c}} & \colhead{Notes\tablenotemark{d}}\\
\colhead{MSX}          & \colhead{2MASS-J}& \colhead{mag} & \colhead{mag}                  & \colhead{$L_\odot$}           & \colhead{}     &  \colhead{}        & \colhead{}     \\}
\startdata
G010.8856$+$00.1221    & 18090796-1927237 & 9.6           &  6                             &  5.5$\times 10^3$            & 2020-09-15     & HR 6378             & \nodata         \\
G012.9090$-$00.2607    & 18143956-1752023 & 9.2           &  6.1                           &  3.2$\times 10^4$            & 2020-09-15     & HR 6378             & W33A      \\
G020.7617$-$00.0638C   & 18291219-1050347 & 9.8           &  2.6                           &  1.3$\times 10^4$            & 2020-07-02     & HR 7236             & 0.3$''$ slit \\ 
G023.6566$-$00.1273    & 18345155-0818214 & 9.9           &  3.7                           &  1.0$\times 10^4$            & 2020-07-03     & HR 6629             & 0.3$''$ slit \\ 
G024.6343$-$00.3233    & 18372271-0731417 & 10            &  5.8                           &  6.8$\times 10^3$            & 2020-07-03     & HR 7236             & 0.3$''$ slit \\ 
G025.4118$+$00.1052A   & 18371700-0638244 & 9.0           &  8.7                           &  9.7$\times 10^3$            & 2021-07-28     & HR 7236             & RMS 2MASS source ``a''\\
                       &                  &               &                                &                              & 2020-08-23     & HR 7236             & RMS 2MASS source ``a''\\
G026.3819$+$01.4057A   & 18342567-0510502 & 9.1           &  4                             &  1.7$\times 10^4$            & 2020-07-02     & HR 7236             & RMS 2MASS source 'b', confused with 'c' (2MASSJ18342562-0510500); 0.3$''$ slit \\ 
G027.7571$+$00.0500    & 18414799-0434532 & 9.3           &  7.4                           &  1.3$\times 10^4$            & 2020-09-15     & HR 7236             & \nodata         \\
G027.7954$-$00.2772    & 18430228-0441491 & 10.7          &  5.9                           &  5.1$\times 10^3$            & 2020-07-03     & HR 7236             & 0.3$''$ slit \\ 
                       &                  &               &                                &                              & 2020-08-23     & HR 7236             & \nodata         \\
                       &                  &               &                                &                              & 2021-07-28     & HR 7236             & \nodata         \\
G029.8620$-$00.0444    & 18455955-0245061 & 9.8           &  5.3                           &  2.8$\times 10^4$            & 2020-07-03     & HR 7236             & 0.3$''$ slit \\
                       &                  &               &                                &                              & 2020-09-15     & HR 7236             & \nodata         \\
G033.5237$+$00.0198    & 18522675+0032088 & 8.9           &  6.9                           &  1.3$\times 10^4$            & 2020-07-03     & HR 7236             & 0.3$''$ slit \\
G034.7123$-$00.5946    & 18564827+0118471 & 9.2           &  9.2                           &  9.7$\times 10^3$            & 2020-07-03     & HR 7236             & 0.3$''$ slit \\
                       &                  &               &                                &                              & 2020-08-23     & HR 7236             & \nodata         \\
G039.5328$-$00.1969    & 19041353+0546547 & 11.6          &  6.3                           &  3.6$\times 10^3$            & 2020-08-03     & HR 7235             & 0.3$''$ slit \\ 
G059.7831$+$00.0648    & 19431121+2344039 & 10.8          &  6.5                           &  2.2$\times 10^4$            & 2020-07-31     & HR 7235             & \nodata         \\
                       &                  &               &                                &                              & 2020-08-03     & HR 7235             & \nodata         \\
G063.1140$+$00.3416    & 19493209+2645151 & 9.7           &  7.2                           &  5.5$\times 10^3$            & 2020-07-31     & HR 7235             & 0.3$''$ slit \\
G084.9505$-$00.6910    & 20553247+4406101 & 9.8           &  5.4                           &  1.3$\times 10^4$            & 2020-09-15     & HR 8585             & 0.3$''$ slit; Eastern of 1.1$''$ double star  \\
                       &                  &               &                                &                              & 2020-12-02     & HR 8371             & Eastern of 1.1$''$ double star\\
G105.5072$+$00.2294    & 22322399+5818583 & 10.6          &  4.9                           &  7.0$\times 10^3$            & 2020-07-23     & HR 8028             & \nodata         \\
                       &                  &               &                                &                              & 2020-08-23     & HR 7236             & \nodata         \\
G108.7575$-$00.9863    & 22584725+5845016 & 10            &  6.9                           &  1.4$\times 10^4$            & 2020-07-23     & HR 8585             & \nodata        \\
                       &                  &               &                                &                              & 2020-08-23     & HR 8585             & \nodata        \\
G110.0931$-$00.0641    & 23052516+6008154 & 9.1           &  3.4                           &  1.7$\times 10^4$            & 2020-08-03     & HR 8585             & \nodata       \\
G111.2552$-$00.7702    & 23161039+5955282 & 11.6          &  6.4                           &  1.1$\times 10^4$            & 2020-08-04     & HR 8585             & \nodata      \\
G111.5671$+$00.7517    & 23140175+6127198 & 10.2          &  8.1                           &  2.3$\times 10^4$            & 2020-08-03     & HR 8585             & NGC 7538 IRS9 \\
G189.0307$+$00.7821    & 06084052+2131004 & 8.6           &  7.6                           &  2.4$\times 10^4$            & 2020-12-22     & HR 2421             & \nodata         \\
G203.3166$+$02.0564    & 06411015+0929336 & 4.92          &  6.59                          &  1.7$\times 10^3$            & 2021-01-30     & HR 2421             & AFGL 989, 0.3$''$ slit \\ 
\enddata
\tablenotetext{a}{MSX name, 2MASS name, K-band magnitude, J-K color, and luminosity ($L$) all taken from the RMS database \citep{lumsden13}.}
\tablenotetext{b}{Date of the IRTF/SpeX observations.}
\tablenotetext{c}{Telluric standard star.}
\tablenotetext{d}{The IRTF/SpeX 0.5$''$ slit is used unless noted otherwise.}
\end{deluxetable}

\section{Source Selection}~\label{sec:sou}

The embedded MYSOs for our 2-5 \mum\ spectroscopic survey were
selected from the Red MSX Survey (RMS; \citealt{lumsden13},
\citealt{pomohaci17}). The RMS survey contains more than 700 YSOs and
candidate YSOs across the Galaxy. We selected those that are visible
from Maunakea at airmass values of $<3$ (Declination $>-50^{o}$). They
must also be MYSOs in an early evolutionary stage, enhancing the
likelihood that strong ice absorption features from their massive
envelopes are detectable, much like previously studied MYSOs
\citep{gibb04}. For this, we selected the most luminous objects
($L>$3000 L$_\odot$) that are radio quiet ($<$0.5 Jy at 5 GHz).  As
such, they likely have not yet started to ionize their surroundings
and produce an HII region. This sample was further reduced by
selecting only those targets with $K-$band magnitudes brighter than
13, which facilitates on-slit guiding with the IRTF. This yielded a
list of 215 targets. For about half of these, 1-5 \mum\ spectra were
obtained with IRTF/SpeX. This final sub-selection was based on the
visibility during the observing runs, and on the $J-K$
color. Preference was given to targets with $J-K>5$, i.e., with the
steepest spectral energy distributions. This further enhanced the
likelihood of high quality spectra in the $M-$band, for which the
IRTF/SpeX sensitivity is strongly background-limited. This increases
the bias towards the most embedded targets with the deepest ice
absorption features.  The full sample will be presented in an upcoming
paper (K. Emerson, in preparation). Here, 23 targets
(Table~\ref{t:sample}) are presented for which the 4.90 \mum\ ice
absorption band of OCS was detected, or hints of it were seen at the
2-3$\sigma$ level.

\section{Observations and Data Reduction}~\label{sec:obs}

All targets were observed with the SpeX spectrometer \citep{rayner03}
at the NASA Infrared Telescope Facility (NASA/IRTF) in a number of
nights in the years 2020 and 2021 (Table~\ref{t:sample}). $K$-band
images were taken with the SpeX guider camera and then compared with
the target designations in the RMS database in order to identify the
MYSO. In several cases, the field was rather confused, and we indicate
the selected target in the last column of
Table~\ref{t:sample}. Subsequently, on-slit $K$-band guiding was
enabled, and spectra were taken with the SpeX LXD\_Long mode, using
the 0.3 or 0.5 arcsec wide slits (Table~\ref{t:sample}). This yields
resolving powers of $R=\lambda/\Delta\lambda=$2,500 and 1,500,
respectively. The instantaneous spectral coverage is 1.95-5.36 \mum.

High S/N $M$-band observations require bright standard stars (M$<$4
mag) for telluric correction. The preferred stars with spectral types
A0V, for which accurate model spectra are available, were usually not
nearby. The standard stars we selected include HR 6378 (A2IV-V), HR
6629 (A1Vn), HR 7236 (B8.5V), HR 7235 (A0IV-Vnn), HR 8585 (A1V), HR
8371 (B8Ib), HR 8028 (A0IIIn), and HR 2421 (A1.5IV), where the
spectral types were obtained from
SIMBAD\footnote{\url{http://simbad.u-strasbg.fr/simbad/}}.

The IRTF/SpeX spectra were reduced using Spextool version 4.1
\citep{cushing04}. Flat fielding was done using the images obtained
with SpeX's calibration unit. The wavelength calibration procedure
uses lamp lines at the shortest wavelengths and sky emission lines in
much of the $L$ and $M$-bands. The telluric correction was done using
the Xtellcor program \citep{vacca03}. This uses a model of Vega to
divide out the stellar photosphere. Although the spectral types of the
standard stars were not exactly of type A0V, no photospheric residuals
were observed that could affect our data analysis. The final continuum
signal-to-noise values near the location of the 4.90 \mum\ absorption
feature is on average 45, determined from the scatter in the data
after the correction for gas phase CO absorption
(\S\ref{sec:gas}). The lowest signal-to-noise value is 15 (G111.2552)
and the highest 105 (G029.8620).


\vspace{40pt}

\section{Results}~\label{sec:res}

\begin{figure*}[p]
  \includegraphics[width=18cm, angle=90, scale=0.54]{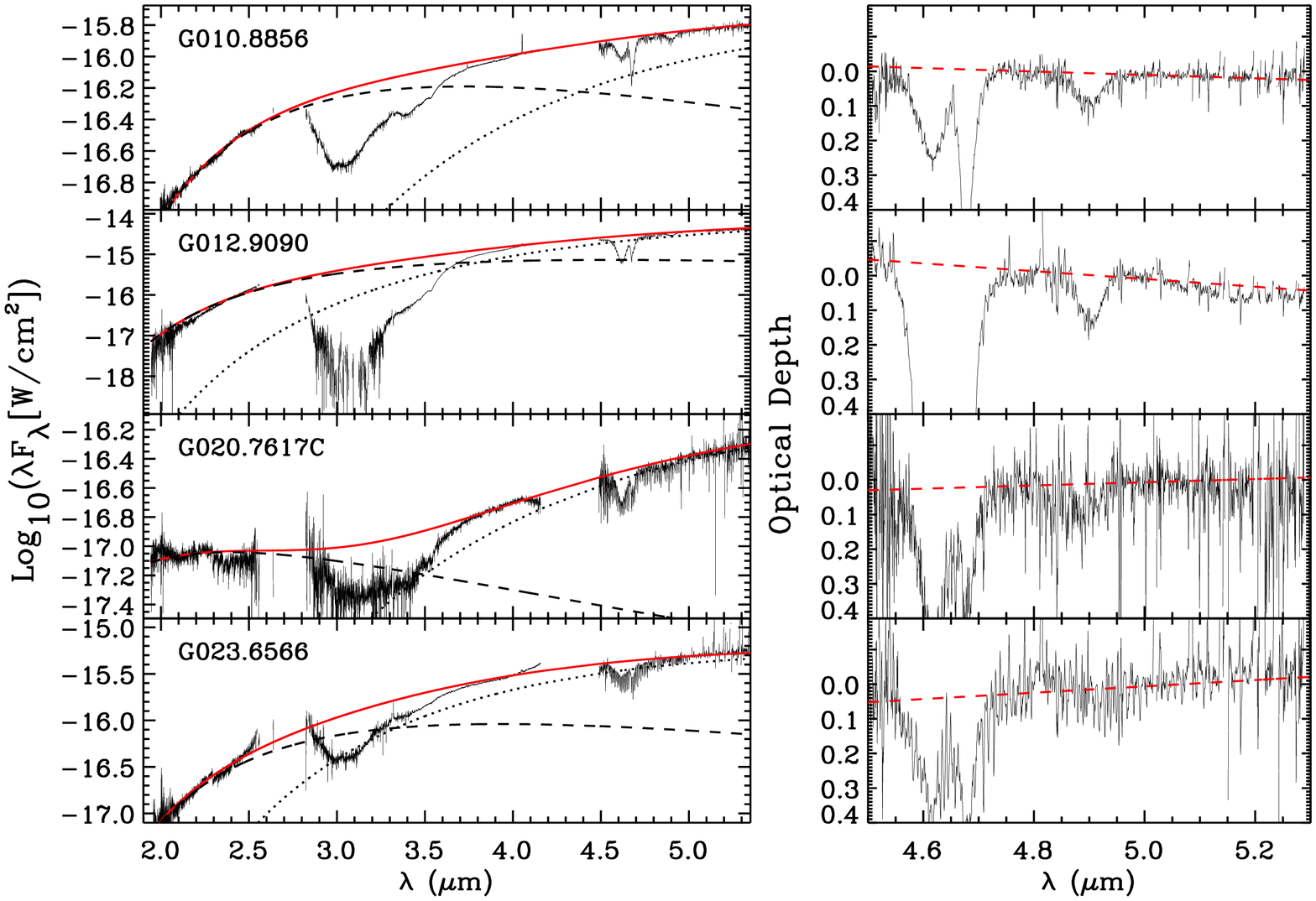}
  \includegraphics[width=18cm, angle=90, scale=0.54]{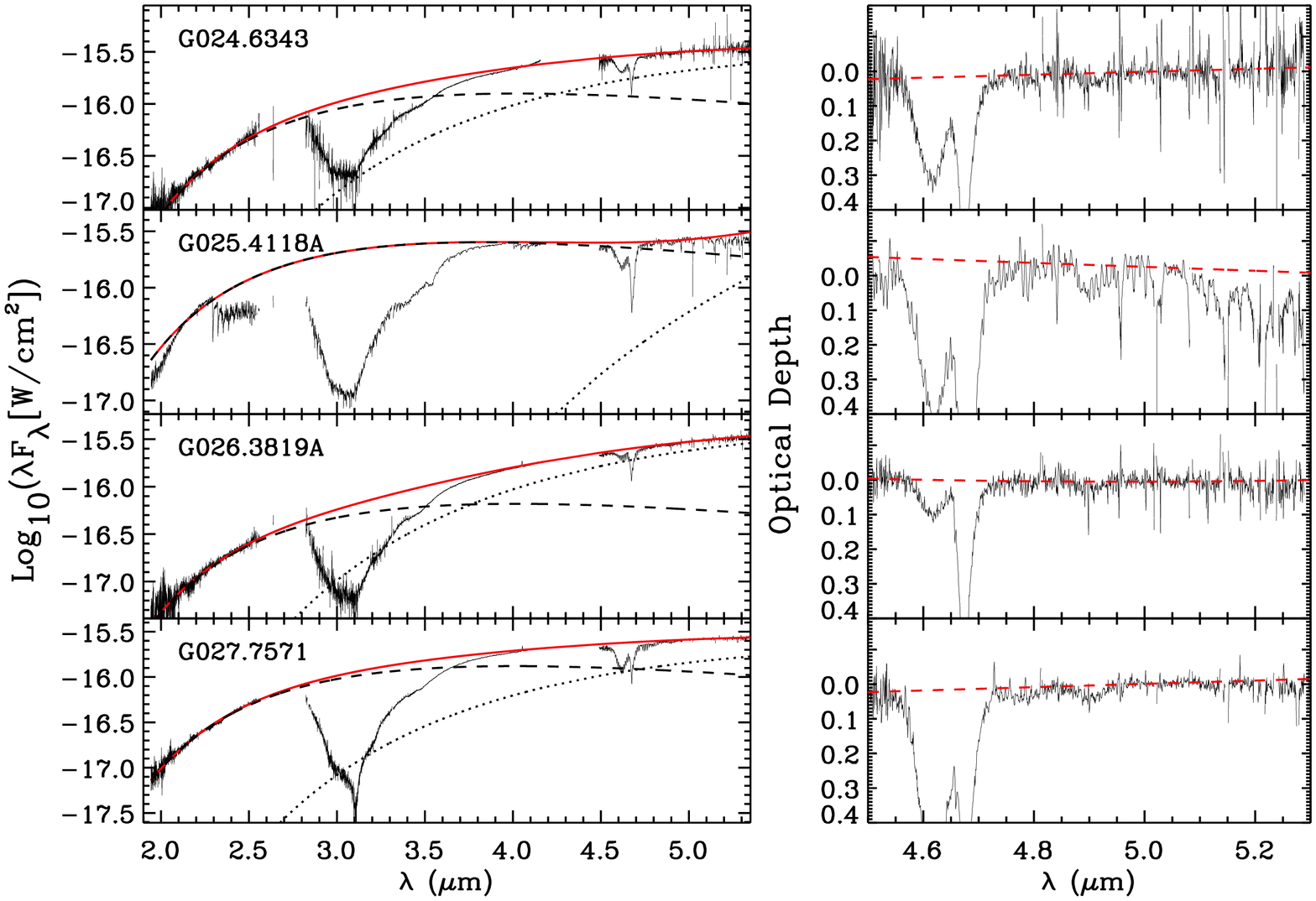}  
    \caption{IRTF/SpeX spectra of MYSO sample. The left panels
      show the flux spectra with the Rayleigh-Jeans continuum of a hot
      star (dashed) and a blackbody continuum spectrum of warm dust
      (dotted), both reddened by the same amount of foreground
      extinction. The sum of the two components, indicated by the red
      line, is used to put the spectra on optical depth scale. The
      right panels zoom in on the $M$-band spectrum, on optical depth
      scale. A local baseline is indicated with the dashed red line,
      which is subtracted before the OCN$^-$, CO, and OCS ice
      absorption features are analyzed.\label{f:flux1}}
\end{figure*}

\begin{figure*}[p]
  \includegraphics[width=18cm, angle=90, scale=0.54]{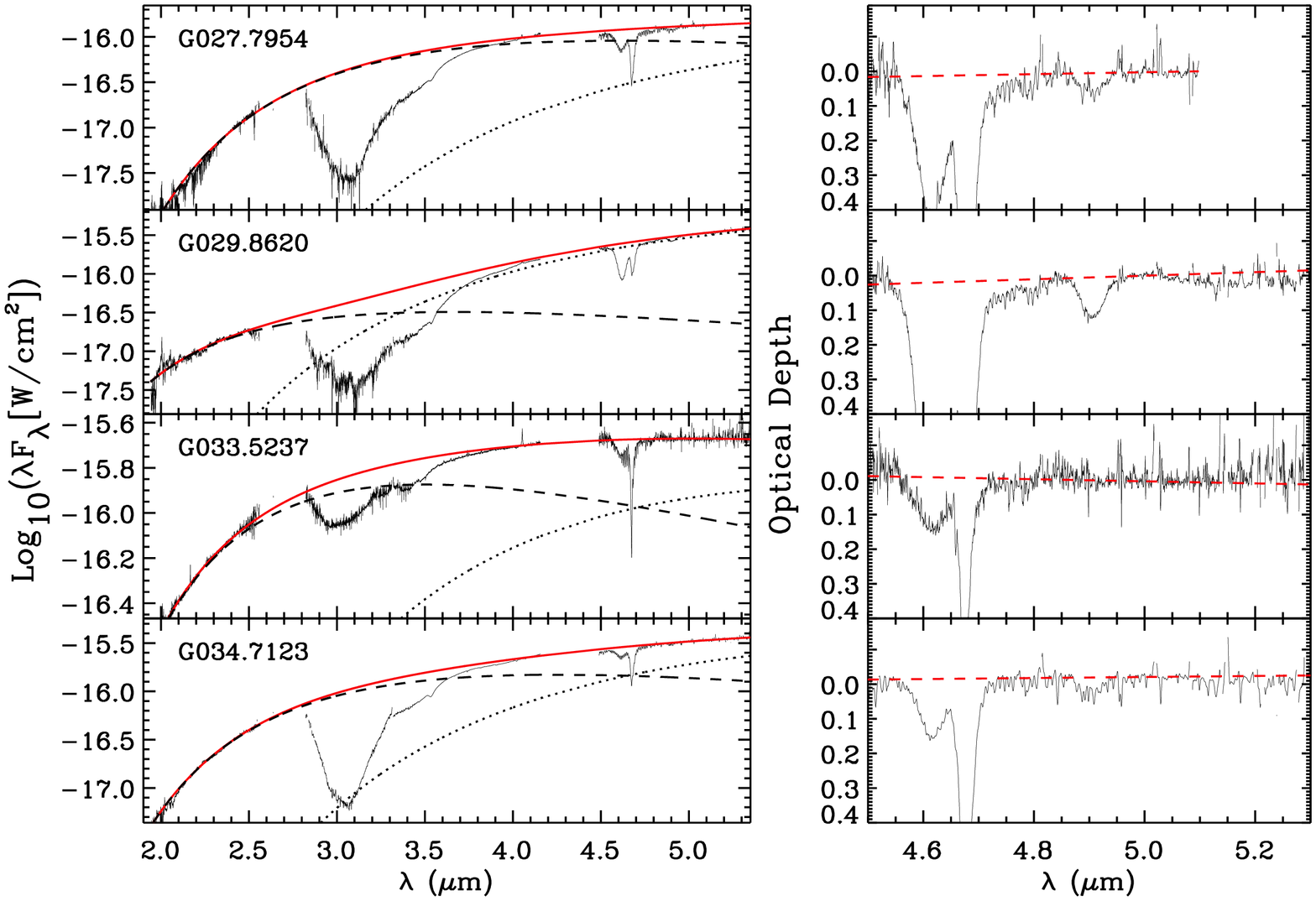}
  \includegraphics[width=18cm, angle=90, scale=0.54]{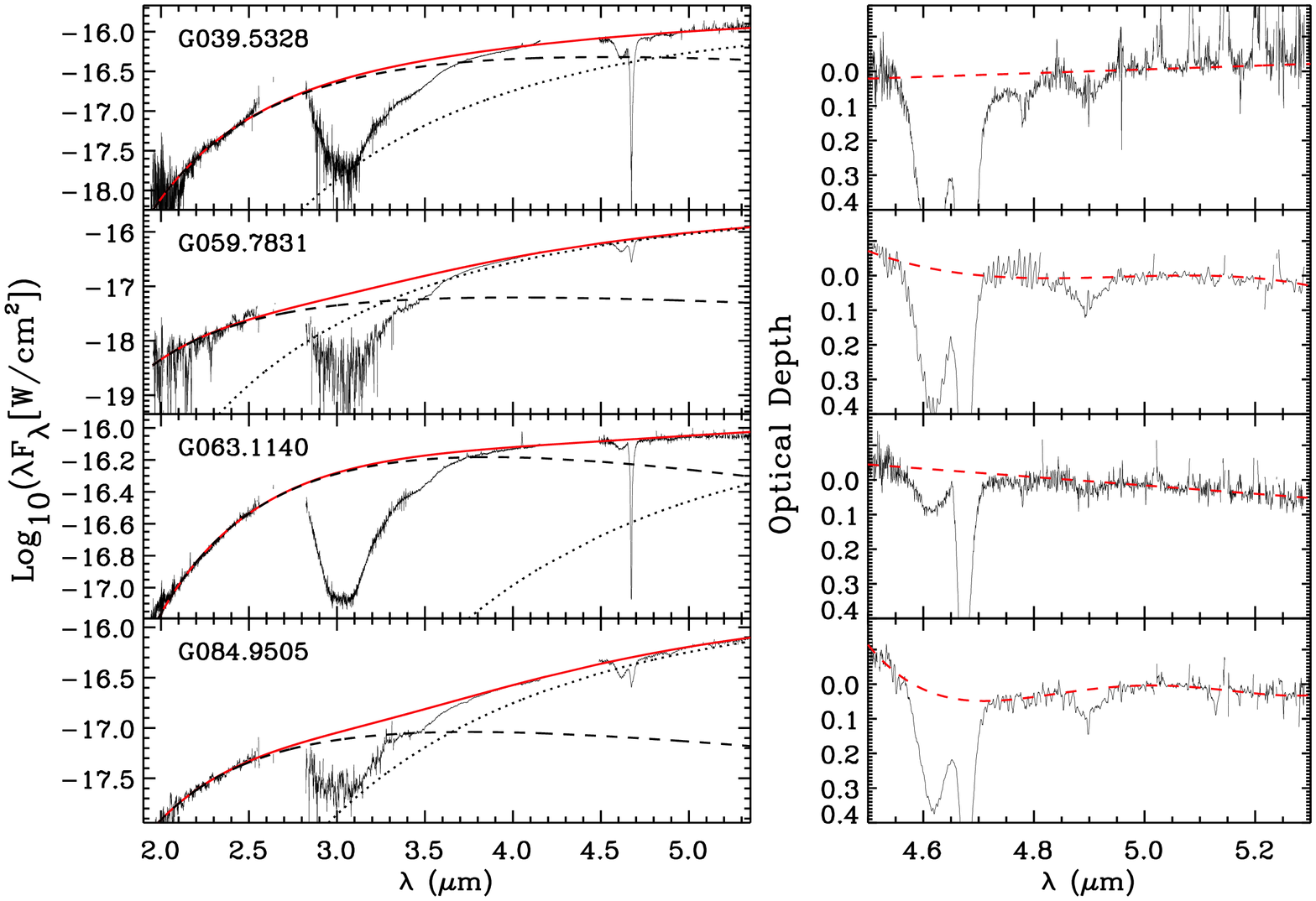}  
    \caption{IRTF/SpeX spectra of MYSO sample. For more details, see
      the caption of Figure~\ref{f:flux1}.\label{f:flux2}}
\end{figure*}

\begin{figure*}[p]
  \includegraphics[width=18cm, angle=90, scale=0.54]{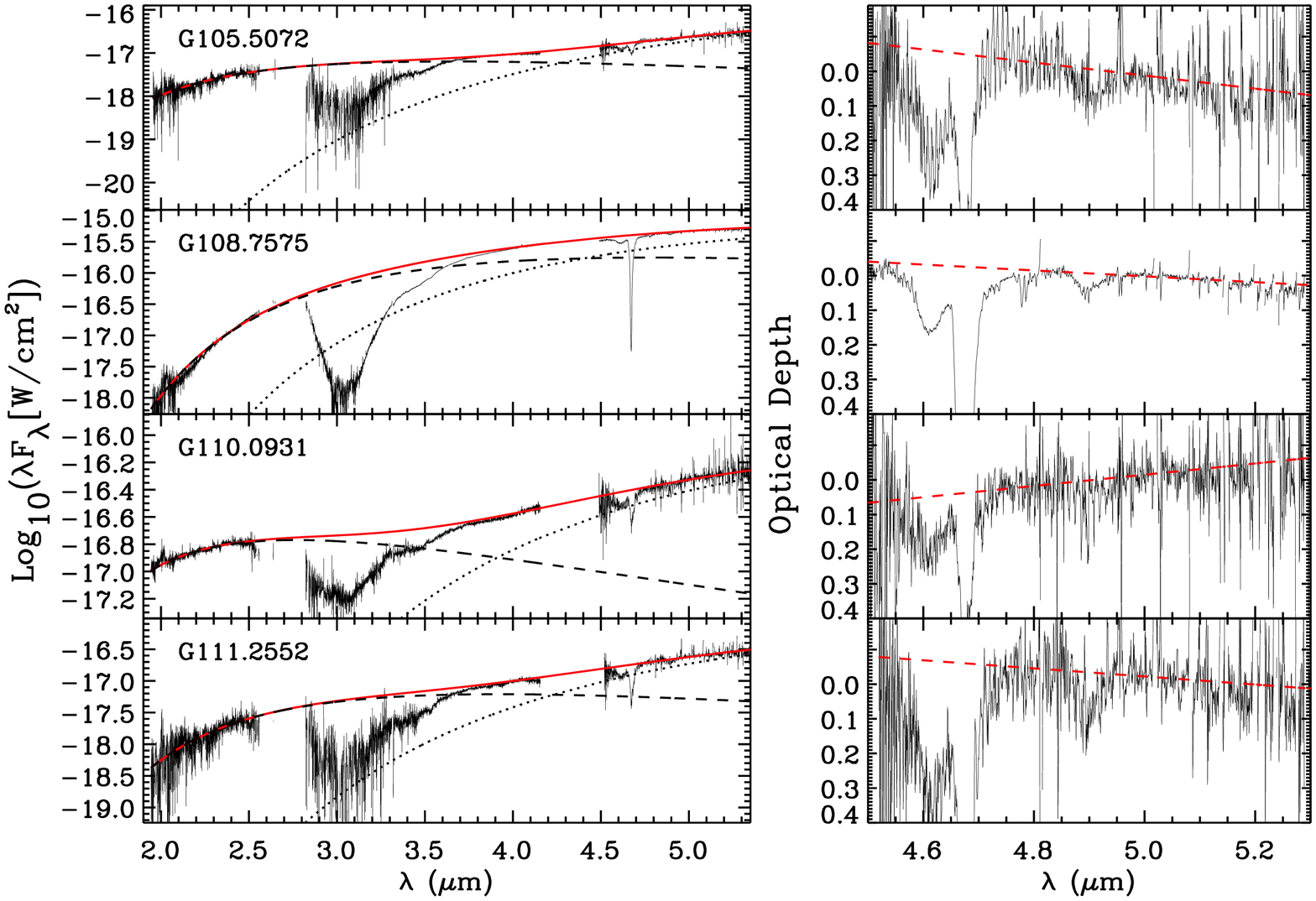}
  \includegraphics[width=18cm, angle=90, scale=0.42]{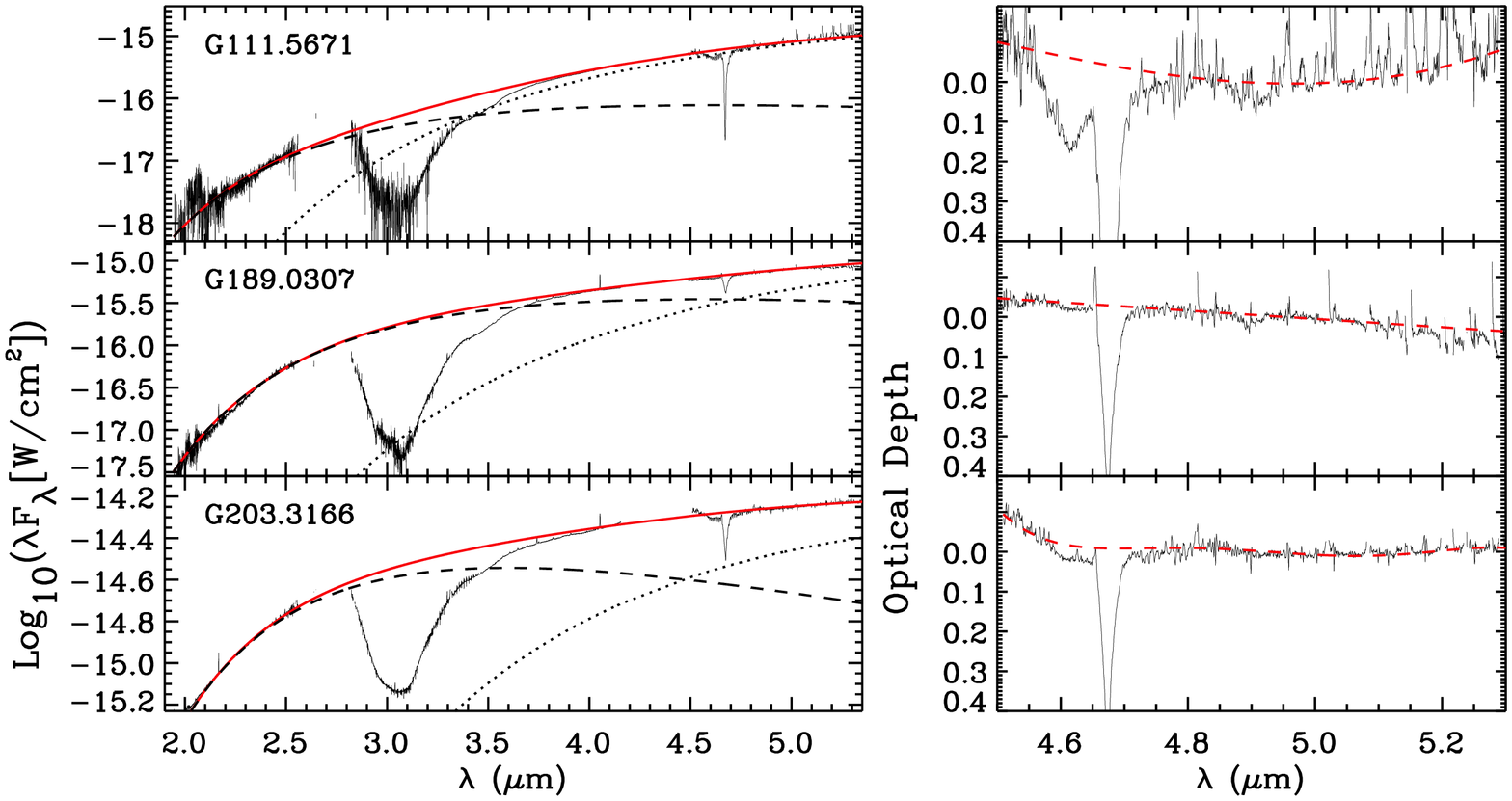}  
    \caption{IRTF/SpeX spectra of MYSO sample. For more details, see
      the caption of Figure~\ref{f:flux1}.\label{f:flux3}}
\end{figure*}

The observed MYSOs have steeply rising spectra in the 1.95-5.36
\mum\ wavelength range (Figs.~\ref{f:flux1}-\ref{f:flux3}). All have
deep absorption features centered on 3.0 \mum, covering the wavelength
range of 2.8 to $\sim$3.7 \mum. At the short wavelength side, the band
is cut off by the opaque Earth's atmosphere. The 3 \mum\ band is
mostly attributed to H$_2$O ice, with minor contributions from NH$_3$,
NH$_3$ hydrates, CH$_3$OH, and likely the O-H and C-H stretch mode of
other, yet unidentified, species.

\begin{figure*}[h!]
  \center
  \includegraphics[width=18cm, angle=90, scale=0.44]{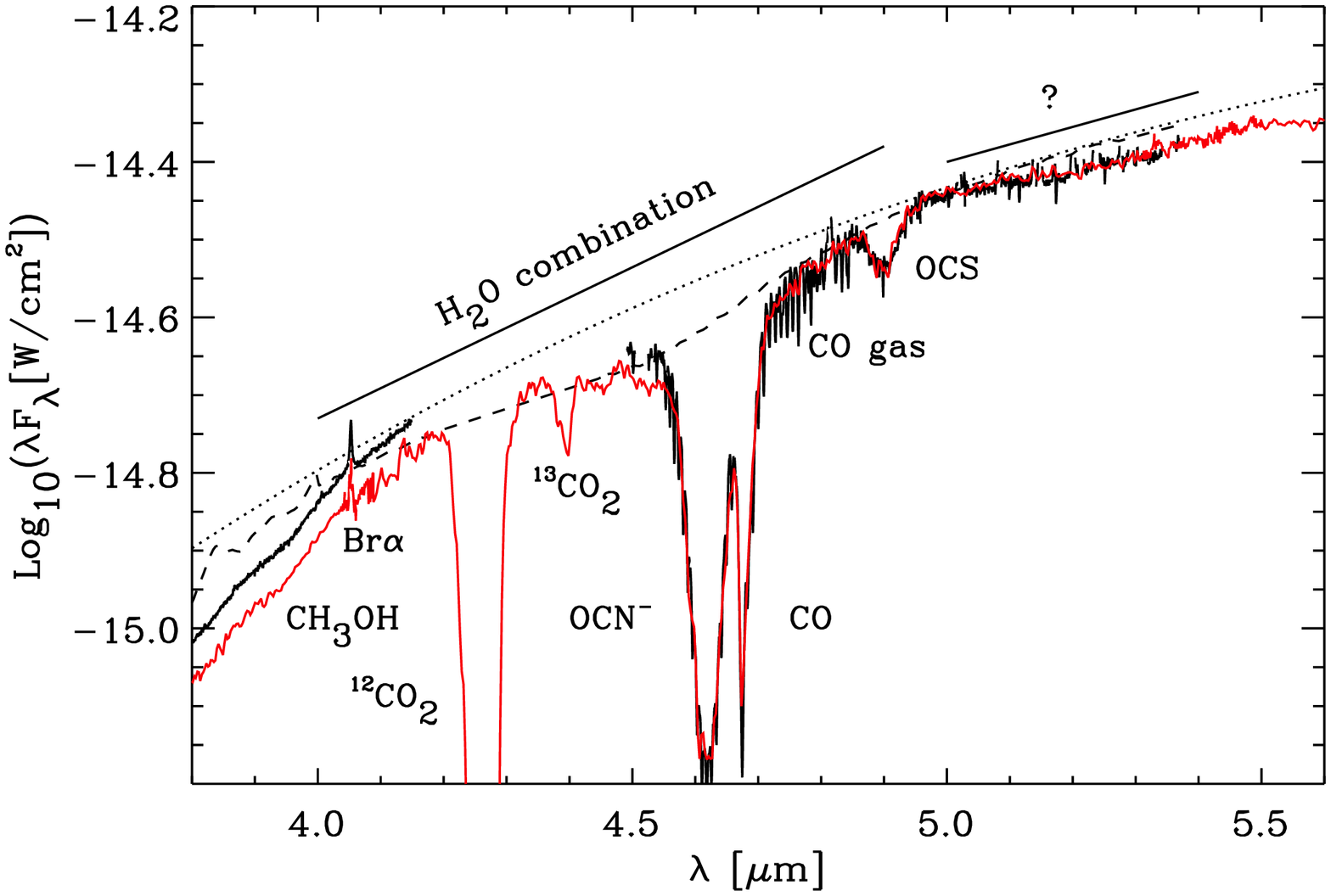}
  \caption{IRTF/SpeX spectrum of the MYSO G012.9090-00.2607 (W33A;
    black solid line) compared to a space based ISO/SWS spectrum
    across the full wavelength range (red; \citealt{gibb04}). The
    IRTF/SpeX and ISO/SWS spectra are overall in good agreement,
    showing the same absorption features and continuum shape. The
    dotted line indicates the continuum level described in
    \S\ref{sec:cont} and Figs.~\ref{f:flux1}-\ref{f:flux3}. The dashed
    line is the continuum level including absorption by H$_2$O ice,
    showing the effect of the broad $\sim$4.5 \mum\ H$_2$O combination
    mode. A change in slope is visible near $\sim$5.0 \mum\ in both
    the IRTF/SpeX and ISO/SWS spectra, due to what appears to be a
    shallow absorption feature centered on $\sim$5.25 \mum. The change
    in slope at $\sim$ 5.5 \mum\ is due to the onset of the prominent
    6.0 \mum\ absorption band. The apparent lack of absorption by gas
    phase CO lines in the ISO/SWS spectrum is due to its low spectral
    resolving power ($R=800$ versus $R=$1,500).\label{f:features}}
\end{figure*}

The $M$-band spectra show relatively narrow absortion features due to
CO (4.67 \mum), OCN$^-$ (4.62 \mum), and OCS (4.90 \mum).  Most MYSOs
also show the presence of the shallow, broad absorption by the
combination mode of H$_2$O ice. In addition, a number of targets show
a weak change in slope at $\sim 5.0$ \mum. This seems to be due to a
new, previously unreported absorption feature centered at $\sim$5.25
\mum. Figure \ref{f:features} illustrates the presence of all these
spectral features by putting them in a wider context using an Infrared
Space Observatory/Short Wavelength Spectrometer (ISO/SWS) spectrum
\citep{gibb04} for comparison. Many targets also show the unresolved
absorption lines of the ro-vibrational transitions of gas phase
CO. For a few targets, these lines are in emission.

Finally, a number of MYSOs show unresolved emission lines due to
hydrogen Br$\gamma$, Br$\alpha$, and Pf$\beta$. These are not analyzed
here, and their wavelength ranges are excluded from the continuum and
ice absorption feature analysis.

\subsection{Continuum Determination}~\label{sec:cont}

To be able to analyze the profile and depth of the ice absorption
features, a global baseline defining the continuum emission was
determined over the full observed wavelength range of 1.95-5.36 \mum.
This baseline was determined by fitting a simple model of two reddened
blackbody functions to the data. The first blackbody is that from a
hot central source. Its temperature is assumed to be at least several
tens of thousand Kelvin, so that it is represented by a
temperature-independent Rayleigh-Jeans tail in the observed wavelength
range. The second blackbody represents the emission by warm dust along
the line of sight, perhaps from a circumstellar disk or the inner
envelope. Both blackbodies are reddened by the same foreground
extinction due to cooler envelope and foreground cloud material. The
extinction curve used for this is that derived for the {\it J}, {\it
  H}, and {\it K}-bands by \citet{indebetouw05}, and at longer
wavelengths it is assumed to be flat. Furthermore, the 3.0 \mum\ ice
band was also included in the model, because H$_2$O absorption covers
such a large portion of the observed wavelength range. For this, we
used template 3.0 \mum\ band spectra derived from previously well
studied MYSOs \citep{gibb04}, covering a range of profiles (AFGL 2136,
MonR2 IRS3, AFGL 989, NGC 7538 IRS9, S140 IRS1). This model thus
contains four variables: the temperature of the warm dust emission,
the foreground extinction $A_{\rm V}$, the peak optical depth of the
3.0 \mum\ ice band $\tau_{3.0}$, and a relative flux scaling factor
for the hot (Rayleigh-Jeans) and warm blackbody emission. This model
is then normalized to the observed spectrum near 5 \mum\ and fitted to
selected wavelength ranges. Those ranges avoid the long-wavelength
wing of the 3.0 \mum\ ice band ($\sim$ 3.2-3.9 \mum), the OCN$^-$, CO,
and OCS ice features (4.5-4.9 \mum), as well as regions with poor
atmospheric transmission.

The global baseline model generally fits the observations well
(Figs.~\ref{f:flux1}-\ref{f:flux3}). The spectra were subsequently
converted to optical depth scale. At a detailed level in the $M$-band
optical depth spectra, small offsets from the zero optical depth level
were observed, however, and these were removed by subtracting an
additional local straight line fit, shown in
Figs.~\ref{f:flux1}-\ref{f:flux3}.

Because it is our primary goal to determine the baseline for the ice
features, we do not present or further interpret the parameters and
their uncertainties for the blackbody fit parameters. Due to the small
width of the ice features in the $M$-band, and the subtraction of an
additional local baseline, the uncertainty due to the baseline is
small. An error of 0.01 is included in the uncertainties derived for
the ice band optical depths. Uncertainties for the 3.0 \mum\ H$_2$O
ice band optical depths are larger ($\sim 10\%$).  This was checked by
using a different method to determine the baseline for the 3.0 $\mu$m
band, applying a spline function (K. Emerson et al., in
preparation). These uncertainties are included in the derived H$_2$O
ice column densities (\S\ref{sec:h2o}).

\subsection{Gas Phase CO}~\label{sec:gas}

\begin{figure*}[h!]
  \includegraphics[width=18cm, angle=90,
    scale=0.44]{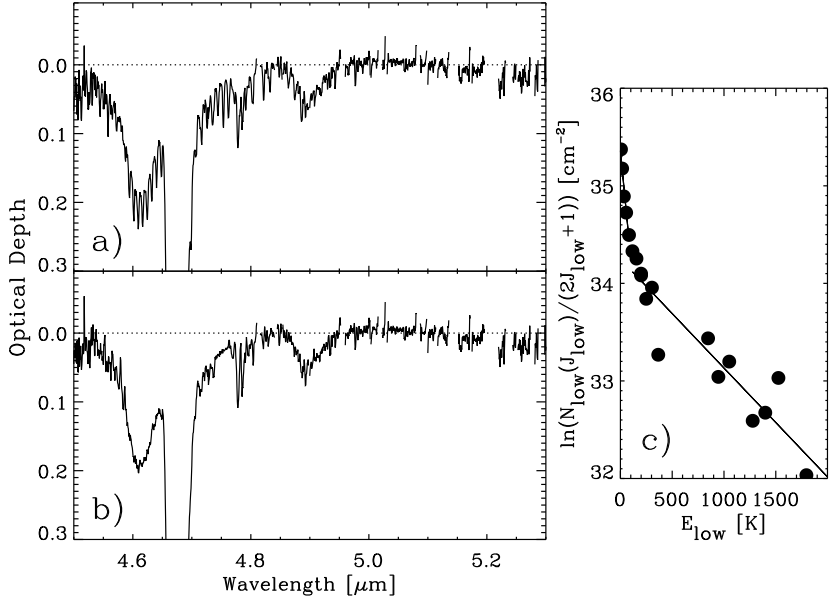}
  \caption{Illustration of the removal of gas phase $^{12}$CO
    ro-vibrational absorption lines for the MYSO
    G108.7575$-$00.9863. {\bf Panel a} is the IRTF/SpeX spectrum
    before removal, and {\bf panel b} after. {\bf Panel c} shows the
    rotation diagram used in this process, with $E_{\rm low}$, $N_{\rm
      low}$, and $J_{\rm low}$ the energy level, column density, and
    rotational quantum number for the lower energy levels,
    respectively. Two slope regimes are discerned, which are fitted
    with two straight lines, corresponding to excitation temperatures
    of 90 and 895 K and column densities of 7.5$\times 10^{16}$ and
    2.5$\times 10^{17}$ \sqcm, respectively. These temperatures are
    likely highly inaccurate, and the column densities strongly
    underestimated due to the lines being optically
    thick. \label{f:gas}}
\end{figure*}

Many targets show absorption, and some emission, by the ro-vibrational
lines of gas phase $^{12}$CO, and sometimes $^{13}$CO. These lines are
unresolved at the IRTF/SpeX resolving power of $R=1,500-2,500$. At the
level of a few percent of the continuum, these lines somewhat affect
the profiles of the ice absorption features. We therefore attempt to
remove them using a rotation diagram analysis
(Fig.~\ref{f:gas}). Integrated absorption line depths were measured,
and converted to lower level column densities assuming optically thin
absorption in Local Thermodynamic Equilibrium (LTE).  The constructed
rotation diagrams typically show a steep linear relation at low energy
levels, and a shallower linear relation at higher energy levels.
Straight lines were fitted to each of these, providing integrated line
depths for all rotational levels. These were then used to divide out
the gas phase CO lines. The detected $^{12}$CO lines are most likely
optically thick, and we do not further use the column densities and
excitation temperatures, derived from the rotation diagrams in the
optically thin limit, in this paper.

\subsection{Ice Band Profiles and Column Densities}~\label{sec:prof}

\begin{figure*}[p]
  \includegraphics[width=18cm, angle=90, scale=0.56]{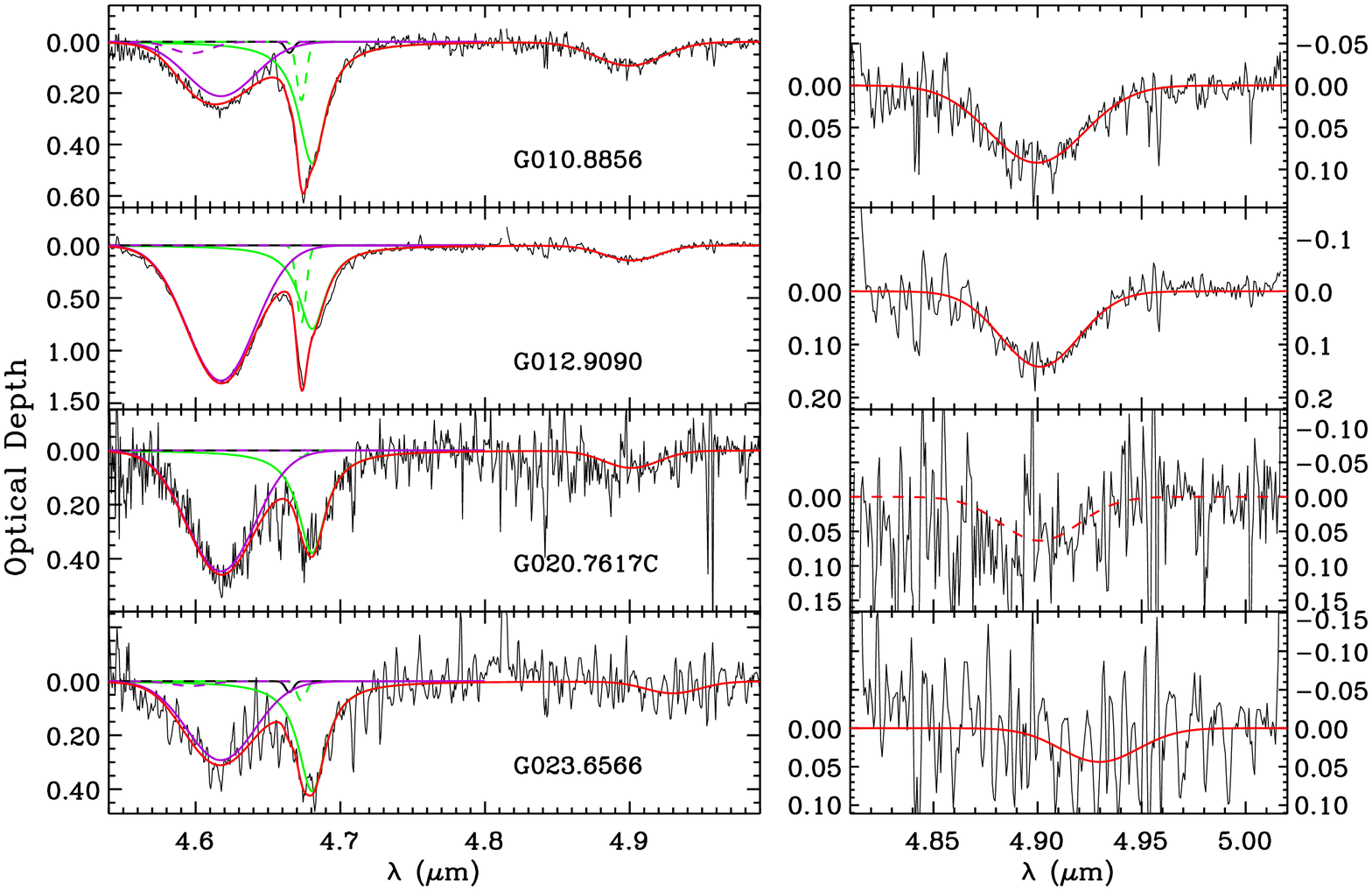}
  \includegraphics[width=18cm, angle=90, scale=0.56]{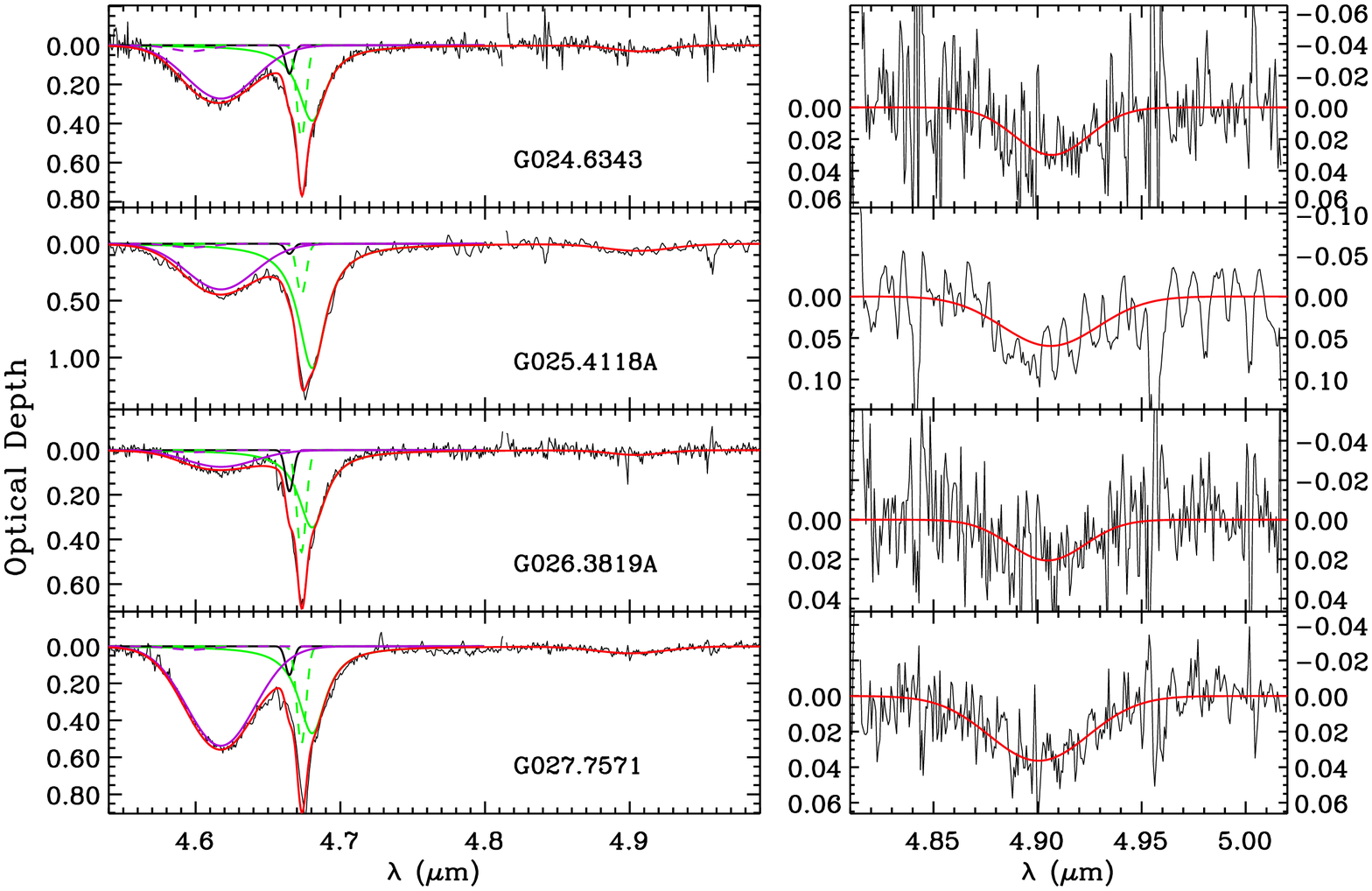}  
  \caption{$M$-band spectra of the MYSO sample on optical depth
    scale. A zoom in on the OCS ice band is shown in the right
    panels. The OCS band is fitted with a single Gaussian. To this are
    added the Gaussian and Lorentzian fits to the CO ice bands (green
    and black, respectively) and Gausian fits to the 2165.7 and 2175.4
    \waven\ components of the OCN$^-$ band (solid purple and dashed
    purple, respectively). A dashed red line in the right panels
    indicates that OCS was not detected and the Gaussian depth should
    be regarded as an upper limit.\label{f:mtau1}}
\end{figure*}

\begin{figure*}[p]
  \includegraphics[width=18cm, angle=90, scale=0.56]{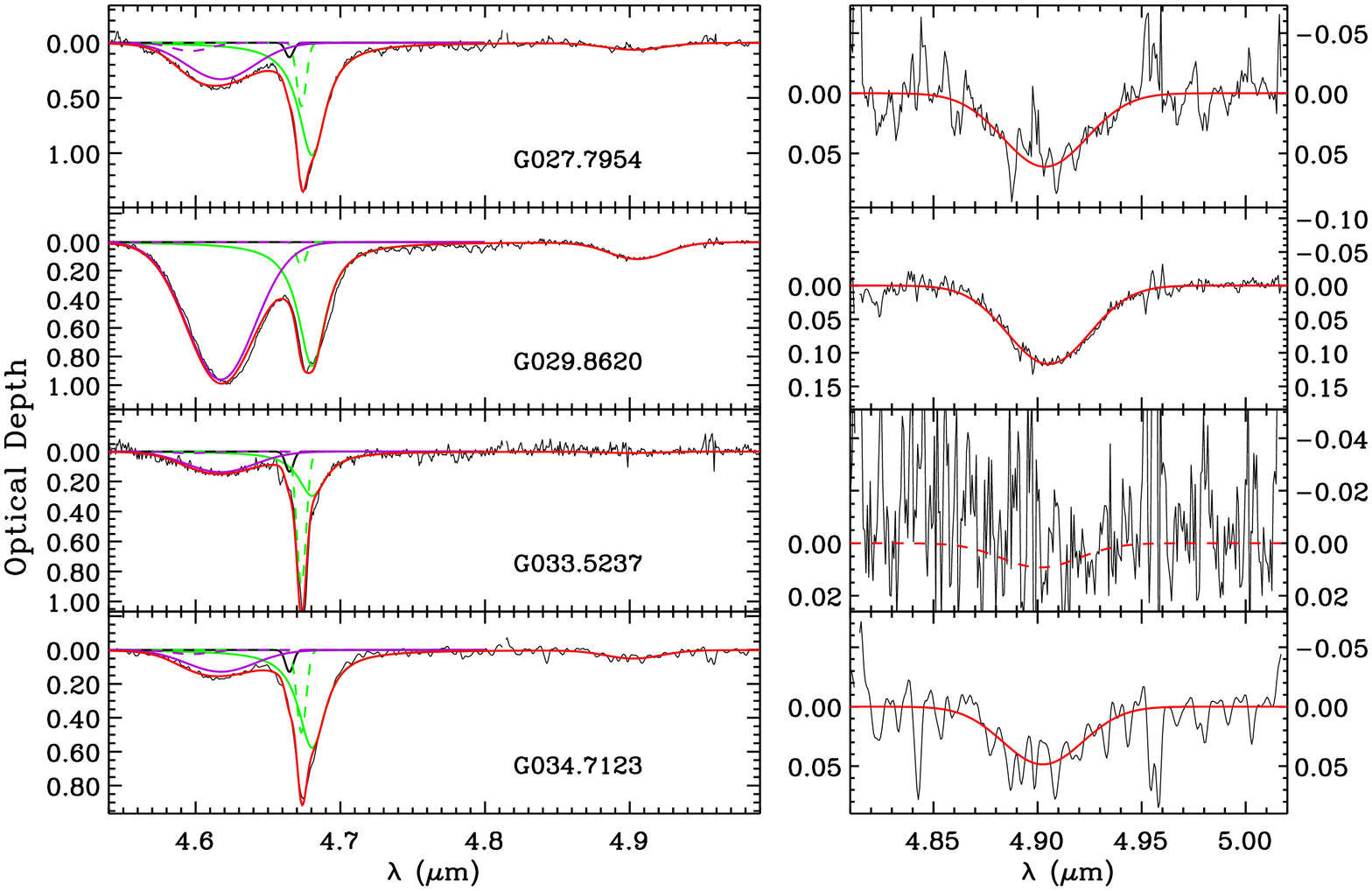}
  \includegraphics[width=18cm, angle=90, scale=0.56]{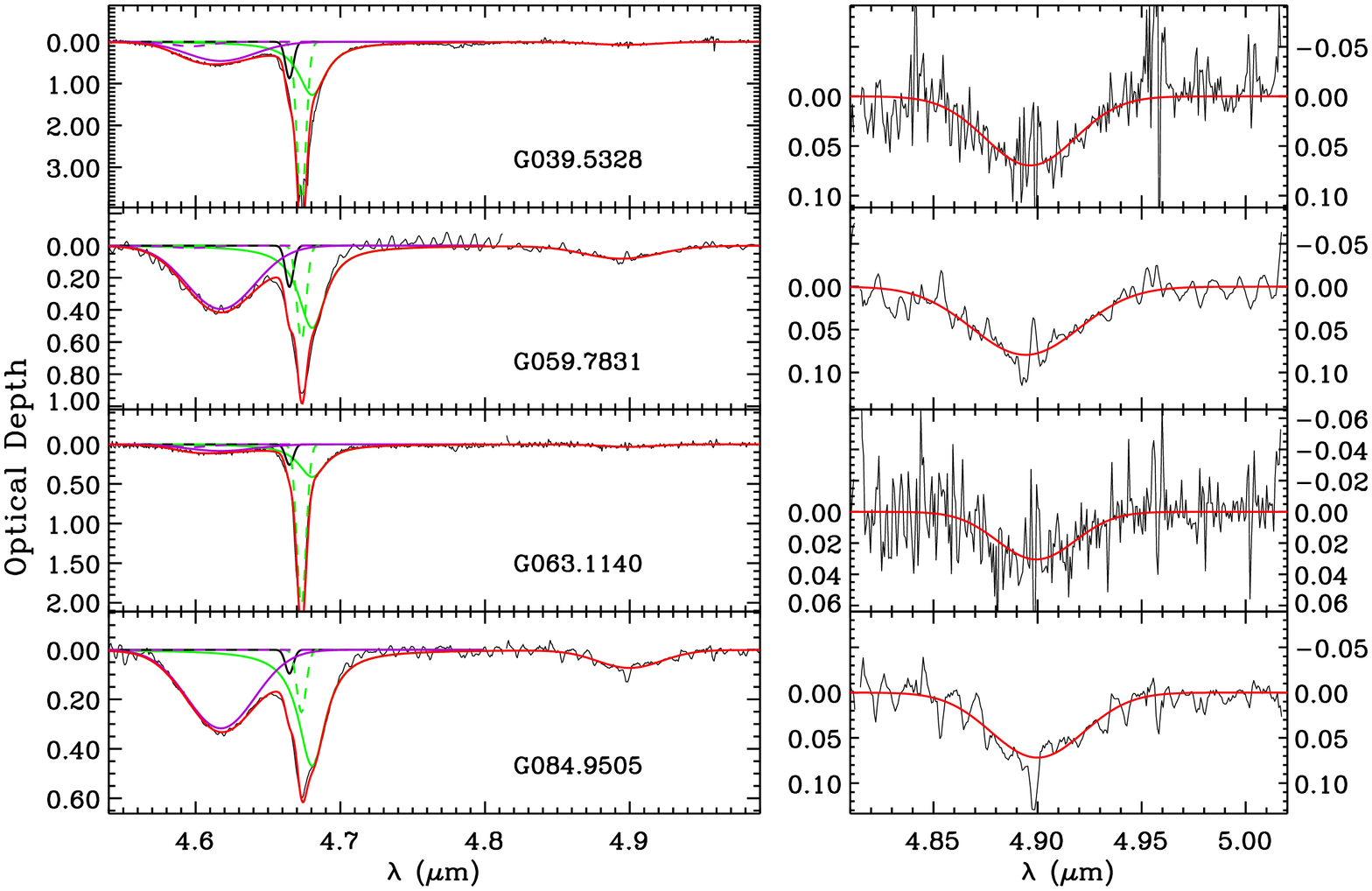}  
  \caption{$M$-band spectra of the MYSO sample on optical depth
    scale. For more details, see the caption of
    Figure~\ref{f:mtau1}.\label{f:mtau2}}
\end{figure*}

\begin{figure*}[p]
  \includegraphics[width=18cm, angle=90, scale=0.56]{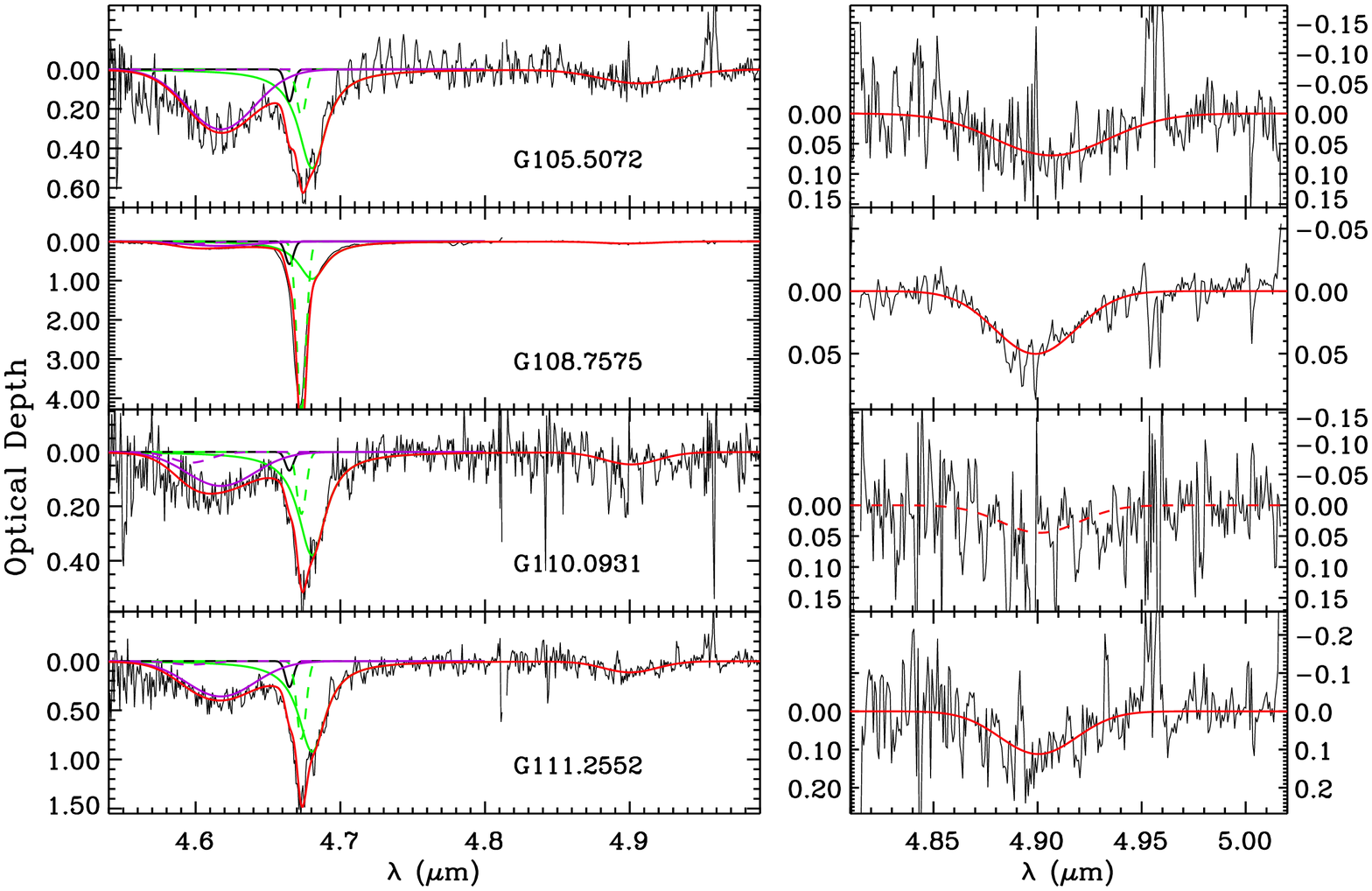}
  \includegraphics[width=18cm, angle=90, scale=0.435]{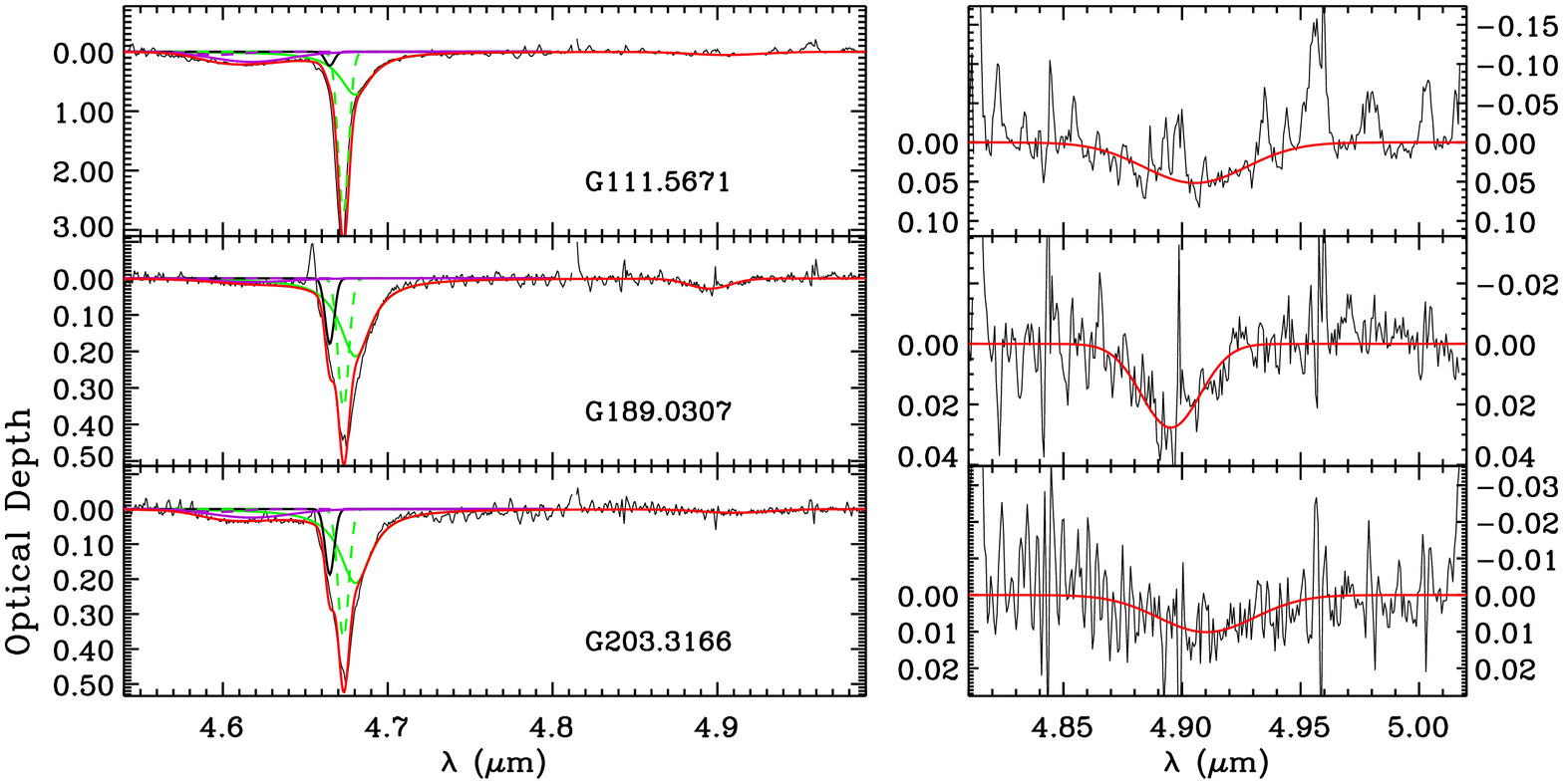}  
  \caption{$M$-band spectra of the MYSO sample on optical depth
    scale. For more details, see the caption of
    Figure~\ref{f:mtau1}.\label{f:mtau3}}
\end{figure*}

The absorption features detected at $\sim$4.62, 4.67, and 4.90
\mum\ are shown in Figures~\ref{f:mtau1}-\ref{f:mtau3} on optical
depth scale. They are ascribed to OCN$^-$, CO, and OCS, respectively,
and are fitted with Gaussian and Lorentzian functions to characterize
their profiles and depths. To put the results in an astrochemical
context, the 3.0 and 3.53 \mum\ absorption bands of H$_2$O and
CH$_3$OH ice are discussed as well. However, for a more detailed
analysis of these two bands, in a much larger sample of MYSOs, we
refer to K. Emerson et al. (in preparation).

\subsubsection{CO}~\label{sec:profco}

The 4.67 \mum\ absorption band of CO ice shows significant profile
variations (Figs.~\ref{f:mtau1}-\ref{f:mtau3}). For some targets the
profile is much narrower (e.g., G063.1140) than for others (e.g.,
G025.4118A). Such variations were previously observed for both high
and low mass YSOs, and were ascribed to the presence of CO in
different molecular environments \citep{tielens91, boogert02,
  pontoppidan03}: a central, narrow absorption due to pure CO (and
traces of other low dipole moment molecules), a broad red-shifted
absorption due to CO mixed with high dipole moment species (H$_2$O,
CH$_3$OH), and a minor blue-shifted absorption, likely due to CO mixed
wth CO$_2$ ice.  By fitting constant profiles for CO$_{\rm apolar}$,
CO$_{\rm polar}$, and CO$_{\rm blue}$, and only varying the peak
optical depths, good fits are obtained across a wide range of low and
high mass YSOs as well as background stars \citep{pontoppidan03}. The
same approach is adopted for the MYSOs in this work. CO$_{\rm apolar}$
was fitted using a Gaussian function (peak wavelength 4.6731 \mum,
FWHM=0.0076 \mum), CO$_{\rm polar}$ with a Lorentzian (4.6806 and
0.0232 \mum), and CO$_{\rm blue}$ with a Gaussian (4.6648 and 0.0065
\mum).  The fitting process was done simultaneously with the 4.62
\mum\ band of OCN$^-$ (\S\ref{sec:profocn}), due to the wavelength
overlap.  Also, all components were convolved with the Gaussian SpeX
instrument profile before fitting them to the data.  Satisfactory fits
are obtained for all bands (Figs.~\ref{f:mtau1}-\ref{f:mtau3}). The
total CO ice column densities and those of the different components
are listed in Table~\ref{t:h2oco}. An integrated band strength of
1.1$\times 10^{-17}$ cm/molecule, as measured in the laboratory
\citep{jiang75}, was used for this.

%
%
\begin{deluxetable}{lllllll}
\tabletypesize{\scriptsize}
\tablecolumns{7}
\tablewidth{0pc}
\tablecaption{H$_2$O and CO Ice Column Densities~\label{t:h2oco}}
\tablehead{
\colhead{Source}       & \colhead{$N$(H$_2$O)\tablenotemark{a}} & \multicolumn{2}{c}{$N$(CO)}                                        & \colhead{$N$(CO$_{\rm apolar}$)} & \colhead{$N$(CO$_{\rm polar}$)} & \colhead{$N$(CO$_{\rm blue}$)} \\
\colhead{MSX}          & \colhead{$10^{18}$ cm$^{-2}$}            & \colhead{$10^{17}$ cm$^{-2}$} & \colhead{\% H$_2$O\tablenotemark{a}} & \colhead{$10^{17}$ cm$^{-2}$}   & \colhead{$10^{17}$ cm$^{-2}$}  & \colhead{$10^{17}$ cm$^{-2}$} \\
}
\startdata
 G010.8856$+$00.1221 &  1.75 (0.10) &  5.40 (0.27) & 30.86 (2.29) &  0.72 (0.11) &  4.56 (0.22) &  0.12 (0.10) \\
 G012.9090$-$00.2607 & 11.50 (1.29) &  9.97 (0.40) &  8.67 (1.03) &  2.32 (0.17) &  7.65 (0.33) &  0.00 (0.15) \\
G020.7617$-$00.0638C & $\sim$ 1.15 &  3.73 (0.55) & $\sim$32.43 &  0.08 (0.23) &  3.65 (0.44) &  0.00 (0.22) \\
 G023.6566$-$00.1273 &  1.76 (0.15) &  4.27 (0.51) & 24.26 (3.52) &  0.23 (0.22) &  3.92 (0.42) &  0.11 (0.20) \\
 G024.6343$-$00.3233 &  2.69 (0.18) &  5.62 (0.26) & 20.89 (1.69) &  1.50 (0.11) &  3.72 (0.22) &  0.40 (0.09) \\
G025.4118$+$00.1052A &  4.88 (0.32) & 12.20 (0.32) & 25.00 (1.78) &  1.43 (0.13) & 10.50 (0.27) &  0.24 (0.12) \\
G026.3819$+$01.4057A &  3.62 (0.32) &  5.29 (0.27) & 14.61 (1.50) &  1.45 (0.11) &  3.34 (0.22) &  0.50 (0.10) \\
 G027.7571$+$00.0500 &  4.22 (0.42) &  6.63 (0.30) & 15.71 (1.73) &  1.68 (0.12) &  4.53 (0.25) &  0.42 (0.11) \\
 G027.7954$-$00.2772 &  4.32 (0.28) & 12.00 (0.30) & 27.78 (1.94) &  1.83 (0.12) &  9.82 (0.25) &  0.36 (0.11) \\
 G029.8620$-$00.0444 &  4.08 (0.82) &  8.88 (0.29) & 21.76 (4.41) &  0.54 (0.12) &  8.34 (0.25) &  0.00 (0.11) \\
 G033.5237$+$00.0198 &  0.76 (0.08) &  6.04 (0.31) & 79.68 (9.39) &  2.82 (0.13) &  2.85 (0.26) &  0.36 (0.11) \\
 G034.7123$-$00.5946 &  4.22 (0.24) &  7.46 (0.27) & 17.68 (1.20) &  1.55 (0.11) &  5.56 (0.23) &  0.35 (0.10) \\
 G039.5328$-$00.1969 &  4.26 (0.40) & 26.70 (1.77) & 62.68 (7.25) & 12.10 (0.77) & 12.30 (1.42) &  2.37 (0.72) \\
 G059.7831$+$00.0648 &  5.81 (1.29) &  7.51 (0.38) & 12.93 (2.94) &  1.87 (0.15) &  4.94 (0.31) &  0.70 (0.15) \\
 G063.1140$+$00.3416 &  3.02 (0.18) & 11.10 (0.32) & 36.75 (2.41) &  6.43 (0.13) &  3.97 (0.27) &  0.71 (0.12) \\
 G084.9505$-$00.6910 &  2.32 (0.40) &  5.58 (0.23) & 24.05 (4.30) &  0.80 (0.09) &  4.52 (0.19) &  0.26 (0.08) \\
 G105.5072$+$00.2294 &  4.44 (1.13) &  6.02 (0.54) & 13.56 (3.66) &  0.75 (0.22) &  4.83 (0.44) &  0.44 (0.21) \\
 G108.7575$-$00.9863 &  6.28 (0.48) & 24.50 (1.24) & 39.01 (3.60) & 13.70 (0.56) &  9.23 (0.99) &  1.58 (0.50) \\
 G110.0931$-$00.0641 &  1.73 (0.24) &  4.57 (0.42) & 26.42 (4.42) &  0.72 (0.18) &  3.66 (0.34) &  0.19 (0.16) \\
 G111.2552$-$00.7702 & $\sim$ 4.56 & 12.10 (0.68) & $\sim$26.54 &  2.51 (0.28) &  8.88 (0.55) &  0.72 (0.28) \\
 G111.5671$+$00.7517 &  5.37 (0.48) & 16.20 (0.36) & 30.17 (2.80) &  8.60 (0.15) &  6.97 (0.30) &  0.63 (0.13) \\
 G189.0307$+$00.7821 &  5.46 (0.48) &  3.68 (0.31) &  6.74 (0.82) &  1.13 (0.13) &  2.06 (0.26) &  0.49 (0.11) \\
 G203.3166$+$02.0564 &  2.24 (0.11) &  3.72 (0.23) & 16.61 (1.32) &  1.18 (0.09) &  2.04 (0.20) &  0.51 (0.08) \\
\enddata
\tablecomments{Values in brackets indicate 1$\sigma$ uncertainties.}
\tablenotetext{a}{Values preceded by $\sim$ refer to targets with
  highly uncertain $N$(H$_2$O) values, due to seemingly saturated 3.0
  \mum\ absorption bands.}
\end{deluxetable}

\clearpage

\subsubsection{OCN$^-$}~\label{sec:profocn}

The 4.62 \mum\ band of OCN$^-$ was detected toward all targets. In a
study of low and high mass YSOs, it was found that this band consists
of two components, centered on 2165.7 and 2175.4 \waven\ (4.6174 and
4.5969 \mum; \citealt{vanbroekhuizen05}). The 2165.7 \waven\ component
was found to be consistent with solid OCN$^-$ in polar environments,
while the other component relates perhaps to CO bonding to the grain
surfaces or to OCN$^-$ in apolar environments \citep{oberg11}. We
adopt the same decomposition procedure, by fitting two Gaussians, one
at 2165.7 \waven\ ($FWHM=$26 \waven) and one at 2175.4
\waven\ ($FWHM=$15 \waven). The fitting was done simultaneously with
those of the CO ice feature (\S\ref{sec:profco}), since they
overlap. Good quality fits were obtained
(Figs.~\ref{f:mtau1}-\ref{f:mtau3}). The contribution by the 2175.4
\waven\ component is usually at the less than 2$\sigma$ confidence
level, except for three targets at 2-3$\sigma$, and three targets at
$>3\sigma$. The OCN$^-$ column densities are listed in
Table~\ref{t:ocnch3oh}. An integrated band strength of 1.3$\times
10^{-16}$ cm/molecule, as measured in the laboratory
\citep{vanbroekhuizen04}, was used for this.

%
%
\begin{deluxetable}{lllllll}
\tablecolumns{7}
\tablewidth{0pc}
\tablecaption{OCN$^-$ and CH$_3$OH Ice Column Densities~\label{t:ocnch3oh}}
\tablehead{
\colhead{Source}  & \multicolumn{2}{c}{$N$(OCN$^-_{\rm red}$)}                            & \multicolumn{2}{c}{$N$(OCN$^-_{\rm blue}$)\tablenotemark{b}}            & \multicolumn{2}{c}{$N$(CH$_3$OH)}                 \\      
\colhead{MSX}     & \colhead{$10^{16}$ cm$^{-2}$} & \colhead{\% H$_2$O\tablenotemark{a}}  & \colhead{$10^{16}$ cm$^{-2}$} & \colhead{\% H$_2$O\tablenotemark{a}}   & \colhead{$10^{16}$ cm$^{-2}$} & \colhead{\% H$_2$O\tablenotemark{a}} \\     
}
\startdata
 G010.8856$+$00.1221 &  4.24 (0.35) &  2.42 (0.24) &  0.52 (0.23) &  0.30 (0.13) &  60.2 (13.5) &  34.4 ( 7.9) \\
 G012.9090$-$00.2607 & 25.70 (0.37) &  2.23 (0.25) &  0.00 (0.24) &  0.00 (0.02) & 235.0 (52.6) &  20.4 ( 5.1) \\
G020.7617$-$00.0638C &  8.93 (0.67) & $\sim$ 7.77 &  0.00 (0.47) & $\sim$ 0.00 & $<$115.0 & $<$100.0 \\
 G023.6566$-$00.1273 &  5.85 (0.63) &  3.32 (0.45) &  0.20 (0.44) &  0.11 (0.25) &  37.3 ( 8.3) &  21.2 ( 5.0) \\
 G024.6343$-$00.3233 &  5.45 (0.36) &  2.03 (0.19) &  0.35 (0.24) &  0.13 (0.09) & $<$ 22.9 & $<$  8.5 \\
G025.4118$+$00.1052A &  8.03 (0.41) &  1.65 (0.14) &  0.38 (0.28) &  0.08 (0.06) & 138.0 (30.8) &  28.3 ( 6.6) \\
G026.3819$+$01.4057A &  1.50 (0.31) &  0.41 (0.09) &  0.12 (0.20) &  0.03 (0.06) &  27.3 ( 6.1) &   7.5 ( 1.8) \\
 G027.7571$+$00.0500 & 10.80 (0.34) &  2.56 (0.27) &  0.22 (0.22) &  0.05 (0.05) & $<$ 15.5 & $<$  3.7 \\
 G027.7954$-$00.2772 &  6.61 (0.35) &  1.53 (0.13) &  0.86 (0.23) &  0.20 (0.06) &  87.1 (19.5) &  20.2 ( 4.7) \\
 G029.8620$-$00.0444 & 19.20 (0.35) &  4.71 (0.95) &  0.00 (0.23) &  0.00 (0.06) & 165.0 (36.8) &  40.4 (12.1) \\
 G033.5237$+$00.0198 &  2.75 (0.34) &  3.63 (0.59) &  0.12 (0.22) &  0.16 (0.29) & $<$ 11.5 & $<$ 15.2 \\
 G034.7123$-$00.5946 &  2.56 (0.29) &  0.61 (0.08) &  0.28 (0.18) &  0.07 (0.04) & 110.0 (24.7) &  26.1 ( 6.0) \\
 G039.5328$-$00.1969 &  9.19 (0.36) &  2.16 (0.22) &  1.24 (0.23) &  0.29 (0.06) &  32.0 ( 7.1) &   7.5 ( 1.8) \\
 G059.7831$+$00.0648 &  7.89 (0.34) &  1.36 (0.31) &  0.16 (0.22) &  0.03 (0.04) & 137.0 (30.5) &  23.6 ( 7.4) \\
 G063.1140$+$00.3416 &  1.69 (0.30) &  0.56 (0.10) &  0.42 (0.19) &  0.14 (0.06) &  34.4 ( 7.7) &  11.4 ( 2.6) \\
 G084.9505$-$00.6910 &  6.34 (0.29) &  2.73 (0.49) &  0.00 (0.18) &  0.00 (0.08) & $<$ 28.7 & $<$ 12.4 \\
 G105.5072$+$00.2294 &  6.08 (0.78) &  1.37 (0.39) &  0.06 (0.56) &  0.01 (0.13) & $<$ 45.9 & $<$ 10.3 \\
 G108.7575$-$00.9863 &  2.33 (0.34) &  0.37 (0.06) &  0.79 (0.22) &  0.13 (0.04) &  64.9 (14.5) &  10.3 ( 2.4) \\
 G110.0931$-$00.0641 &  2.50 (0.76) &  1.45 (0.48) &  0.45 (0.54) &  0.26 (0.31) & $<$ 23.7 & $<$ 13.7 \\
 G111.2552$-$00.7702 &  7.19 (0.98) & $\sim$ 1.58 &  0.41 (0.71) & $\sim$ 0.09 & 149.0 (33.3) & $\sim$ 32.7 \\
 G111.5671$+$00.7517 &  3.41 (0.39) &  0.64 (0.09) &  0.60 (0.26) &  0.11 (0.05) &  43.0 ( 9.6) &   8.0 ( 1.9) \\
 G189.0307$+$00.7821 &  0.20 (0.25) &  0.04 (0.05) &  0.00 (0.15) &  0.00 (0.03) &  39.2 ( 8.8) &   7.2 ( 1.7) \\
 G203.3166$+$02.0564 &  0.49 (0.27) &  0.22 (0.12) &  0.10 (0.17) &  0.04 (0.07) & $<$ 14.3 & $<$  6.4 \\
\enddata
\tablecomments{Values in brackets indicate 1$\sigma$ uncertainties.}
\tablenotetext{a}{Values preceded by $\sim$ refer to targets with
  highly uncertain $N$(H$_2$O) values, due to seemingly saturated 3.0
  \mum\ absorption bands.}
\tablenotetext{b}{Assuming that this 2175.4 \waven\ (4.597 \mum) 'blue' component
  is indeed caused by OCN$^-$ absorption \citep{vanbroekhuizen05}.}
\end{deluxetable}

\subsubsection{OCS}~\label{sec:profocs}

\begin{figure*}[p]
  \includegraphics[width=18cm, angle=90, scale=0.56]{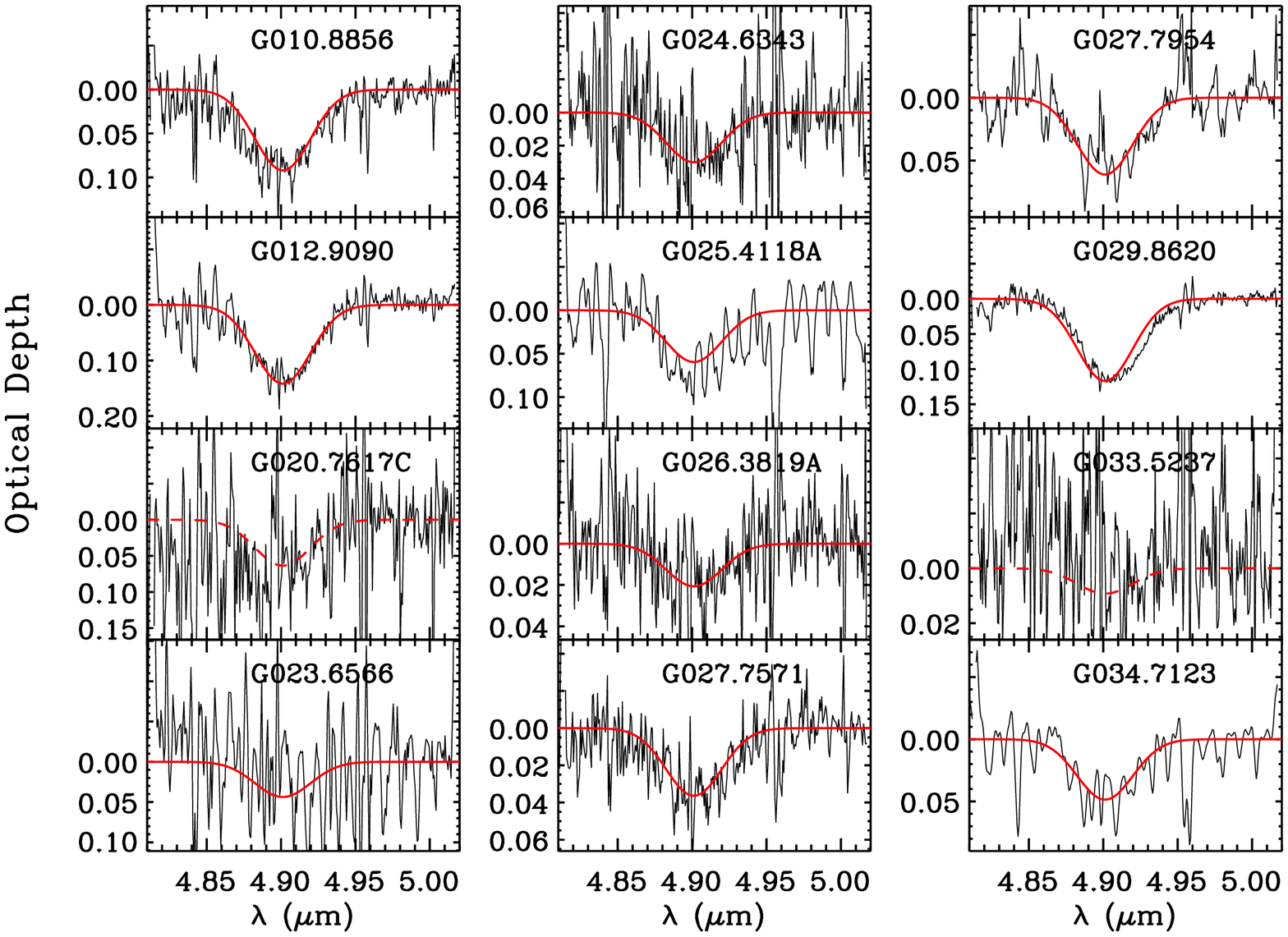}
  \includegraphics[width=18cm, angle=90, scale=0.56]{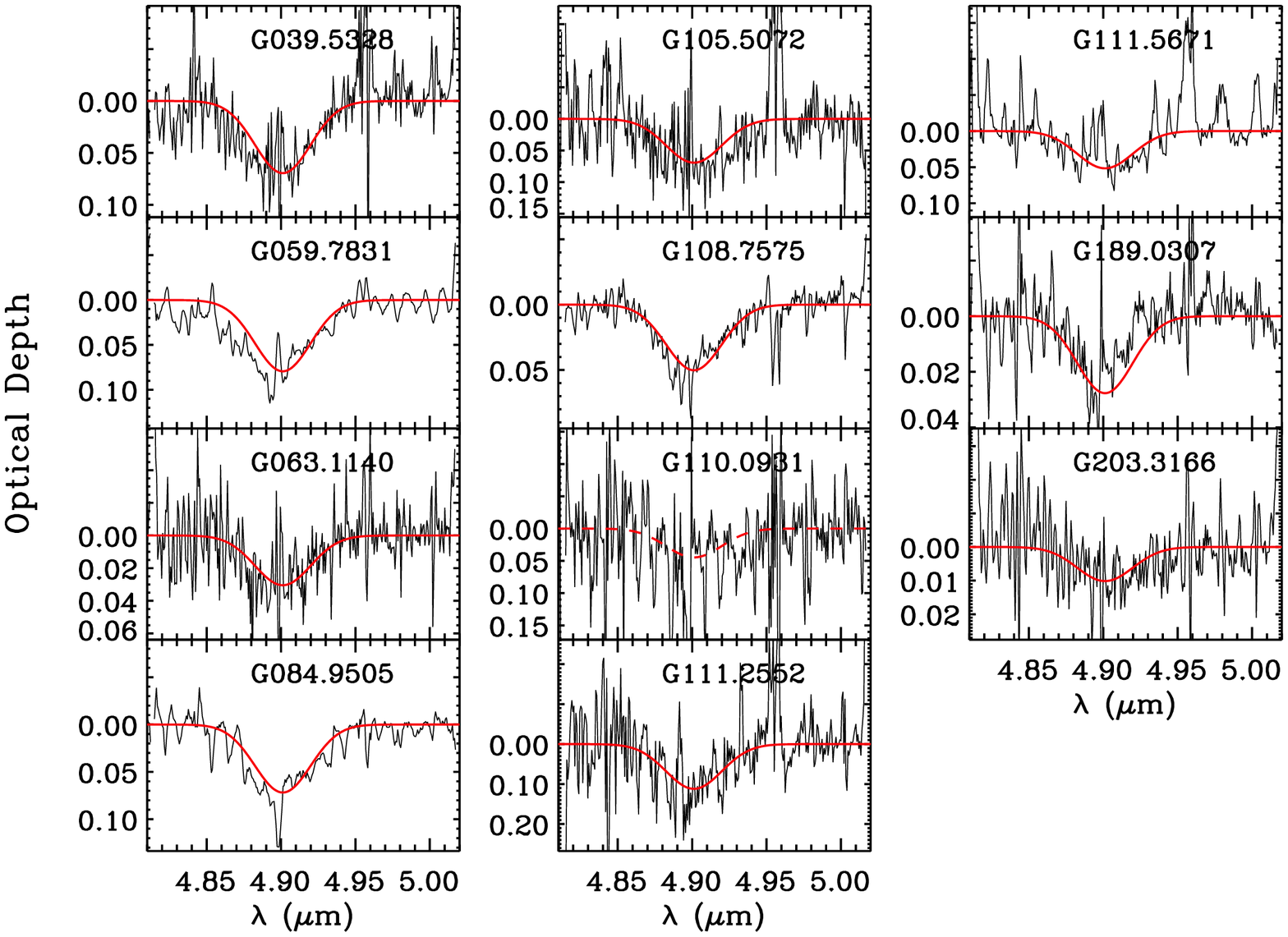}  
  \caption{The OCS ice band towards the sample of MYSOs compared to
    the same Gaussian, scaled to the best-fit peak optical depth of
    Table~\ref{t:ocs} in order to highlight any variations in the
    sample. The Gaussian represents the best fit to the target
    G012.9090 (W33A; $\lambda=$4.9012 \mum, FWHM=0.0441
    \mum).\label{f:ocsref}}
\end{figure*}

The absorption feature at 4.90 \mum, attributed to the C-O stretch
mode of solid OCS, is well fitted with a single Gaussian for all MYSOs
(Figs.~\ref{f:mtau1}-\ref{f:mtau3}). Upper limits are derived for
three of the targets.  The fit parameters are listed in
Table~\ref{t:ocs}.  Solid OCS column densities were derived using an
integrated band strength of 1.5$\times 10^{-16}$ cm/molecule, as
measured in the laboratory \citep{palumbo97}. In order to highlight
similarities and differences among the targets, each target is
compared to the same Gaussian in Figure \ref{f:ocsref}. Clearly, any
profile variations are small.

%
\begin{deluxetable}{llllllll}
\tablecolumns{6}
\tablewidth{0pc}
\tablecaption{OCS Band Parameters and Column Densities~\label{t:ocs}}
\tablehead{
\colhead{Source}       & \multicolumn{2}{c}{$\lambda_{\rm peak}$} & \multicolumn{2}{c}{FWHM}              & \colhead{$\tau_{\rm peak}$} & \multicolumn{2}{c}{$N$(OCS)}\\
\colhead{MSX}          & \colhead{$\mu$m}   & \colhead{cm$^{-1}$}& \colhead{$\mu$m}   & \colhead{cm$^{-1}$}&                          & \colhead{$10^{16}$ cm$^{-2}$}  & \colhead{\% H$_2$O\tablenotemark{a}}\\
}
\startdata
 G010.8856$+$00.1221 & 4.8996 (0.0008) & 2041.0 (0.4) & 0.0535 (0.0017) &  22.3 (0.7) & 0.092 (0.000) &  1.37 (0.04) &  0.78 (0.05) \\
 G012.9090$-$00.2607 & 4.9012 (0.0016) & 2040.3 (0.7) & 0.0441 (0.0034) &  18.4 (1.4) & 0.142 (0.002) &  1.74 (0.13) &  0.15 (0.02) \\
G020.7617$-$00.0638C & \dots & \dots & \dots & \dots & \dots & $<$ 0.78 & $<$ 0.68 \\
 G023.6566$-$00.1273 & 4.9297 (0.0072) & 2028.5 (3.0) & 0.0442 (0.0159) &  18.2 (6.6) & 0.044 (0.002) &  0.53 (0.19) &  0.30 (0.11) \\
 G024.6343$-$00.3233 & 4.9068 (0.0079) & 2038.0 (3.3) & 0.0417 (0.0163) &  17.3 (6.8) & 0.030 (0.002) &  0.35 (0.14) &  0.13 (0.05) \\
G025.4118$+$00.1052A & 4.9061 (0.0032) & 2038.3 (1.3) & 0.0542 (0.0069) &  22.5 (2.9) & 0.060 (0.001) &  0.89 (0.12) &  0.18 (0.03) \\
G026.3819$+$01.4057A & 4.9048 (0.0030) & 2038.8 (1.2) & 0.0426 (0.0060) &  17.7 (2.5) & 0.021 (0.000) &  0.24 (0.03) &  0.07 (0.01) \\
 G027.7571$+$00.0500 & 4.9005 (0.0038) & 2040.6 (1.6) & 0.0562 (0.0081) &  23.4 (3.4) & 0.036 (0.001) &  0.57 (0.08) &  0.13 (0.02) \\
 G027.7954$-$00.2772 & 4.9035 (0.0006) & 2039.4 (0.3) & 0.0475 (0.0013) &  19.8 (0.5) & 0.061 (0.000) &  0.80 (0.02) &  0.19 (0.01) \\
 G029.8620$-$00.0444 & 4.9053 (0.0003) & 2038.6 (0.1) & 0.0465 (0.0005) &  19.3 (0.2) & 0.117 (0.000) &  1.50 (0.02) &  0.37 (0.07) \\
 G033.5237$+$00.0198 & \dots & \dots & \dots & \dots & \dots & $<$ 0.11 & $<$ 0.15 \\
 G034.7123$-$00.5946 & 4.9025 (0.0024) & 2039.8 (1.0) & 0.0454 (0.0051) &  18.9 (2.1) & 0.049 (0.001) &  0.61 (0.07) &  0.14 (0.02) \\
 G039.5328$-$00.1969 & 4.8965 (0.0026) & 2042.3 (1.1) & 0.0508 (0.0054) &  21.2 (2.3) & 0.069 (0.001) &  0.98 (0.11) &  0.23 (0.03) \\
 G059.7831$+$00.0648 & 4.8943 (0.0012) & 2043.2 (0.5) & 0.0609 (0.0025) &  25.4 (1.1) & 0.079 (0.001) &  1.34 (0.06) &  0.23 (0.05) \\
 G063.1140$+$00.3416 & 4.8994 (0.0044) & 2041.1 (1.9) & 0.0442 (0.0091) &  18.4 (3.8) & 0.030 (0.001) &  0.37 (0.08) &  0.12 (0.03) \\
 G084.9505$-$00.6910 & 4.8999 (0.0005) & 2040.8 (0.2) & 0.0523 (0.0011) &  21.8 (0.5) & 0.072 (0.000) &  1.04 (0.02) &  0.45 (0.08) \\
 G105.5072$+$00.2294 & 4.9067 (0.0117) & 2038.0 (4.9) & 0.0641 (0.0259) &  26.6 (10.8) & 0.069 (0.004) &  1.23 (0.50) &  0.28 (0.13) \\
 G108.7575$-$00.9863 & 4.8990 (0.0010) & 2041.2 (0.4) & 0.0460 (0.0019) &  19.2 (0.8) & 0.050 (0.000) &  0.64 (0.03) &  0.10 (0.01) \\
 G110.0931$-$00.0641 & \dots & \dots & \dots & \dots & \dots & $<$ 0.55 & $<$ 0.32 \\
 G111.2552$-$00.7702 & 4.9004 (0.0063) & 2040.7 (2.6) & 0.0434 (0.0128) &  18.1 (5.3) & 0.112 (0.006) &  1.35 (0.40) & $\sim$ 0.30 \\
 G111.5671$+$00.7517 & 4.9055 (0.0022) & 2038.5 (0.9) & 0.0508 (0.0046) &  21.1 (1.9) & 0.052 (0.001) &  0.73 (0.07) &  0.14 (0.02) \\
 G189.0307$+$00.7821 & 4.8953 (0.0011) & 2042.8 (0.5) & 0.0292 (0.0021) &  12.2 (0.9) & 0.028 (0.000) &  0.22 (0.02) &  0.04 (0.00) \\
 G203.3166$+$02.0564 & 4.9103 (0.0048) & 2036.5 (2.0) & 0.0477 (0.0098) &  19.8 (4.1) & 0.010 (0.000) &  0.13 (0.03) &  0.06 (0.01) \\
\enddata
\tablenotetext{a}{Values preceded by $\sim$ refer to targets with
  highly uncertain $N$(H$_2$O) values, due to seemingly saturated 3.0
  \mum\ absorption bands.}
\end{deluxetable}

\subsubsection{CH$_3$OH}~\label{sec:profch3oh}

\begin{figure*}[h!]
  \includegraphics[width=18cm, angle=90, scale=0.56]{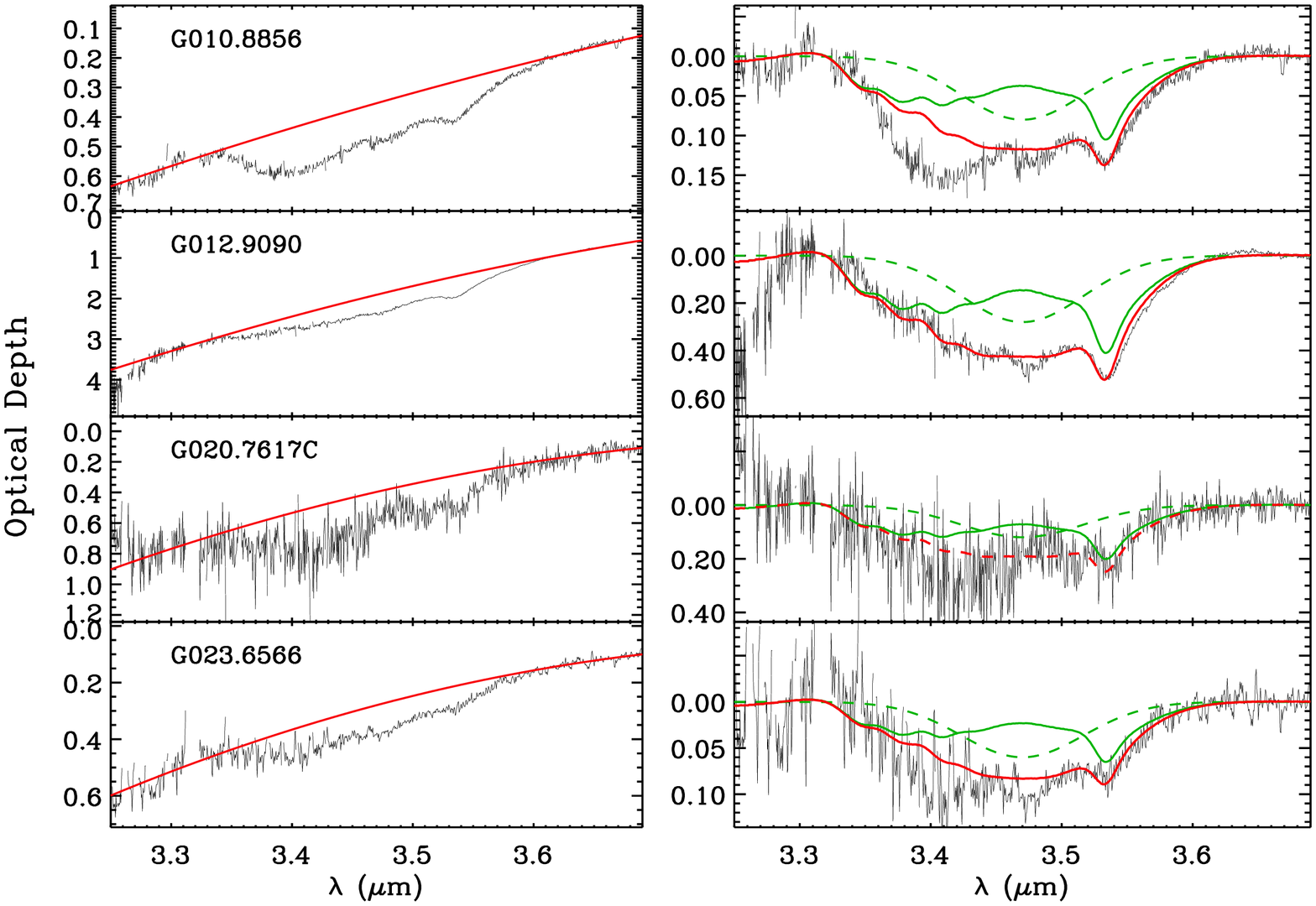}
  \includegraphics[width=18cm, angle=90, scale=0.56]{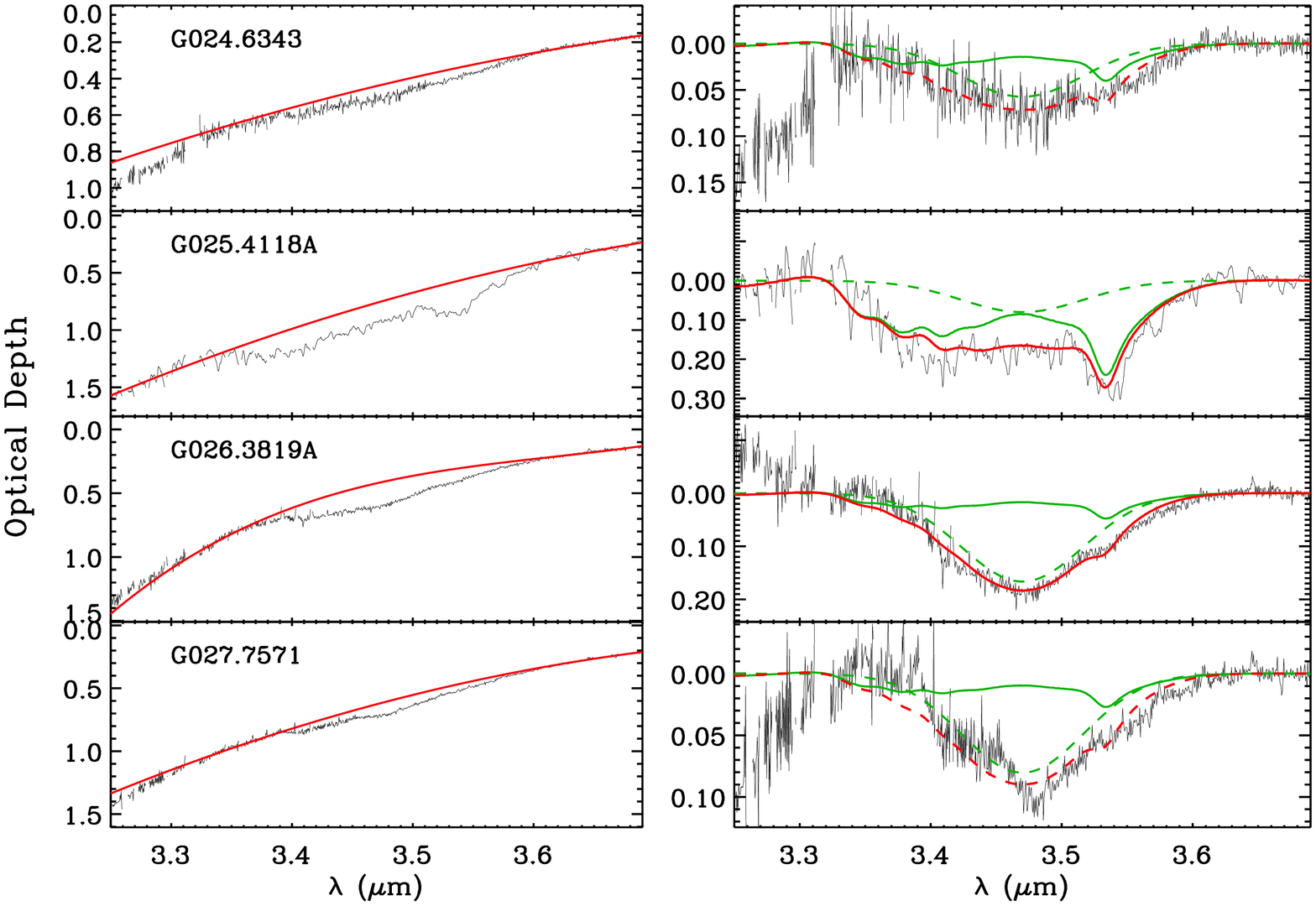}  
  \caption{Portion of the $L-$band spectra of the MYSO sample. All
    targets show the 3.47 \mum\ absorption band, likely caused by
    NH$_3$ hydrates, and some show signs of hydrocarbon absorption as
    well near 3.4 \mum\ (e.g., G033.5237). The feature at 3.53
    \mum\ is attributed to CH$_3$OH ice. For each target, the left
    panel shows the adopted local polynomial baseline, and the right
    panel the baseline subtracted spectrum. In the right panels, the
    solid green line is a laboratory spectrum of CH$_3$OH ice at a
    temperature of 10 K (H$_2$O:CH$_3$OH:CO:NH$_3$=100:50:1:1;
    \citealt{hudgins93}) and the dashed green line a Gaussian. The sum
    of these fits is shown as a solid red line. For targets without a
    CH$_3$OH ice detection, a dashed red line is shown
    instead. \label{f:ch3oh1}}
\end{figure*}

\begin{figure*}[p]
  \includegraphics[width=18cm, angle=90, scale=0.56]{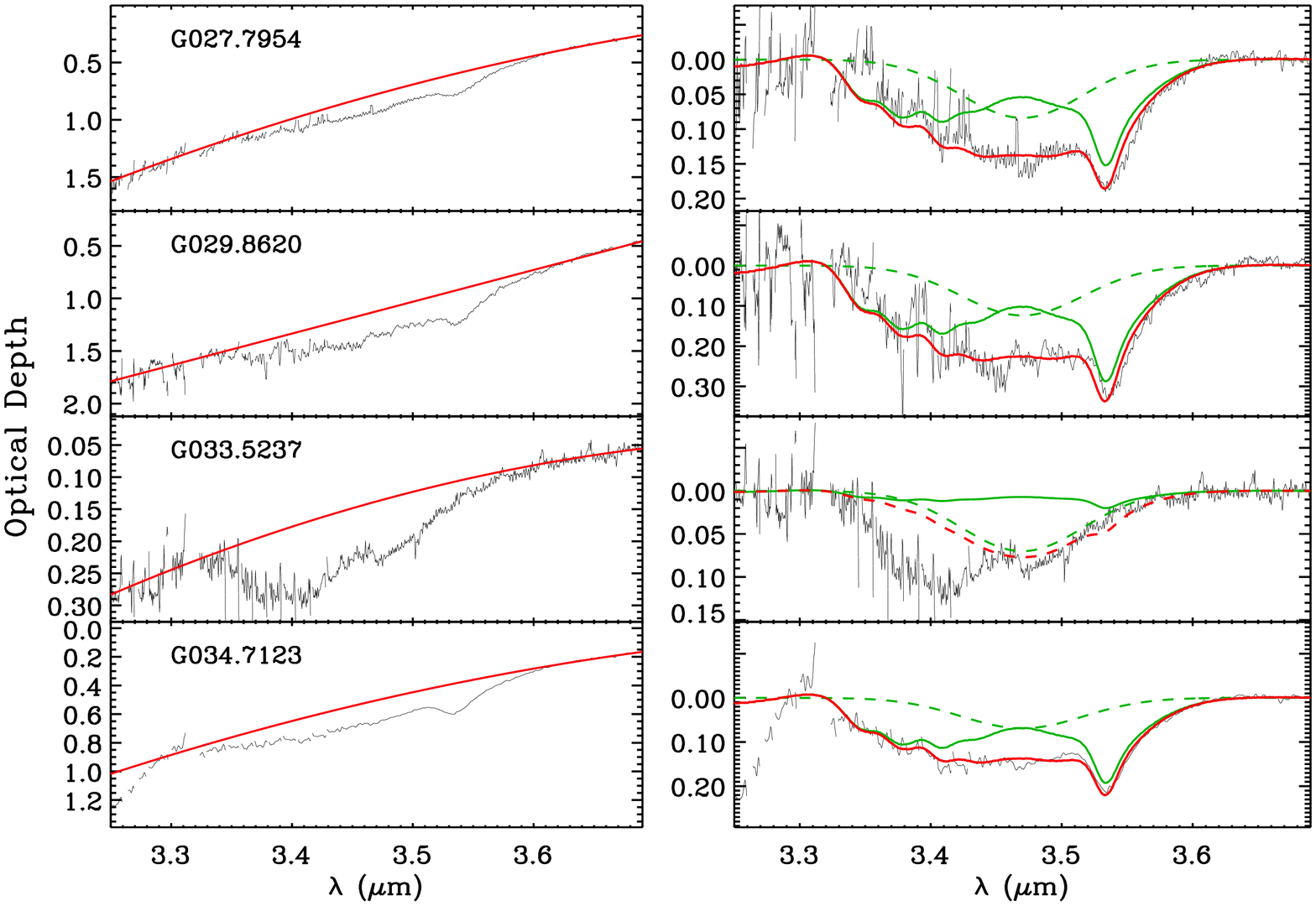}
  \includegraphics[width=18cm, angle=90, scale=0.56]{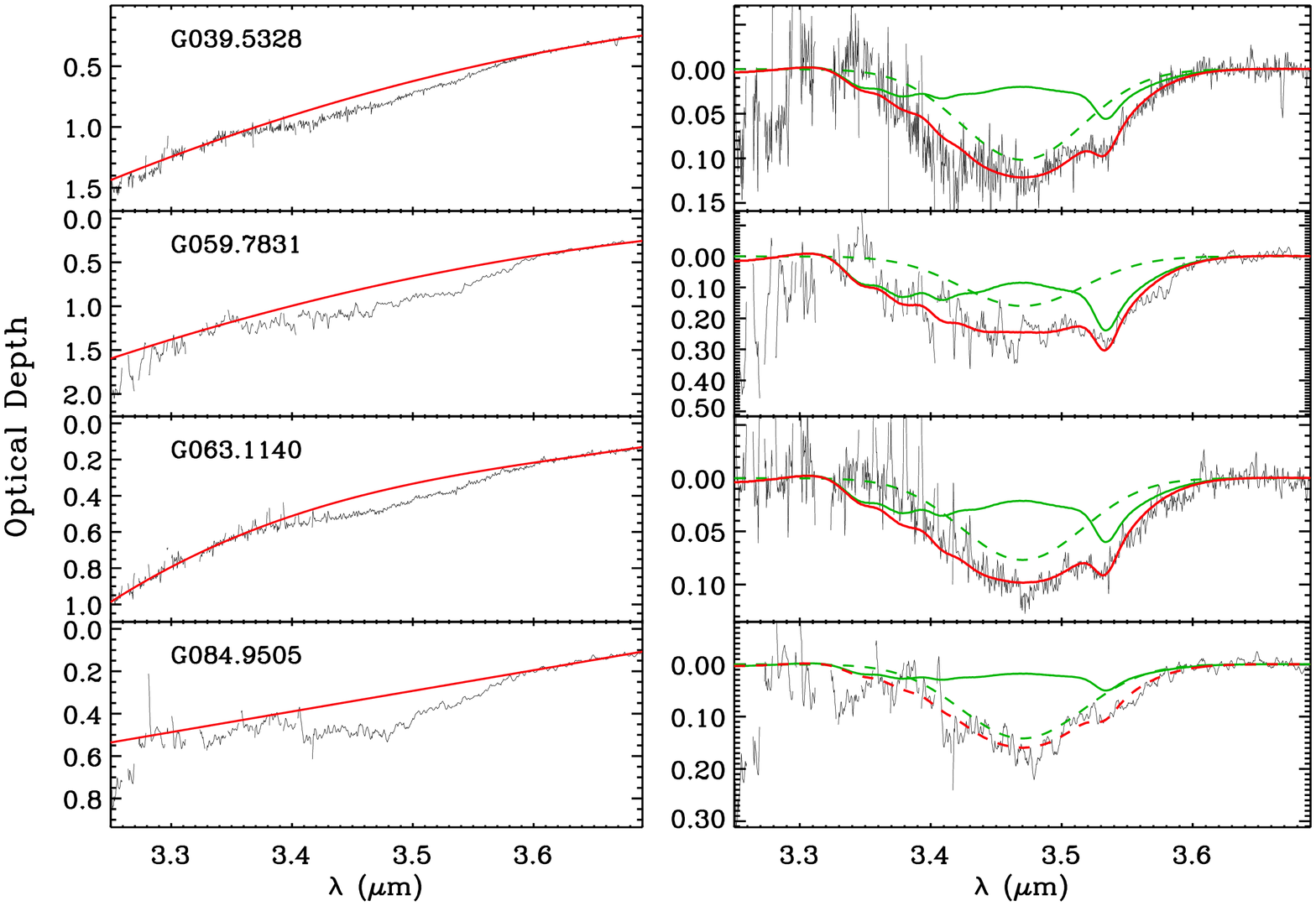}  
  \caption{Extraction of the CH$_3$OH ice absorption band (solid green
    line) from the $L$-band spectra of the MYSO sample. For more
    details, see the caption of
    Figure~\ref{f:ch3oh1}.\label{f:ch3oh2}}
\end{figure*}

\begin{figure*}[p]
  \includegraphics[width=18cm, angle=90, scale=0.56]{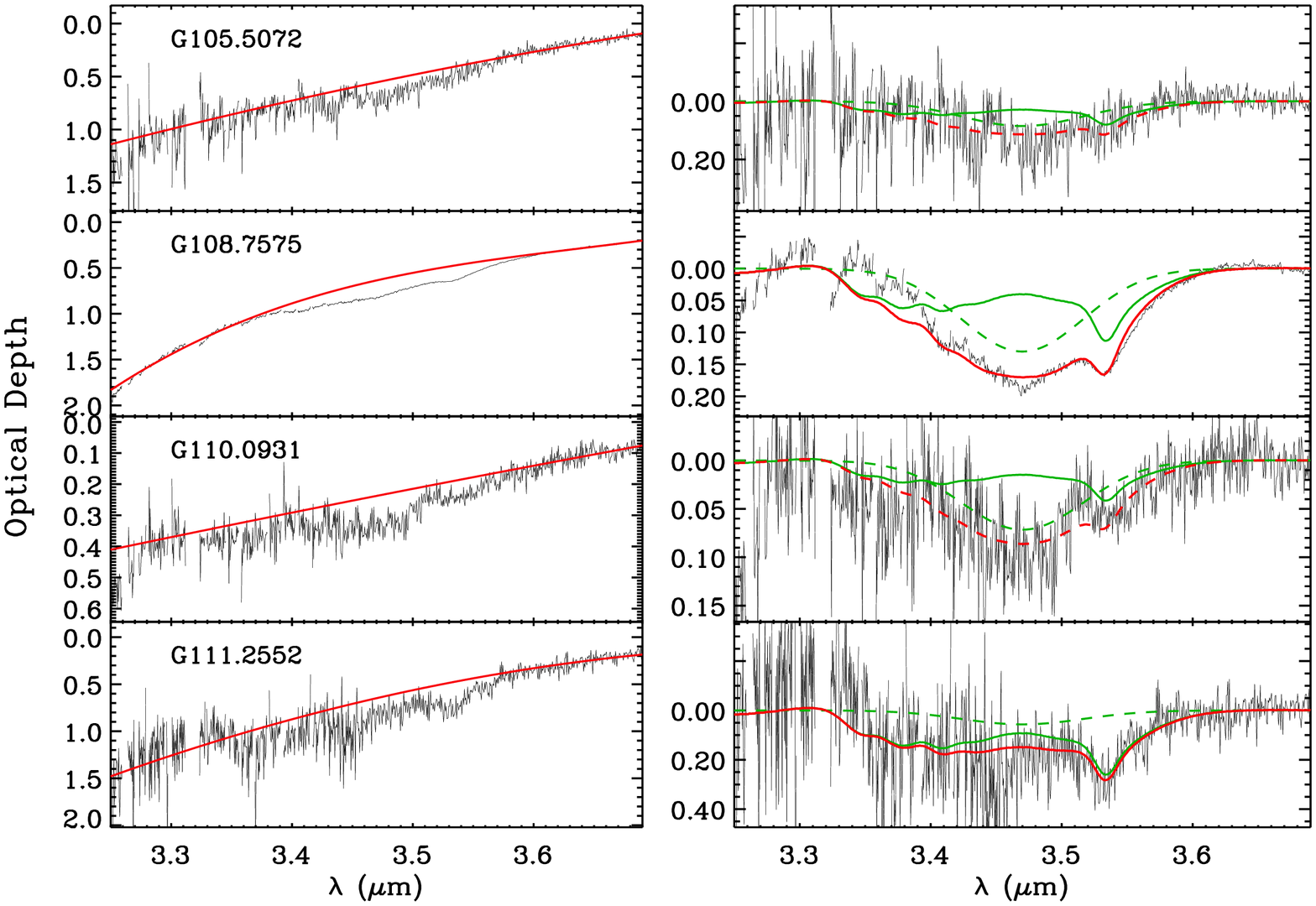}
  \includegraphics[width=18cm, angle=90, scale=0.435]{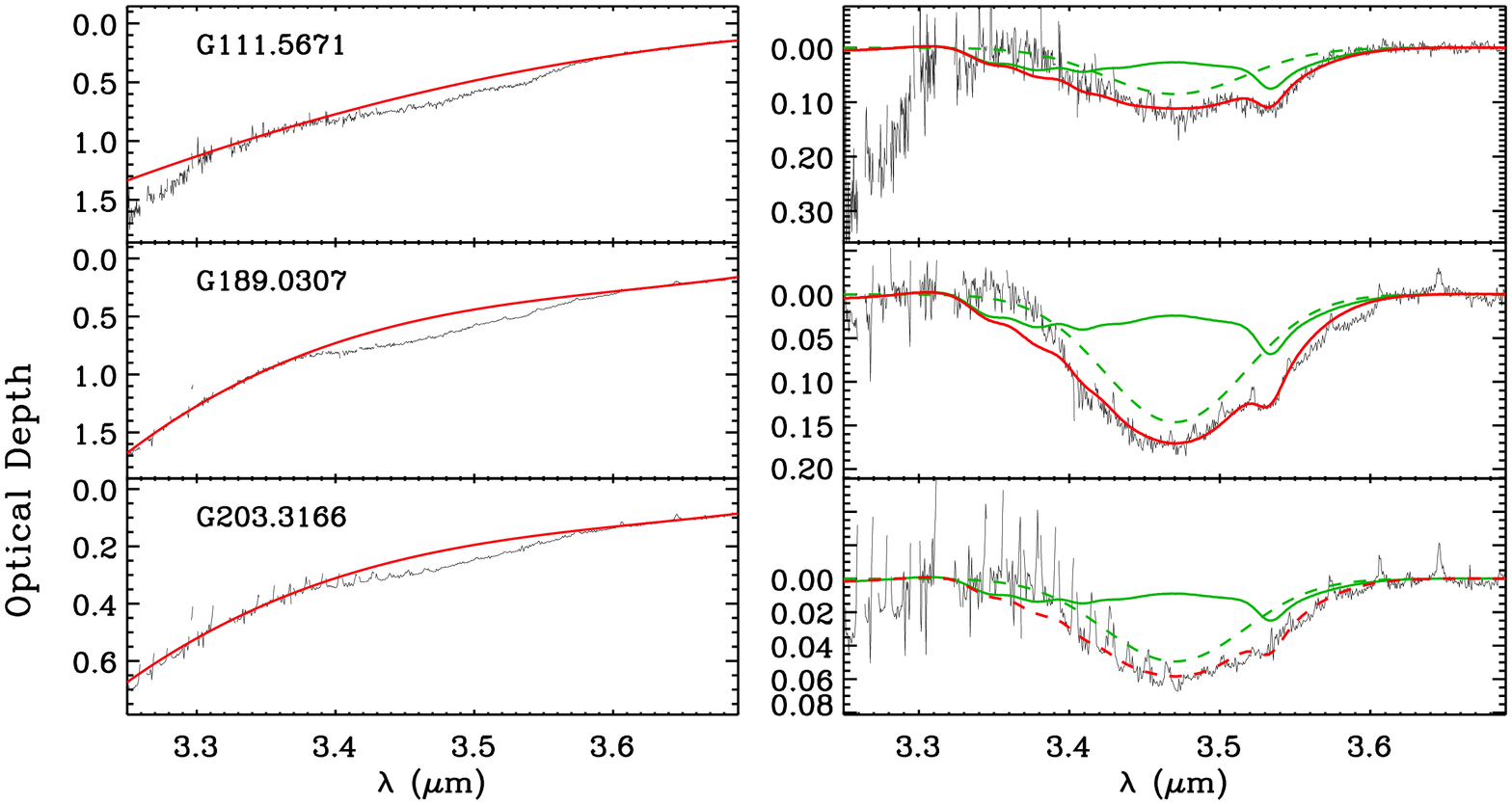}  
  \caption{Extraction of the CH$_3$OH ice absorption band (solid green
    line) from the $L$-band spectra of the MYSO sample.  For more
    details, see the caption of
    Figure~\ref{f:ch3oh1}.\label{f:ch3oh3}}
\end{figure*}

To put the OCS, OCN$^-$, and CO observations in an astrochemical
context, CH$_3$OH ice column densities were derived as well.  The 3.53
\mum\ C-H stretch mode of solid CH$_3$OH is detected towards 15 of the
23 MYSOs. As shown in Figures~\ref{f:ch3oh1}-\ref{f:ch3oh3}, the
CH$_3$OH band overlaps with the 3.47 \mum\ absorption feature, which
is possibly due to NH$_3$ hydrates \citep{dartois01}.  These
absorption features are located on the wing of the deep 3.0 \mum\ band
of H$_2$O ice.  A low order polynomial was used to define and subtract
a local baseline.  The absorption features were then disentangled by
fitting the combination of a Gaussian and a laboratory spectrum.  The
laboratory spectrum used is the mixture
H$_2$O:CH$_3$OH:CO:NH$_3$=100:50:1:1 at a temperature of 10 K
\citep{hudgins93}. To provide more stable solutions, the Gaussian peak
position and $FWHM$ were kept fixed at 3.470 \mum\ and 0.11 \mum,
respectively.  CH$_3$OH column densities were derived by using an
integrated band strength of 5.6$\times 10^{-18}$ cm/molecule
\citep{kerkhof99} for the C-H stretch mode, a $FWHM$ of 0.04 \mum, and
the peak optical depth provided by the fits.  The uncertainties were
estimated to be $\sim$20\% due to the baseline and $\sim$10\% due to
the $FWHM$.  The column densities are listed in
Table~\ref{t:ocnch3oh}.

\vspace{60pt}

\subsubsection{H$_2$O}~\label{sec:h2o}

To be able to derive relative ice abundances, column densities for the
most abundant ice species, H$_2$O, were derived as well.  This was
done as part of the continuum fitting process (\S\ref{sec:cont}). An
integrated band strength of 2.0$\times 10^{-16}$ cm/molecule
\citep{hagen81} was used. The column densities reflect those of the
small grains portion of the 3.0 \mum\ band, excluding the
long-wavelength wing, which may have a varied and uncertain
origin. This approach was also adopted for low mass YSOs and
background stars in previous work (e.g., \citealt{boogert08}).  For a
more in depth analysis of the profile of the 3.0 \mum\ band of this
and a larger sample of MYSOs, we refer to K. Emerson et al. (in
preparation).  The H$_2$O column densities are listed in
Table~\ref{t:h2oco}.

\vspace{60pt}

\subsection{OCS Laboratory Spectroscopy}~\label{sec:labocs}

\begin{figure*}
  \center
  \includegraphics[angle=90, scale=0.50]{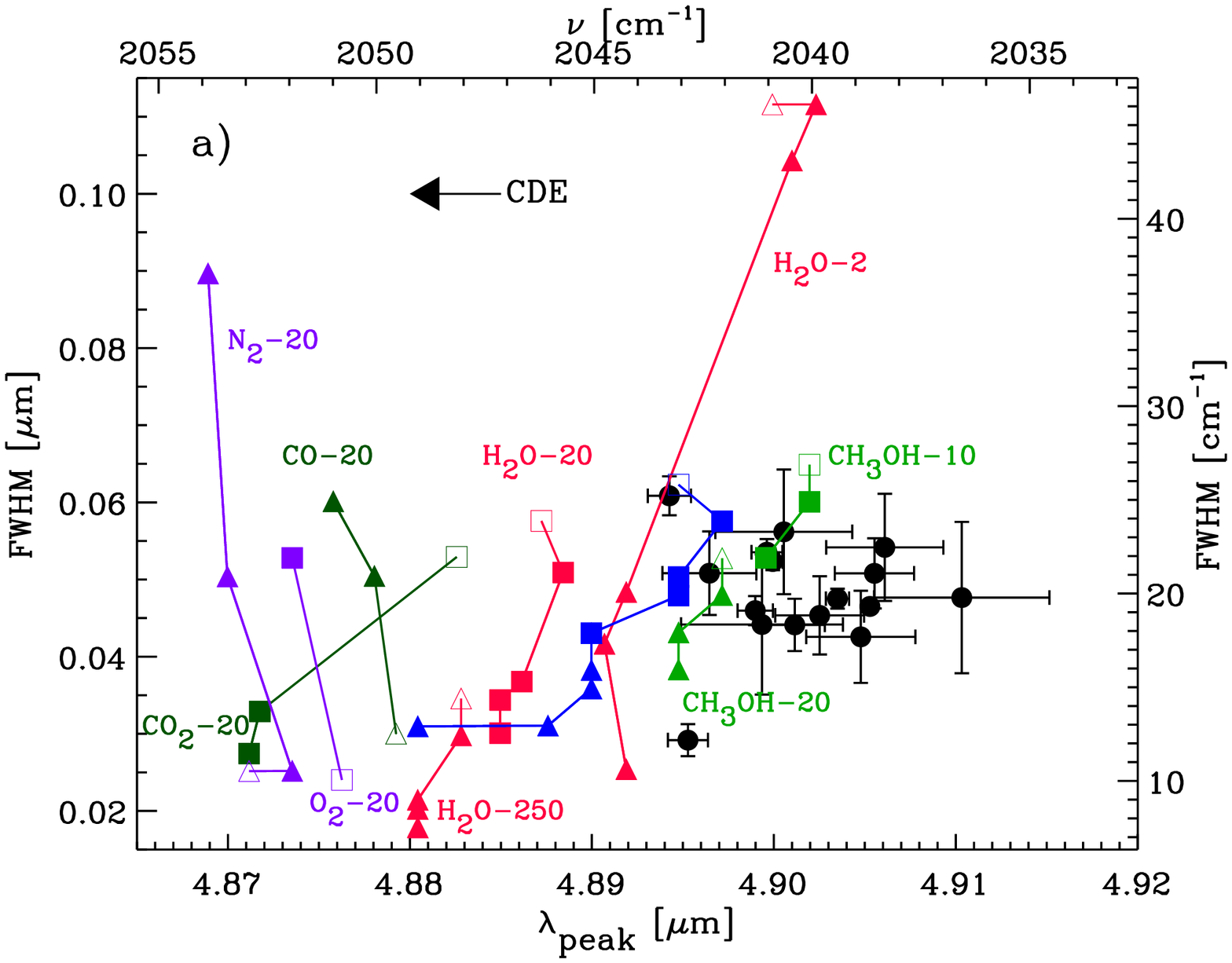}
\includegraphics[angle=90, scale=0.50]{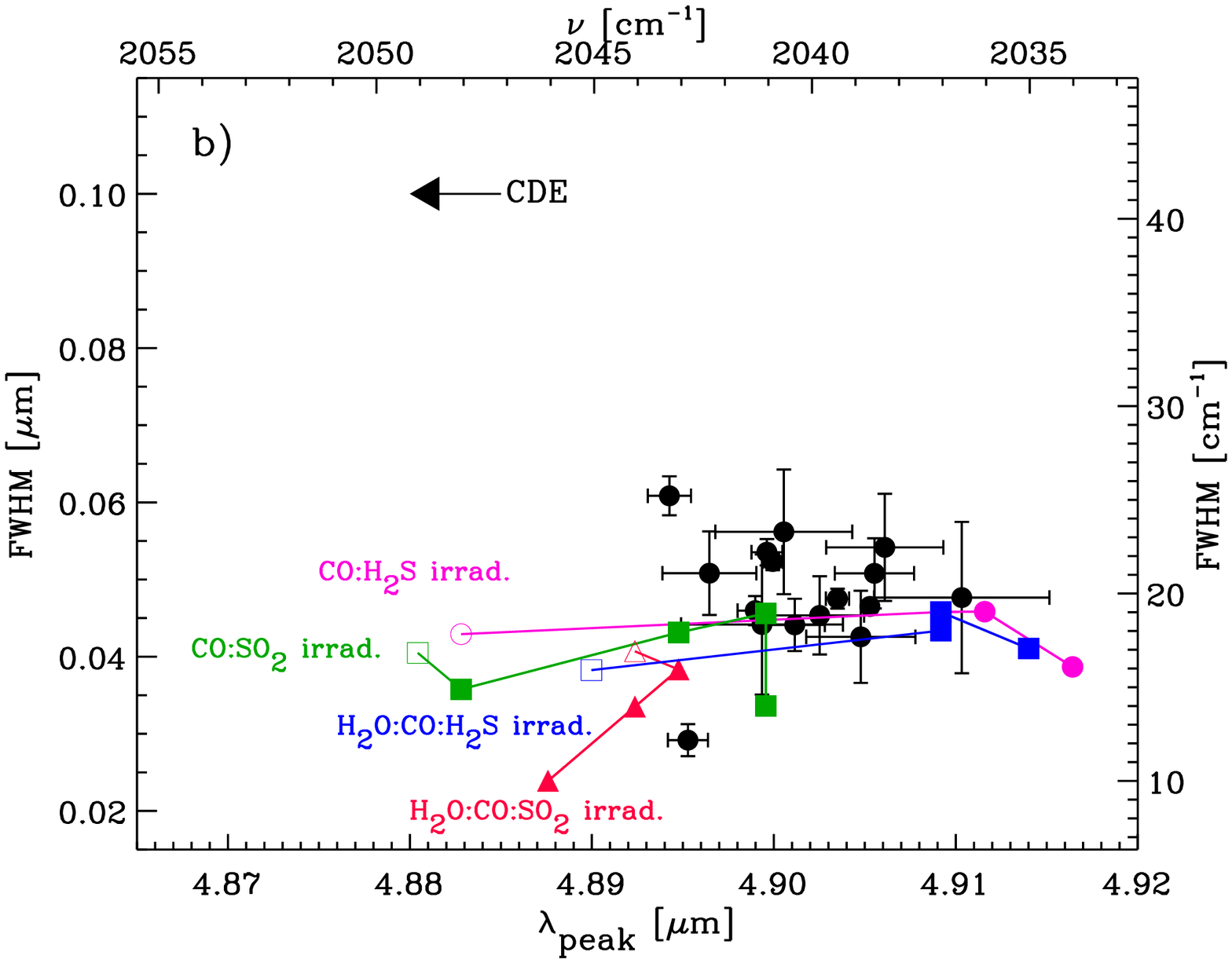}  
\caption{{\bf Panel (a):} Peak position versus width for the profile
  of the C-O stretch mode of solid OCS as a function of mixing ratio
  and temperature for non-irradiated mixtures, taken from
  \citet{hudgins93}, \citet{palumbo95}, and
  \citet{ferrante08}. Different ice mixtures are distinguished by
  different symbols and colors. Mixing ratios are labeled relative to
  OCS (e.g., "CH$_3$OH-10'' means a mixing ratio of
  CH$_3$OH:OCS=10:1). For a giving ice mixture, different temperatures
  are connected with a line. For each temperature series, the lowest
  temperature mixture is indicated with an open symbol. The unlabeled
  dark blue squares indicate a mixture of H$_2$O:CH$_3$OH:OCS=3:10:1,
  and the dark blue triangles of H$_2$O:CH$_3$OH:OCS=100:50:1. The
  peak position changes typically by the amount indicated by the black
  arrow if the OCS is present at concentrations $>10\%$ on grains with
  shapes of a Continuous Distribution of Ellipsoids (CDE), rather than
  on a flat surface as in the laboratory transmission spectra. The
  black bullets with error bars indicate the peak positions and widths
  observed toward the sample of MYSOs. {\bf Panel (b):} Similar to
  panel (a), but here the observations (black bullets) are compared to
  the proton-irradiated mixtures of
  \citet{ferrante08}. }~\label{f:nudnuocs}
\end{figure*}

The spectra of solid OCS in a range of astrophysically relevant ice
mixtures and temperatures were measured by \citet{hudgins93},
\citet{palumbo95}, \citet{ferrante08}, and \citet{garozzo10}.  Peak
positions and FWHM values of the C-O stretch mode of OCS for
non-irradiated mixtures are displayed in Figure~\ref{f:nudnuocs}a.
The peak position depends strongly on the ice composition: mixtures
with apolar species peak near 4.87-4.88 \mum, mixtures with H$_2$O
peak at 4.88-4.89 \mum, and mixtures with CH$_3$OH near 4.90 \mum.
The profiles of the apolar mixtures become much broader (factor of
2-3) upon heating, while the polar mixtures are very broad at
temperatures of 10 K, and then dramatically narrow at higher
temperatures.

The proton-irradiated ices studied in \citet{ferrante08}, generally
exhibit a different behavior (Figure~\ref{f:nudnuocs}b). This is
confirmed by the proton-irradiation experiments of
\citet{garozzo10}. Upon heating, the peak position shifts to much
longer wavelength (up to 4.915 \mum, the longest of all experiments),
while its width does not change much, except for a narrowing for the
H$_2$O:CO:SO$_2$ ice mixture. The strong shift must at least in part
be related to the formation of new species in the ice, such as CO$_3$,
CS$_2$, and CO$_2$, changing the dipole interactions.

Figure~\ref{f:nudnuocs} also shows a comparison to the
observations. Of the non-irradiated mixtures, ices rich in CH$_3$OH
typically give the best match of peak position and FWHM. Of the
mixtures without CH$_3$OH, only those with relatively high OCS/H$_2$O
($\sim$2) provide a match to some of the observed MYSOs, but this
deteriorates when grain shape effects are taken into account, which
typically shift the absorption band to shorter wavelengths. Grain
shape effects are significant if OCS has a relatively high
concentration ($\geq$10\%). The best match in peak position and FWHM
is in fact provided by the proton-irradiated CO:H$_2$S ices, as those
are the only ones that cover the longest observed peak positions near
4.905 \mum. This might point to a relatively complex molecular
environment for OCS.

%

\subsection{Ice Correlations}~\label{sec:correl}

\begin{figure*}
\includegraphics[angle=90, scale=0.42]{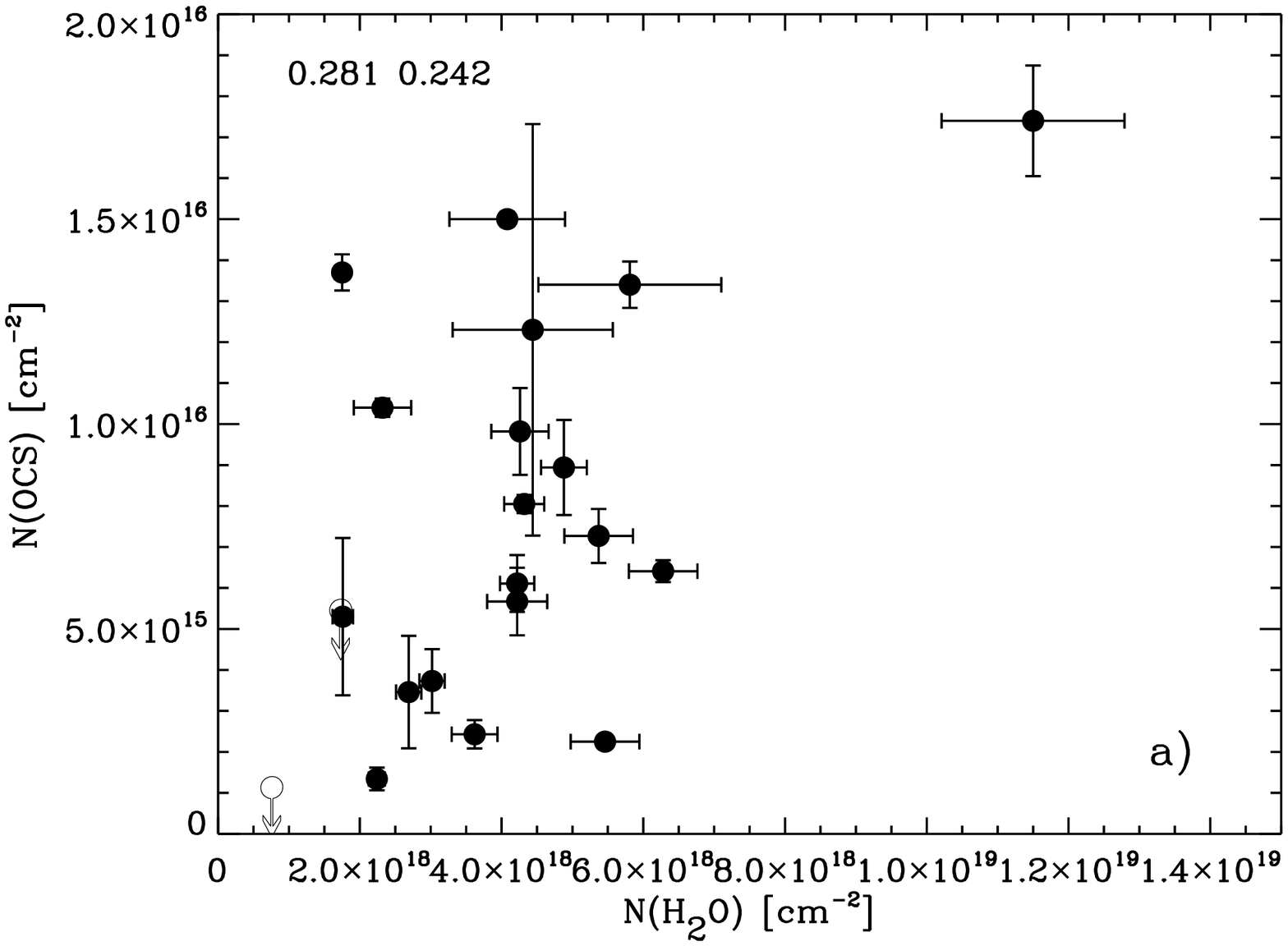}
\includegraphics[angle=90, scale=0.42]{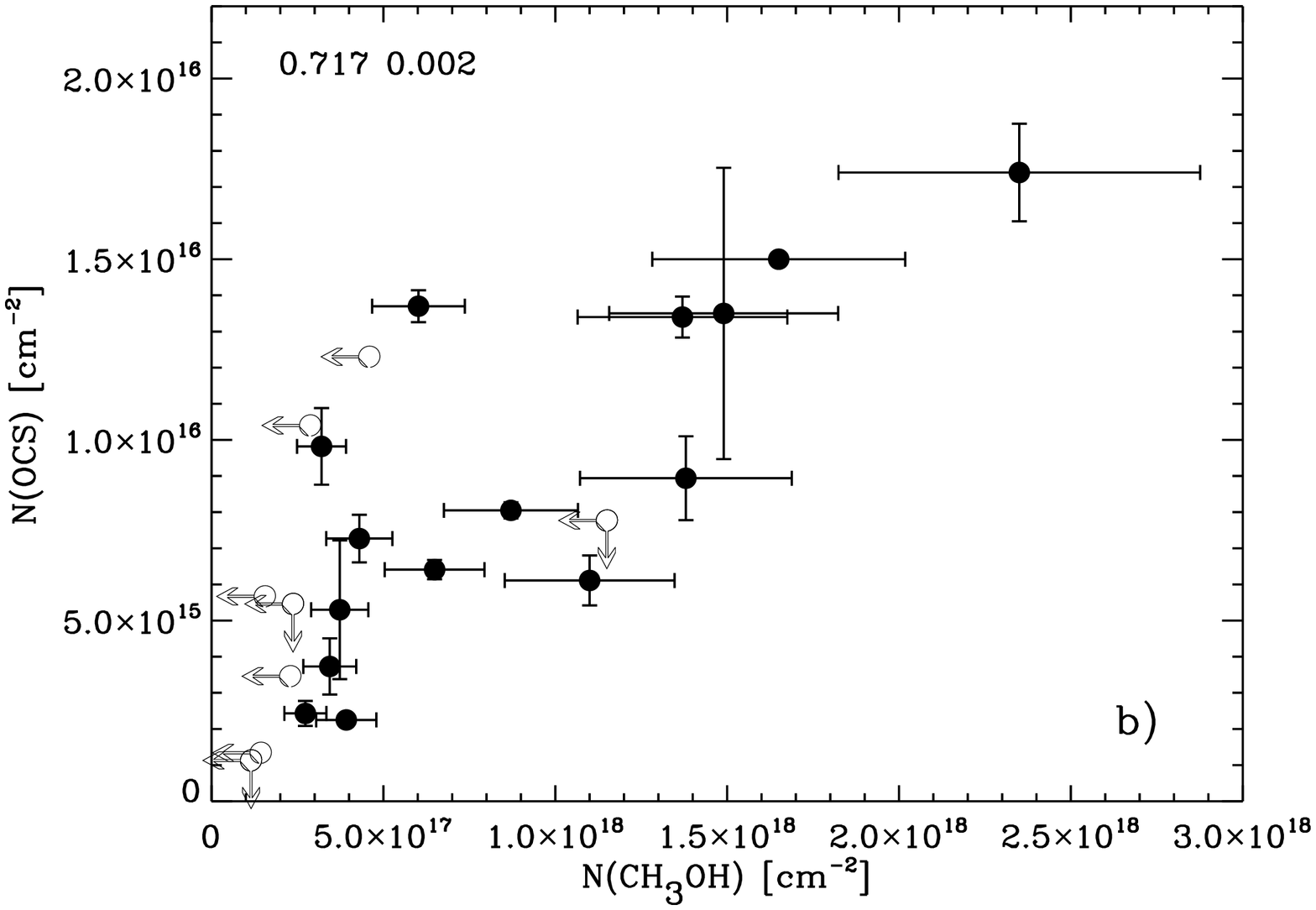}
\includegraphics[angle=90, scale=0.42]{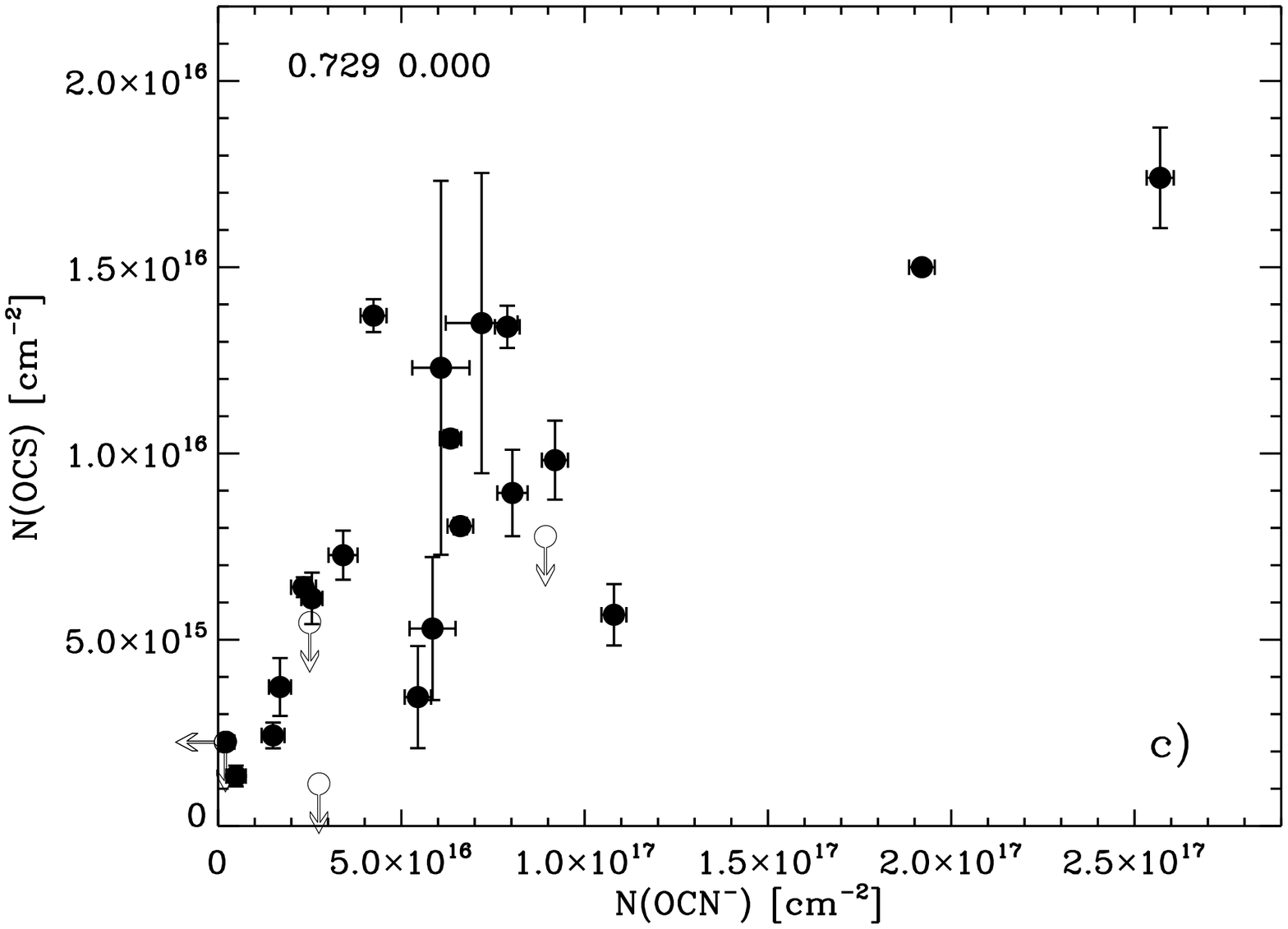}
\includegraphics[angle=90, scale=0.42]{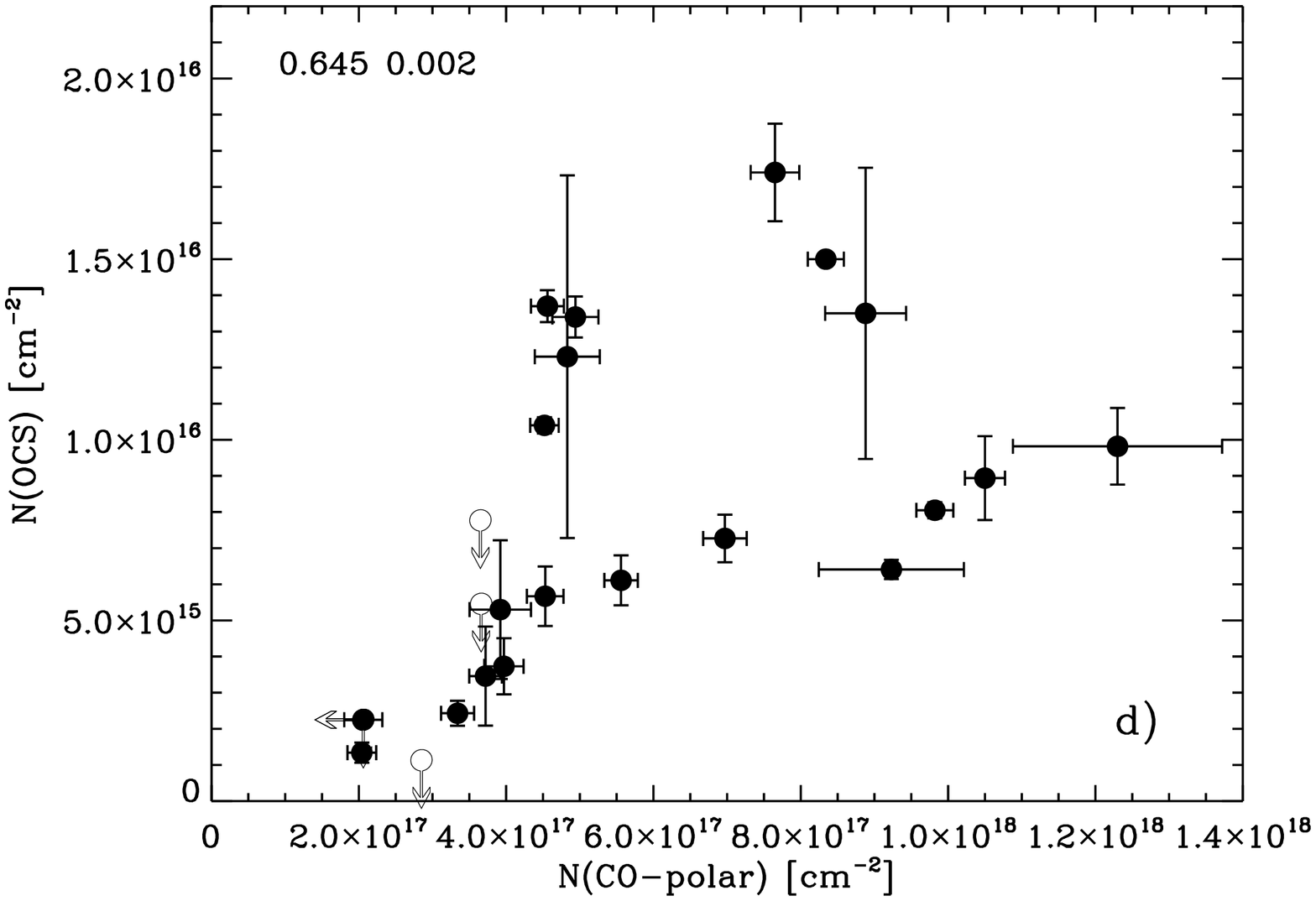}
\includegraphics[angle=90, scale=0.42]{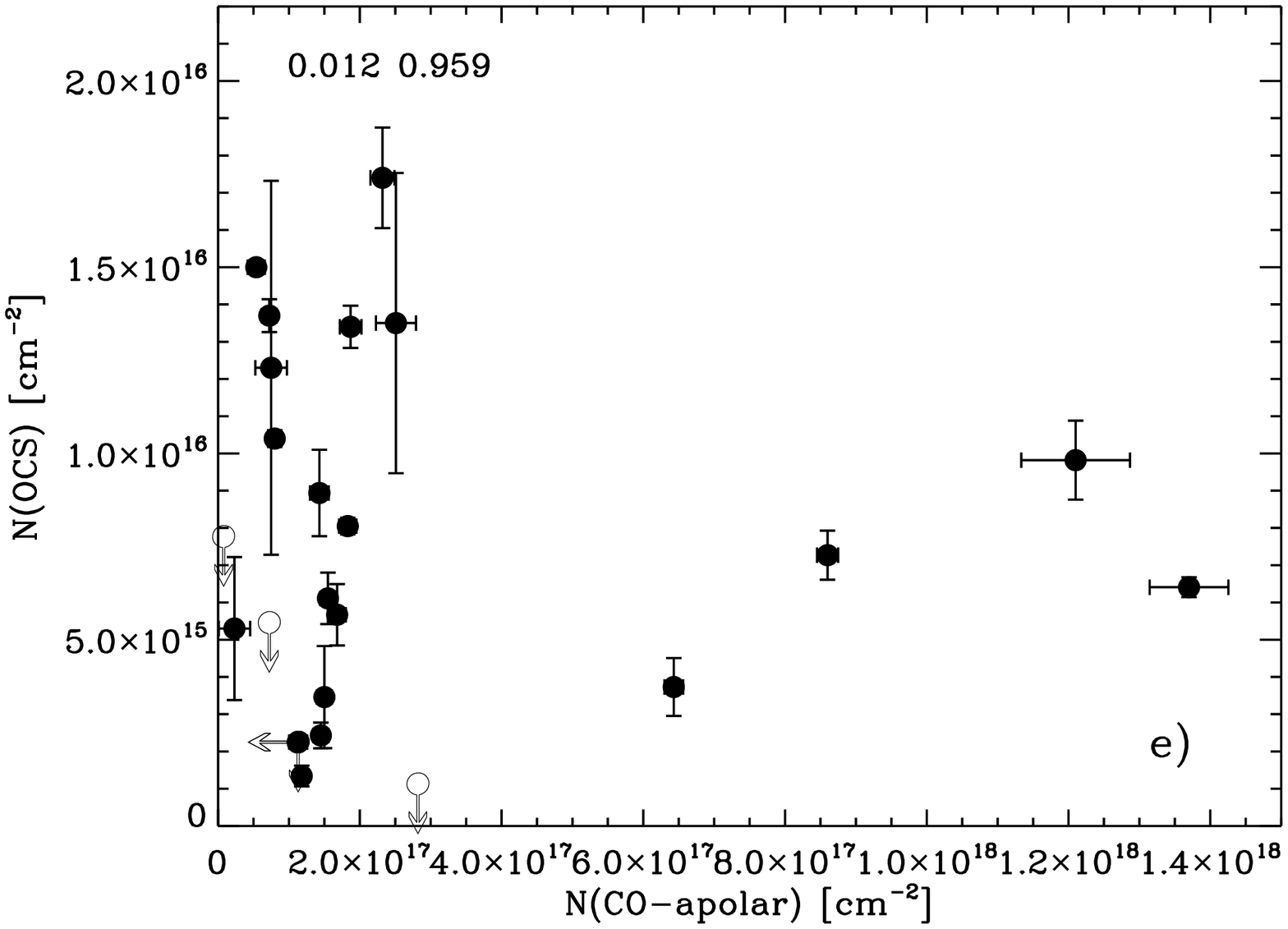}
\caption{Correlation plots of the OCS column density versus that of
  other ice species towards MYSOs. The values in each top left corner
  are the Spearman's Rank coefficient and each second number is the
  significance of there being no correlation.}~\label{f:correl}
\end{figure*}

To further constrain the molecular environment and formation history
of OCS, its column density is plotted versus that of other species
(Figure~\ref{f:correl}), and correlation coefficients are derived.  A
positive Spearman's Rank coefficient ($\rho$) indicates the probability
that there is a monotonically increasing relation, with a maximum of
1.0.  This value is indicated in each panel, with the second number
the significance of there being no correlation. The best correlations
are found for OCS with the OCN$^-$ and CH$_3$OH column densities
($\rho\sim$0.72), followed by CO$_{\rm polar}$ ($\rho\sim$0.65). The
correlations are rather poor for OCS with the H$_2$O column
($\rho\sim$0.28) and there is no correlation with CO$_{\rm apolar}$.

We investigated whether the rather poor correlation of OCS with H$_2$O
could be an artifact related to underestimated H$_2$O column densities
(and uncertainties) due to saturated 3.0 \mum\ absorption
bands. Inspection of Figures~\ref{f:flux1}-\ref{f:flux3} shows that
this could be the case for five MYSOs. But of those, G020.7617 was not
included in the OCS version H$_2$O corelation due to the non-detection
of OCS, G111.2552 was not included due to severe saturation of the 3.0
\mum\ band, and G012.9090 is a well studied MYSO (W33A) for which
$N$(H$_2$O) is quite reliable (e.g., \citealt{gibb04}). Excluding the
remaining two targets (G105.5072 and G059.7831) from the correlation
plot reduces the Spearman's Rank coefficient from 0.28 to
0.21. Including those two targets, and artificially increasing their
column density with a factor of 1.6 to force an improved correlation,
improves the Spearman's Rank coefficient somewhat (0.34), but it is
still much poorer than the correlations of OCS with CH$_3$OH, OCN$^-$,
and CO$_{\rm polar}$.

\begin{figure*}
\includegraphics[angle=90, scale=0.42]{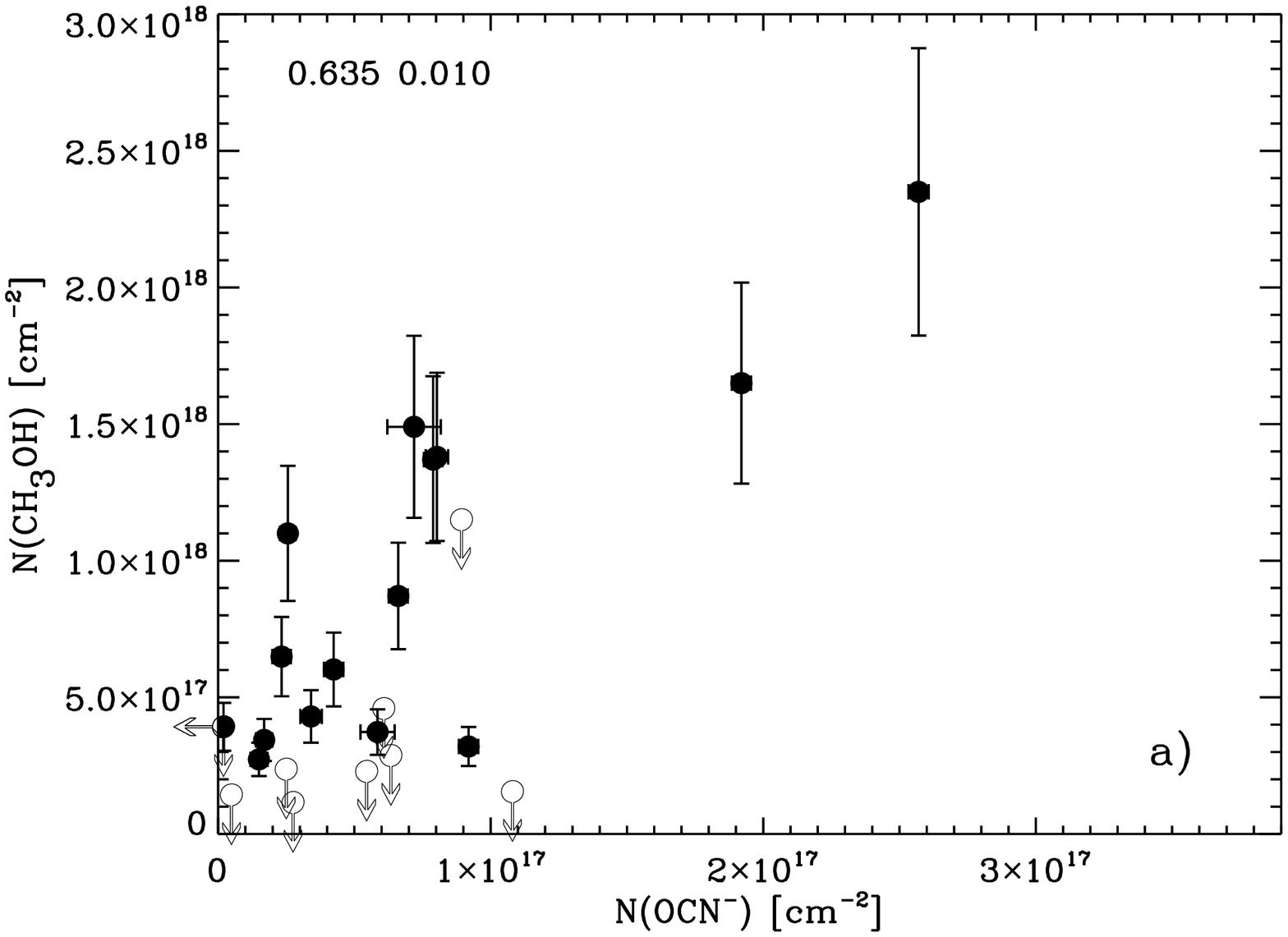}
\includegraphics[angle=90, scale=0.42]{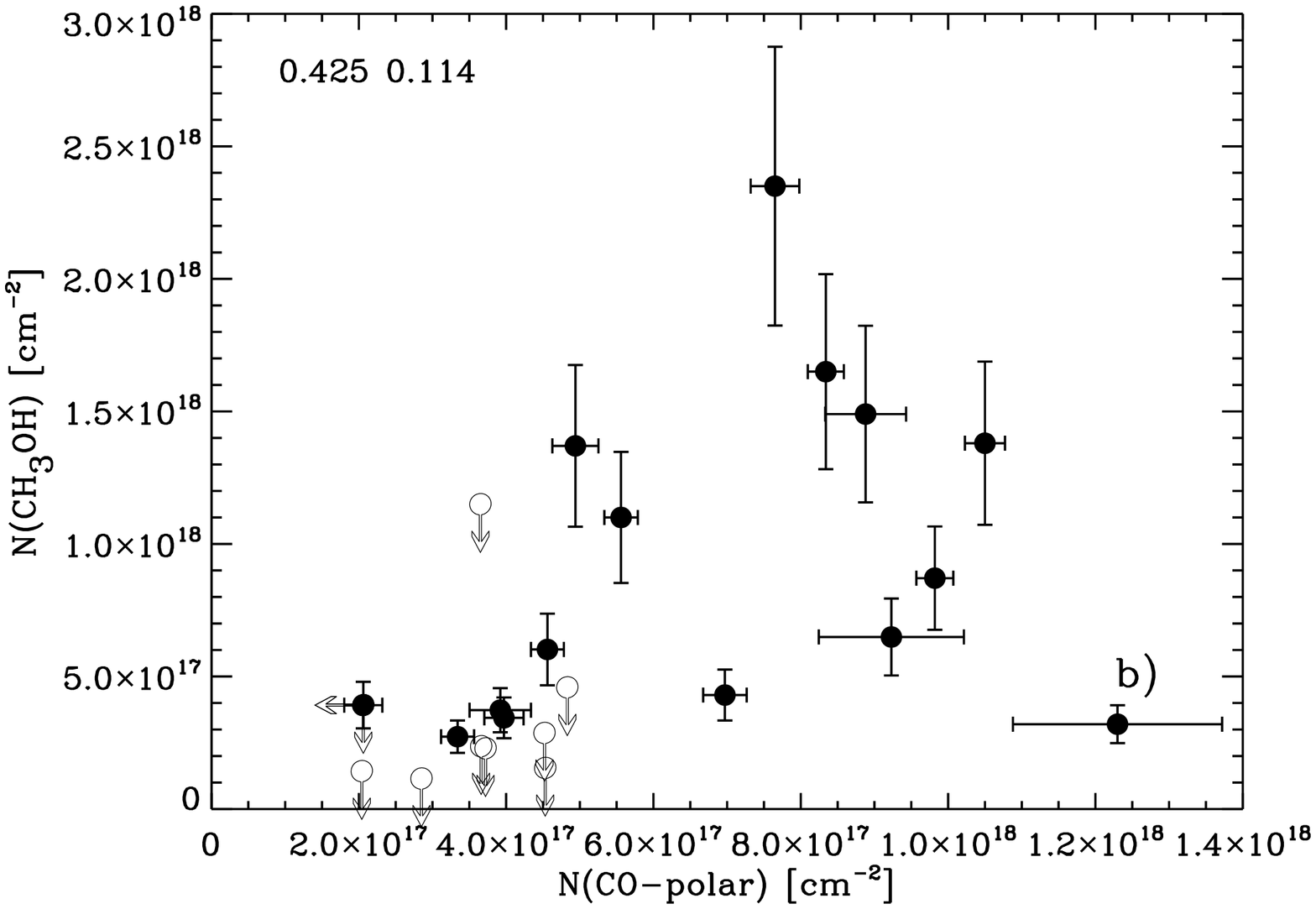}
\includegraphics[angle=90, scale=0.42]{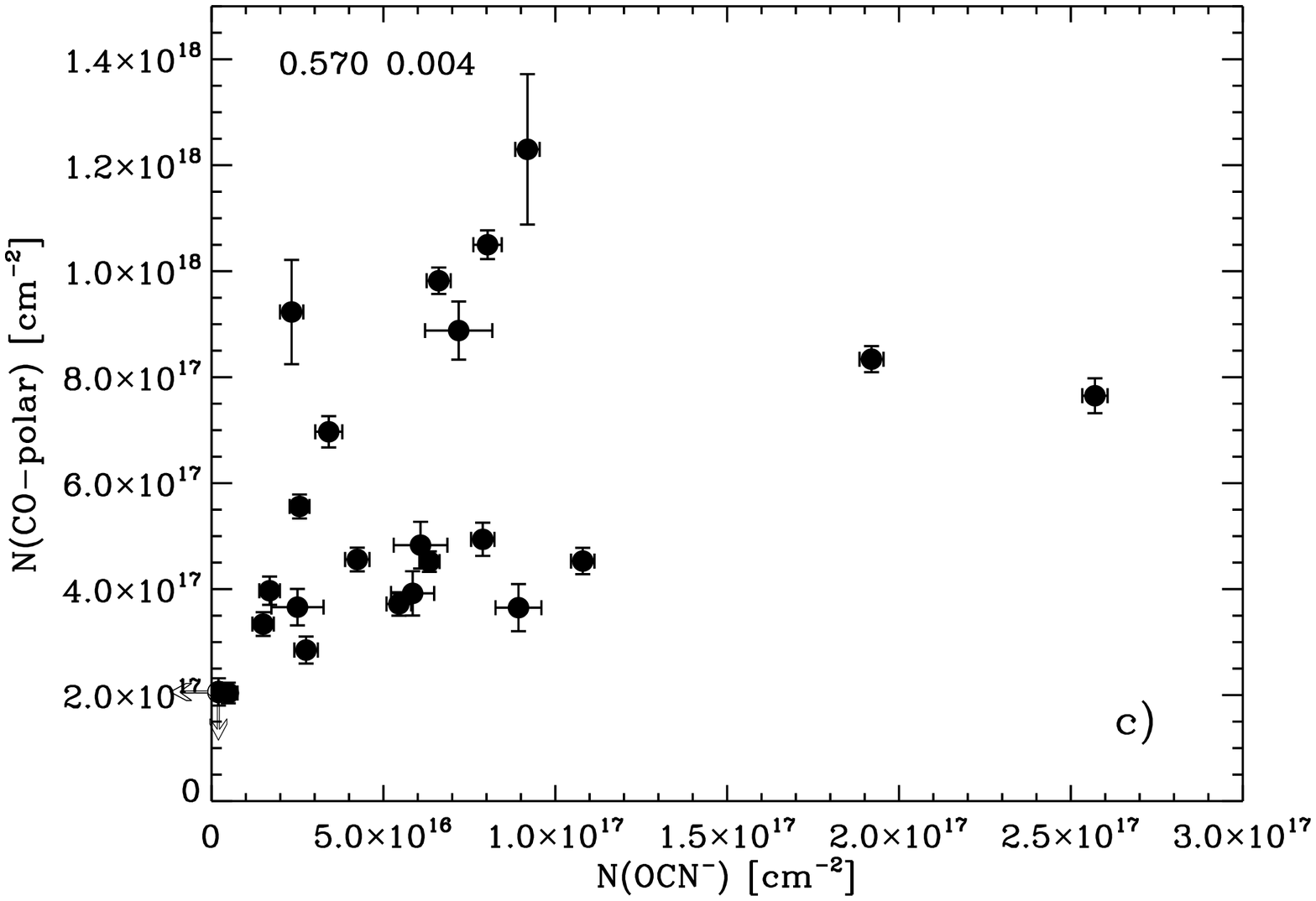}
\caption{Correlation plots of CH$_3$OH with OCN$^-$ ({\bf panel (a)})
  and CO$_{\rm polar}$ ({\bf panel (b)}), and of CO$_{\rm polar}$ with
  OCN$^-$ ({\bf panel (c)}). towards MYSOs. The values in each top
  left corner are the Spearman's Rank coefficient and each second
  number is the significance of there being no
  correlation.}~\label{f:correlother}
\end{figure*}

The correlations among CH$_3$OH, OCN$^-$, and CO$_{\rm polar}$ are
also investigated (Fig.~\ref{f:correlother}). Indeed, as expected, the
best correlation is that between CH$_3$OH and OCN$^-$
($\rho\sim$0.64). The correlations of CO$_{\rm polar}$ with CH$_3$OH
($\rho\sim$0.43) and with OCN$^-$ ($\rho\sim$0.57) are poorer than
that of OCS with CO$_{\rm polar}$ ($\rho\sim$0.65).

In summary, OCS, CH$_3$OH, and OCN$^-$ correlate best with eachother,
and somewhat poorer with CO$_{\rm polar}$. There is no correlation of
these species with H$_2$O and CO$_{\rm apolar}$. Despite the
reasonably good correlations found for OCS with the OCN$^-$ and
CH$_3$OH column densities, a linear model does not describe the
observations well. The scatter in the datapoints is much larger than
the uncertainties. At any given CH$_3$OH or OCN$^-$ column density,
the OCS column density varies by a factor of 2-3.

\subsection{Abundance Distributions}~\label{sec:histo}

\begin{figure*}
\includegraphics[angle=90, scale=0.42]{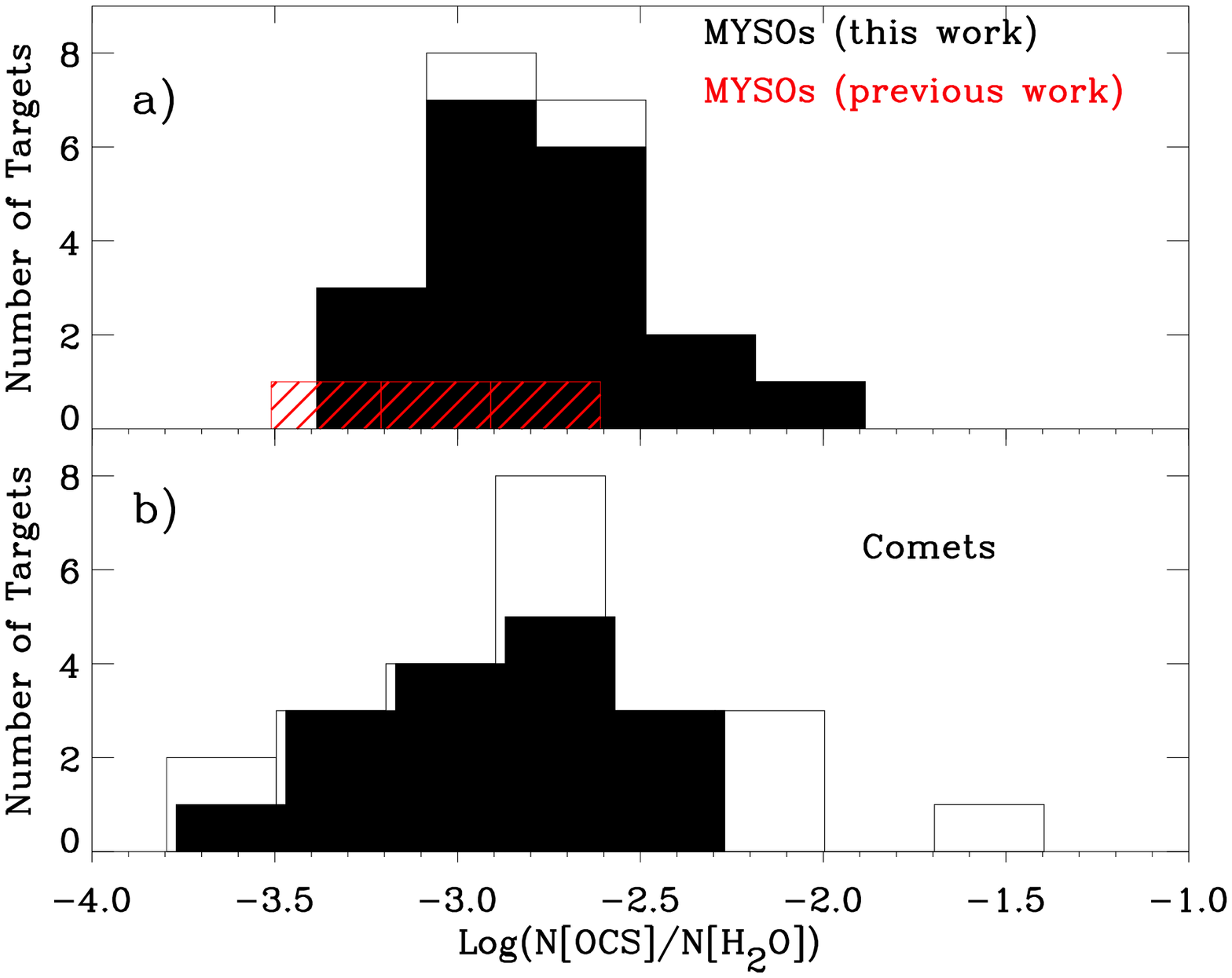}
\includegraphics[angle=90, scale=0.39]{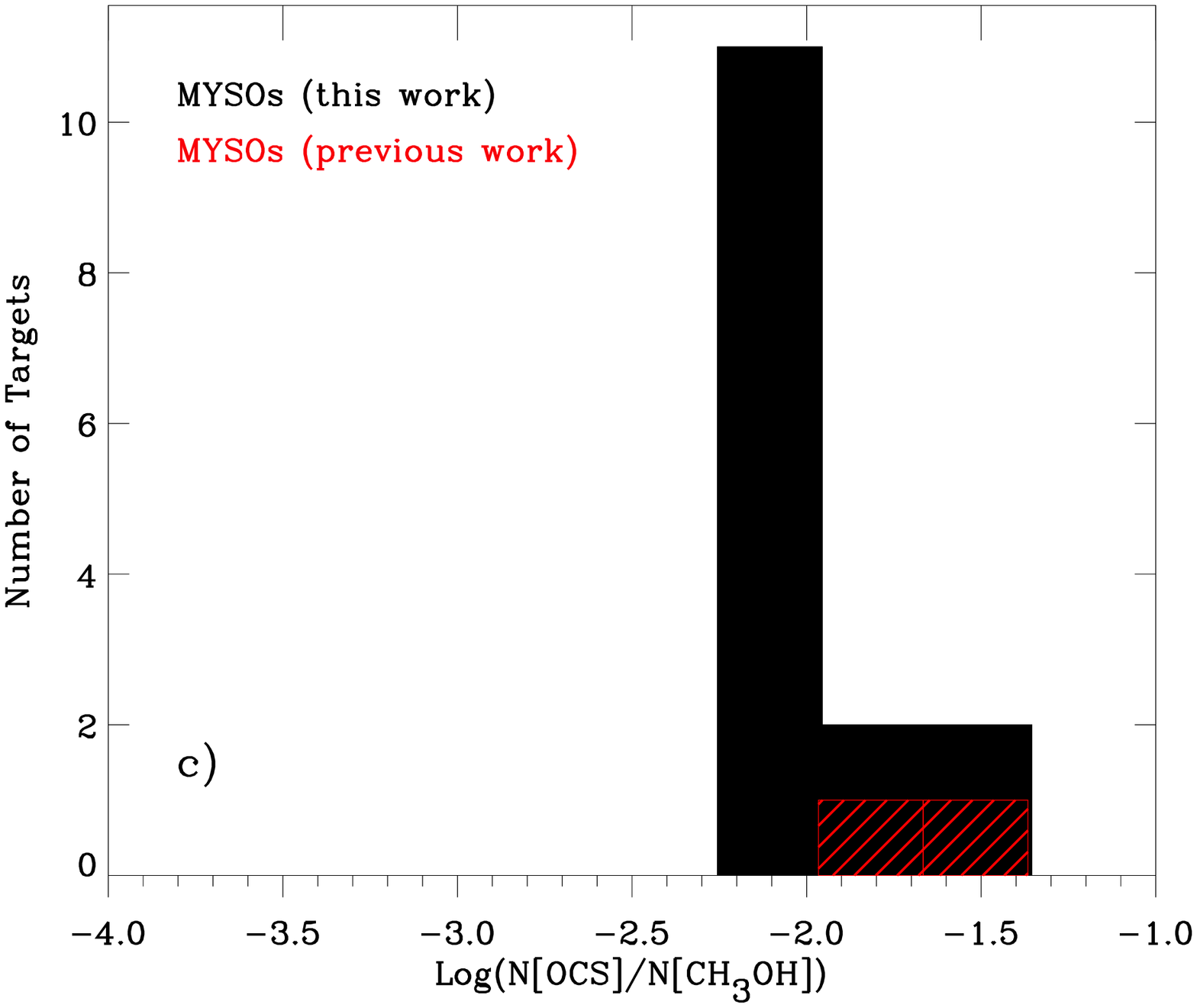}
\caption{Distribution of the OCS ice abundance relative to H$_2$O ice
  towards MYSOs ({\bf panel (a)}). The solid black histogram is for
  MYSO detections presented in this work. The unfilled histogram
  includes abundance upper limits. The hatched red histogram is for
  all OCS ice detections of previous work. {\bf Panel (b)} shows the
  OCS abundance in comets \citep{saki20}, with the open histogram
  including upper limits. {\bf Panel (c)} shows the OCS ice abundances
  relative to CH$_3$OH ice for the MYSOs presented in this (black) and
  previous (red) work.}~\label{f:histo1}
\end{figure*}

\begin{figure*}
\includegraphics[angle=90, scale=0.42]{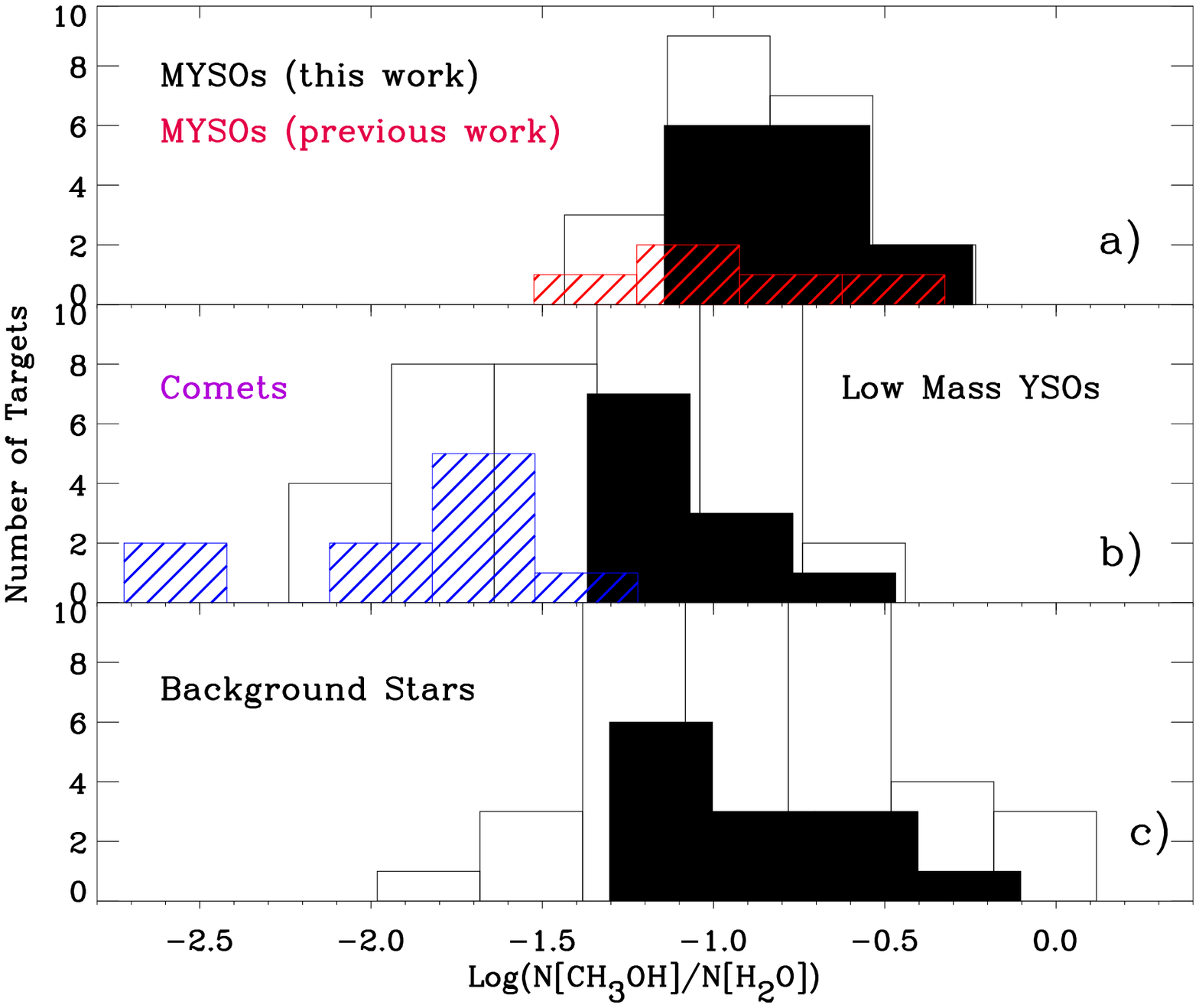}
\includegraphics[angle=90, scale=0.42]{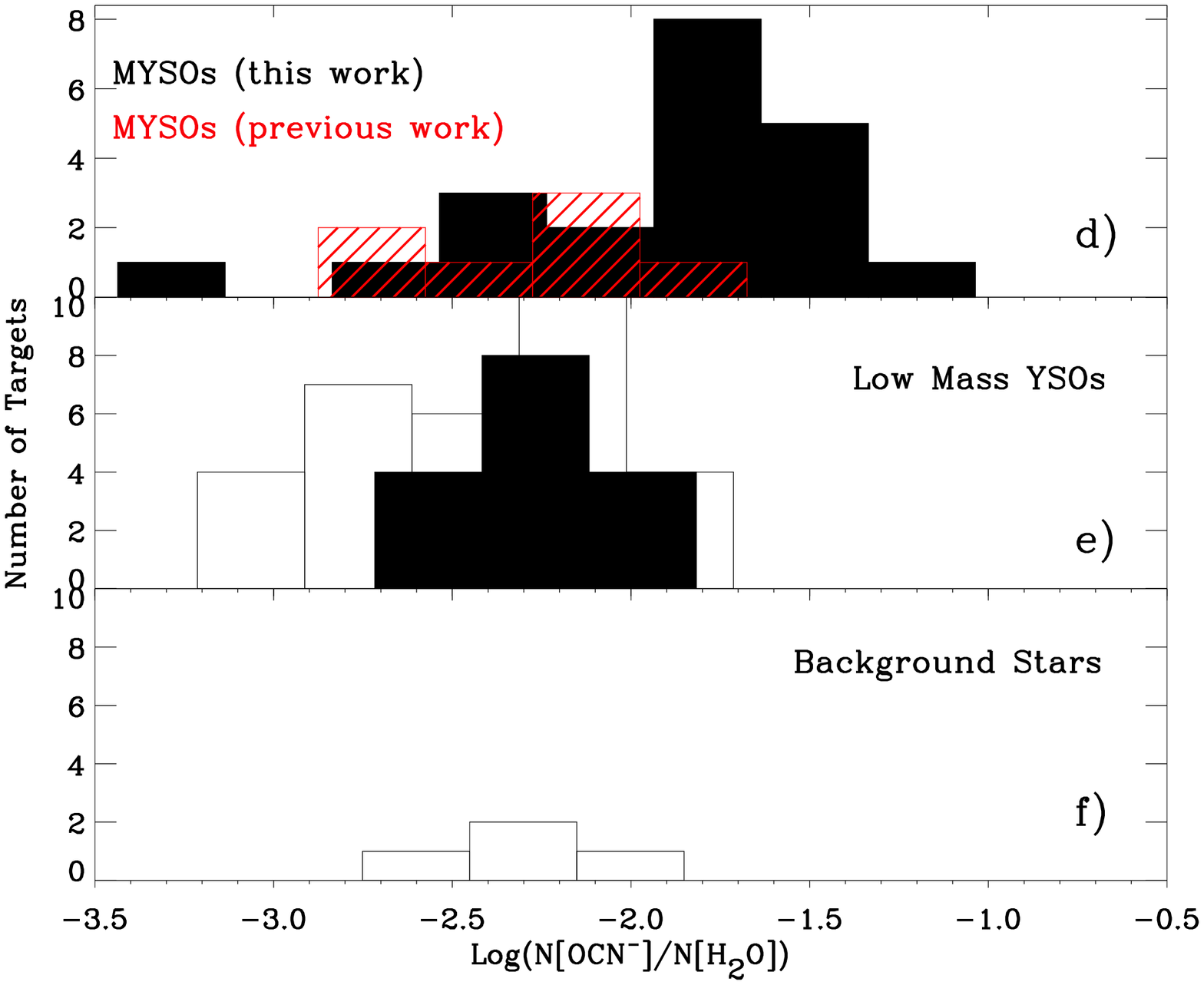}
\caption{{\bf Panel (a):} Distribution of the CH$_3$OH ice abundance
  relative to H$_2$O ice towards MYSOs. The solid black histogram is
  for MYSO detections presented in this work.  The hatched red
  histogram is for all CH$_3$OH ice detections of previous work.  {\bf
    Panel (b)} shows the CH$_3$OH ice abundances toward low mass YSOs
  (black) and comets (blue hatched). {\bf Panel (c) } shows CH$_3$OH
  ice abundances relative to H$_2$O ice for lines of sight tracing
  quiescent dense cloud material. In all panels the unfilled histogram
  includes abundance upper limits. {\bf Panels (d), (e), and (f)} show
  the abundances of OCN$^-$ ice relative to H$_2$O ice for MYSOs, low
  mass YSOs, and background stars, respectively.}~\label{f:histo2}
\end{figure*}

\begin{figure*}
\includegraphics[angle=90, scale=0.32]{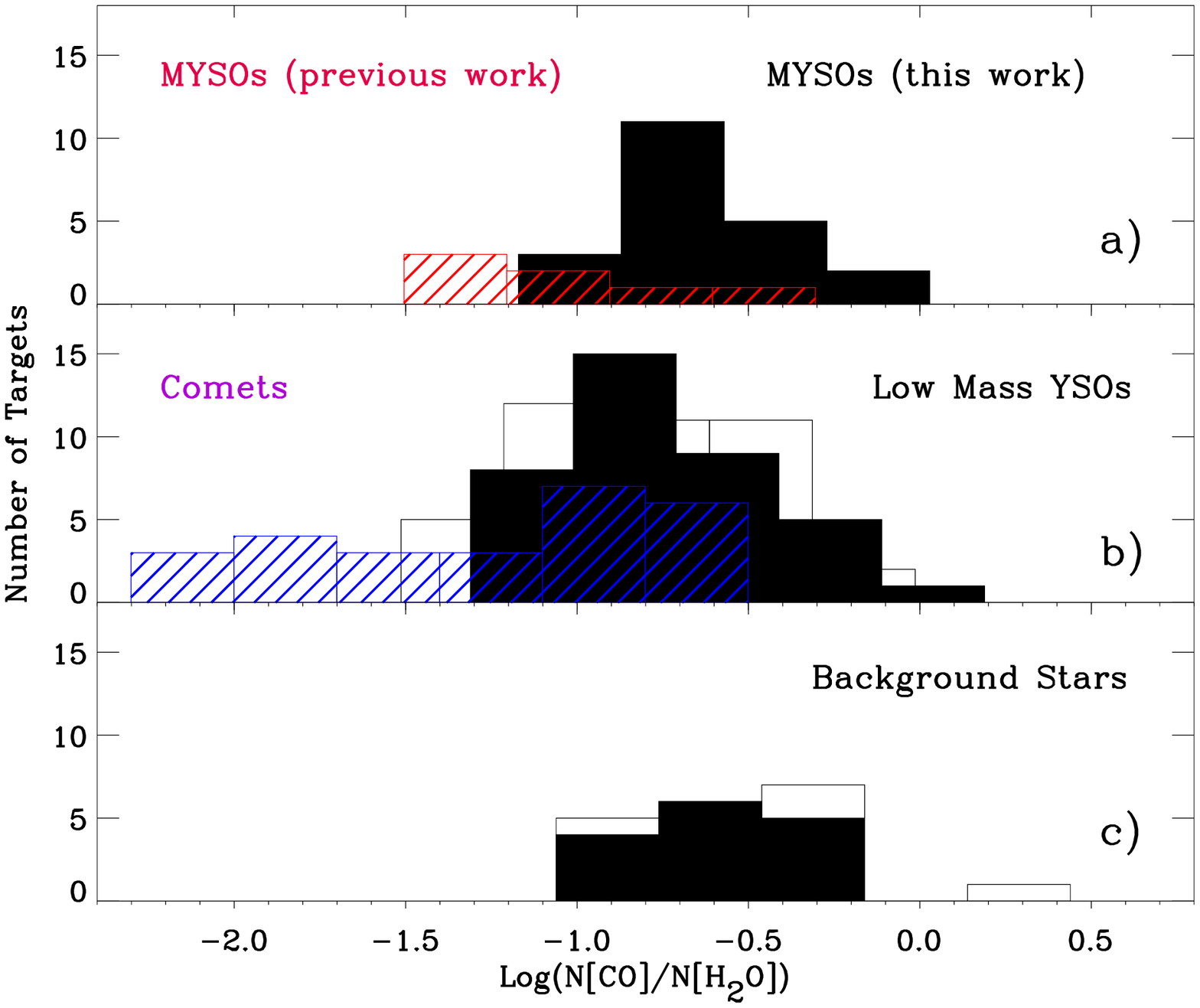}
\includegraphics[angle=90, scale=0.32]{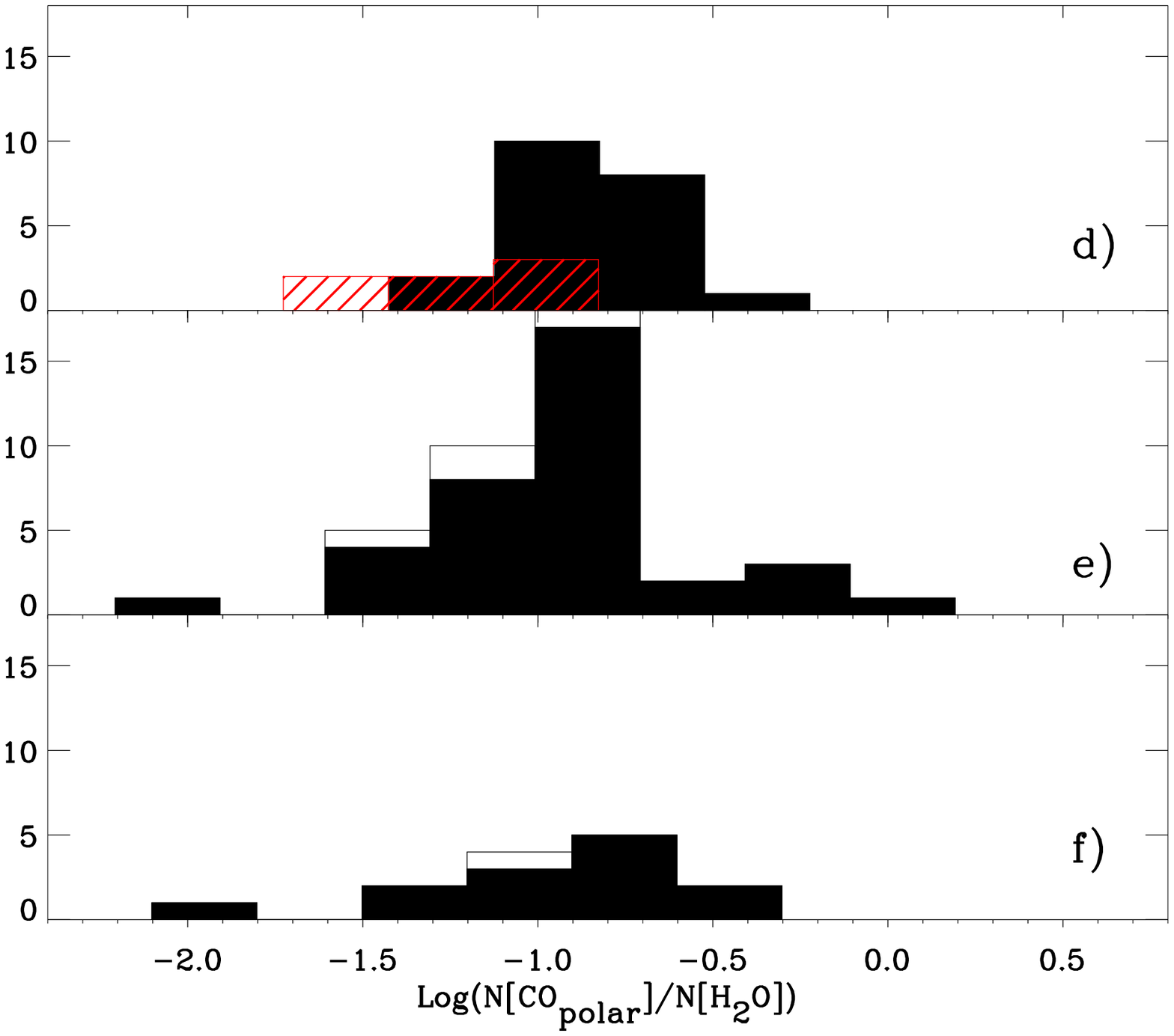}
\includegraphics[angle=90, scale=0.32]{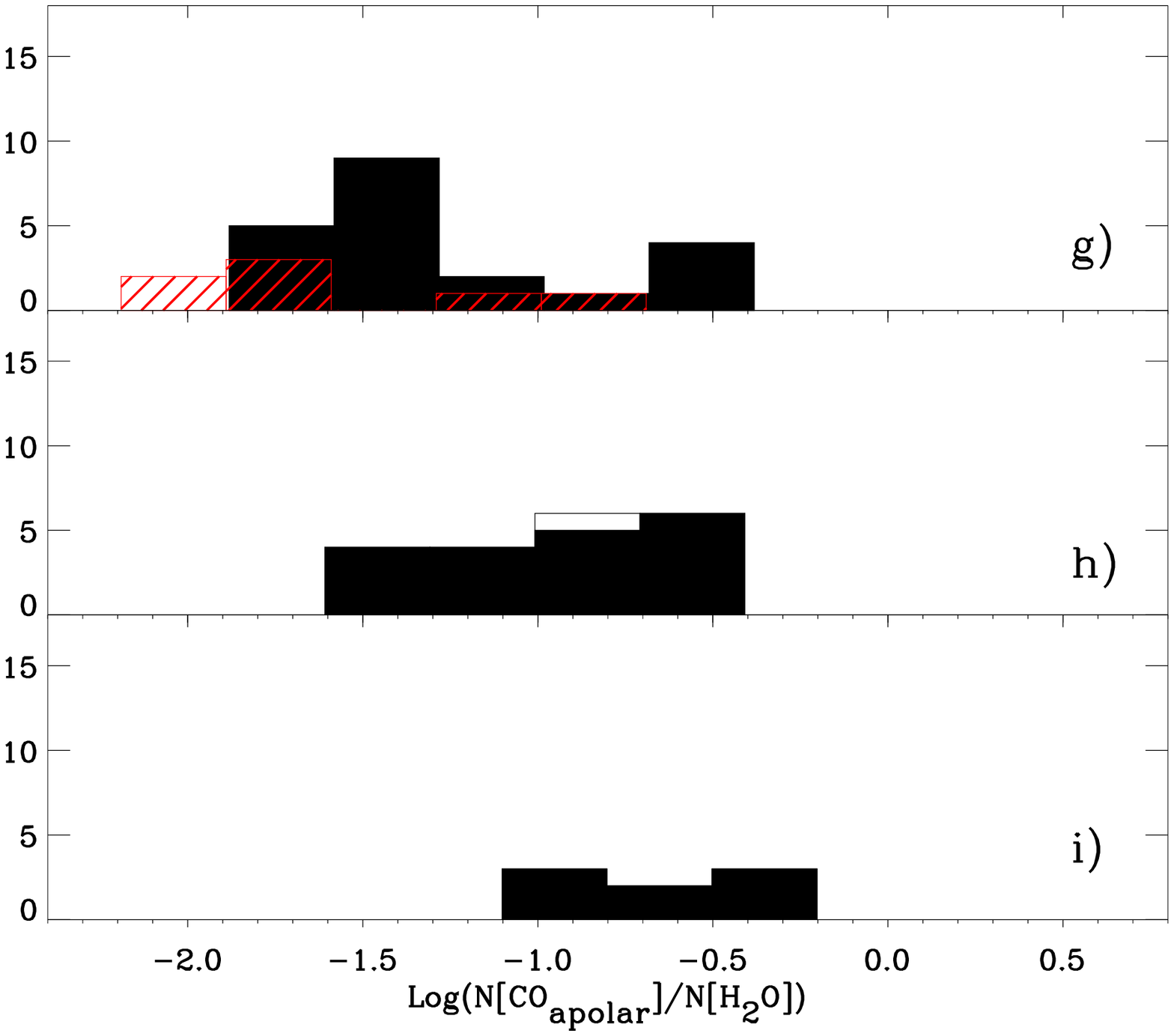}
\caption{{\bf Panel (a):} Distribution of the total CO ice abundance
  relative to H$_2$O ice towards MYSOs. The solid black histogram is
  for MYSO detections presented in this work.  The hatched red
  histogram is for all CO ice detections of previous work. {\bf Panel
    (b)} shows the CO ice abundances toward low mass YSOs (black) and
  comets (blue hatched). {\bf Panel (c)} shows CO ice abundances
  relative to H$_2$O ice for lines of sight tracing quiescent dense
  cloud material. In all panels the unfilled histogram includes
  abundance upper limits. {\bf Panels (d), (e), and (f)} show the
  abundances of CO ice in the polar phase relative to H$_2$O ice for
  MYSOs, low mass YSOs, and background stars, respectively. {\bf
    Panels (g), (h), and (i)} show the abundances of CO ice in the
  apolar phase towards these same target categories.}~\label{f:histo3}
\end{figure*}

The distribution of the ice abundances for the MYSOs and other target
categories is shown in histograms
(Figs.~\ref{f:histo1}-\ref{f:histo3}). The corresponding median
abundances and upper and lower quartiles are listed in
Table~\ref{t:median}, following the format in \citep{oberg11} and
\citep{boogert15}. The distribution of the OCS ice abundances relative
to CH$_3$OH is much narrower compared to abundances relative to H$_2$O
ice (Fig.~\ref{f:histo1}), consistent with the much better Spearman's
Rank coefficient (Fig.~\ref{f:correl}). It also shows the large
improvement in sample size relative to previous work. No detections or
significant upper limits of OCS ice abundances were made towards low
mass YSOs and dense cloud background stars \citep{palumbo97,
  boogert00, gibb04}. Towards comets, OCS abundances relative to
H$_2$O where published by \citet{saki20} and Figure~\ref{f:histo1}
shows a distribution quite similar to that for MYSOs.

The abundance of CH$_3$OH relative to H$_2$O ice towards MYSOs appears
to be fairly similar, or perhaps a little larger, compared to low mass
YSOs and dense cloud background stars (Fig.~\ref{f:histo2}). For
OCN$^-$ relative to H$_2$O ice, a fairly broad distribution is
observed for MYSOs, with a peak at higher abundances compared to low
mass YSOs. Towards dense cloud background stars, OCN$^-$ has not yet
been detected, with upper limits comparable to the detections towards
low mass YSOs \citep{whittet01b, chu20}.

For CO ice, the abundances relative to H$_2$O ice are quite similar
for MYSOs, low mass YSOs, and dense cloud background stars
(Fig.~\ref{f:histo3}).  But while the abundances of the polar
component of CO relative to H$_2$O ice are also quite similar across
the different classes of objects, this is not the case for the apolar
CO phase. Previous work on MYSOs (red hatched histograms in
Fig.~\ref{f:histo3}) showed a deficiency of the total CO and polar CO
ices relative to low mass YSOs and background stars. In the current
sample, this is no longer the case, their abundances are
comparable. However, the deficiency of the apolar CO ice component in
the current MYSO sample is consistent with previous work.

\begin{deluxetable}{lllll}
\tablecolumns{5}
\tablewidth{0pc}
\tablecaption{Median Ice Abundances~\label{t:median}}
\tablehead{
\colhead{Quantity\tablenotemark{a}}& \colhead{MYSOs} & \colhead{LYSOs} & \colhead{BG Stars} & \colhead{Comets} \\
\colhead{}                       & \colhead{     } & \colhead{     } & \colhead{        } & \colhead{      } \\}
\startdata
$N$(OCS)/$N$(H$_2$O)$\times$100\%               & 0.15$_{0.12}^{0.28}$ & \nodata          & \nodata      & 0.14$_{0.08}^{0.21}$\tablenotemark{b} \\
$N$(OCS)/$N$(CH$_3$OH)$\times$100\%             & 0.92$_{0.74}^{1.42}$ & \nodata          & \nodata      & \nodata \\
$N$(CH$_3$OH)/$N$(H$_2$O)$\times$100\%          & 20$_{8}^{26}$       & 5.6$_{4.8}^{11.8}$ & 10$_{7}^{21}$  & 1.7$_{0.6}^{2.5}$\tablenotemark{c} \\
$N$(OCN$^-$)/$N$(H$_2$O)$\times$100\%           & 1.53$_{0.58}^{2.49}$ & 0.63$_{0.39}^{0.77}$& $<0.2-0.9$  & \nodata \\
$N$(CO$_{\rm total}$)/$N$(H$_2$O)$\times$100\%    & 24$_{15}^{31}$      & 15$_{10}^{32}$     & 25$_{12}^{43}$ & 7$_{2}^{15}$\tablenotemark{d}  \\
$N$(CO$_{\rm polar}$)/$N$(H$_2$O)$\times$100\%    & 14$_{10}^{22}$      & 11$_{8}^{17}$      & 13$_{6}^{19}$ & \nodata \\
$N$(CO$_{\rm apolar}$)/$N$(H$_2$O)$\times$100\%   & 4.0$_{2.5}^{10.8}$   & 11$_{5}^{24}$      & 21$_{12}^{35}$ & \nodata \\
\enddata
\tablenotetext{a}{For each molecular column density ratio, the median
  and lower and upper quartile values of the detections are given.}
\tablenotetext{b}{\citet{saki20}}
\tablenotetext{c}{\citet{disanti08}}
\tablenotetext{d}{\citet{disanti08} and \citet{saki20}}
\end{deluxetable}

\vspace{60pt}

\section{Discussion}~\label{sec:dis}

The presented infrared spectroscopic survey of solid OCS towards 23
MYSOs provides observational constraints to the formation environment
of sulfur-bearing species in interstellar and circumstellar ices. When
compared with un-irradiated laboratory ice mixtures, the observed 4.90
\mum\ C-O stretching mode band profiles of OCS are most consistent
with CH$_3$OH-rich OCS ice mixtures, and do not favor H$_2$O, CO$_2$,
and CO-rich environments.  This confirms the early work by
\citet{palumbo97} on a sample of three MYSOs. In our much larger
sample, the identification of OCS in CH$_3$OH-rich mixtures is
supported by a good correlation of the OCS and CH$_3$OH column
densities, indeed much better than that with the H$_2$O and CO column
densities.

Observations and models have shown that CH$_3$OH formation is strongly
enhanced deep into dense clouds ($A_{\rm V}>9$, densities $\geq 10^5$
cm$^{-3}$), once much of the CO freezes out \citep{pontoppidan04,
  cuppen09, boogert11, chu20}. The good correlation of the OCS and
CH$_3$OH column densities would then indicate that OCS is preferably
formed in this later, dense stage of molecular clouds, or the
envelopes of YSOs, where similar conditions occur.  A direct accretion
of OCS from the gas phase can be excluded, considering that gas phase
OCS abundances are $\sim 1\times 10^{-5}$ relative to CO, orders of
magnitude smaller than observed in the ices
(Table~\ref{t:median}). The route to solid state OCS might then be the
oxidation of CS on the grains. At the same time, the oxidation of CO
would likely produce CO$_2$, and indeed, the presence of CO$_2$
heavily diluted in CH$_3$OH ice is indicated by a long-wavelength
shoulder on the 15 \mum\ bending mode of solid CO$_2$ ice
\citep{dartois99, gerakines99}.  The abundance of this CO$_2$ relative
to the CH$_3$OH column density varies between 5-30\%
\citep{pontoppidan08}, indicating that such oxidation reactions in the
CO-rich ice phase are significant.

CS and CO are formed early in the cloud evolution, by gas phase
ion-molecule reactions, when at low densities, C, S, and O are mostly
in atomic form (e.g., \citealt{palumbo97}).  Later gas phase chemistry
does not modify CS, but converts S to SO.  On the grain surfaces,
early oxidation of CO and CS would make CO$_2$ and OCS and oxidation
of S and SO makes SO$_2$. Evidently, at the same time, abundant H$_2$O
ices are produced, and some CH$_3$OH formation is expected as well.
Therefore, if OCS formation takes place early in the cloud's lifetime,
at relatively low densities ($\sim 10^3$ cm$^{-3}$), it is expected to
be mixed with H$_2$O, CO$_2$, and CH$_3$OH. However, while the CO$_2$
abundance correlates very well with that of H$_2$O ice
($N$(CO$_2$)/$N$(H$_2$O)$\sim$ 20-30\%, e.g., \citealt{gerakines99}),
CH$_3$OH and OCS ice have not been detected at the level of $<1\%$ in
these early ices. Our observations are consistent with the formation
of the bulk of these species much deeper in the cloud.

A key question is if there is sufficient CS to produce the OCS
abundance observed for the MYSO sample.  In diffuse clouds and dense
clouds, the gas phase CS/CO abundance ratio is $\sim 5\times 10^{-5}$
(\citealt{laas19} and references therein).  The observed median
abundance of OCS ice relative to CO ice is two orders of magnitude
larger at 6$\times 10^{-3}$ (Table~\ref{t:median}).  Much of the CO
was converted to CH$_3$OH and CO$_2$, however. The ice abundances of
CO, CH$_3$OH, and CO$_2$ are comparable (Table~\ref{t:median},
\citealt{gerakines99}), resulting in a OCS/(CO+CH$_3$OH+CO$_2$) ice
ratio of 2$\times 10^{-3}$. This is still a factor of $\sim 40$ larger
than the CS/CO ratio produced by gas phase chemistry.  In addition,
much like CO ice is converted to H$_2$CO and CH$_3$OH, CS would be
converted to H$_2$CS and CH$_3$SH as well
\citep{lamberts18}. Observations toward a comet \citep{calmonte16,
  altwegg19} and the envelopes of Class 0 protostar
\citep{drozdovskaya19} show $N$(CH$_3$SH)/$N$(CS)$\sim 1$.  The order
of magnitude enhancement of CS due to grain surface chemistry seen in
the models of \citet{laas19}, would not fully remedy this.  Overall,
CS appears to be only a minor precursor to OCS ice.

Sulfurization of CO ice is a more likely route to OCS. \citet{laas19}
find that S atom addition reactions are dominant on the grains.
Sulfur allotropes, such as suggested for the formation of abundant gas
phase SO$_2$ in a hot core \citep{dungee18}, could also play a role.
H$_2$S seems a reasonable candidate as well. The upper limits on the
H$_2$S ice abundance ($<$0.3-1\% relative to H$_2$O;
\citealt{smith91}) are larger than the OCS ice abundances and thus the
observations currently do not constrain this possibility.

The production of OCS by energetic process may play a role as well.
The peak positions and widths of the 4.90 \mum\ band toward many of
the observed targets are consistent with the proton-irradiated
CO:H$_2$S=5:1 and CO:SO$_2$=5:1 ices, with or without H$_2$O, after a
modest amount of heating to temperatures of $<$50 K
(Fig.~\ref{f:nudnuocs}; \citealt{ferrante08}; see also
\citealt{garozzo10}). Such irradiation would produce much CO$_2$, and
minor amounts of C$_3$O$_2$, C$_2$O, HCO, CS$_2$, and for the
SO$_2$-containing ices also O$_3$. Other than CO$_2$, which is also
abundantly produced by grain surface chemistry, the latter species are
not detected in the MYSO spectra, but not with significant upper
limits. Indeed, in the irradiation experiments, the absorption
strengths of the other species observed is much less than that of the
4.90 \mum\ OCS band. A deep search with JWST is needed to further
constrain the irradiation history of the interstellar ices.

The starting mixture in these irradiation experiments is CO:H$_2$S=5:1
and CO:SO$_2$=5:1. The presence of H$_2$S in CO-rich ices locates the
ices at lower densities, where H$_2$S would be formed via
hydrogenation reactions on the grain surface, while the good
correlation of OCS with CH$_3$OH indicates the formation during a high
density phase.  SO$_2$, on the other hand, was suggested to be present
in a CH$_3$OH-rich environment, following a tentative detection by
\citet{boogert97}. At an SO$_2$ abundance of 0.3\% relative to H$_2$O
towards the MYSO W33A (G012.9090-00.2607), the abundance relative to
CO ice is 3.5\%. This is within an order of magnitude of the
composition of the initial laboratory ice, and thus irradiation of
SO$_2$-containing ices could be the source of the observed OCS ices. A
more quantitative assessment is beyond the scope of this work, and
would also benefit from SO$_2$ ice measurements in a wider variety of
sight-lines. Also, other S-bearing species in the CO-rich ice, such as
CS and S allotropes may produce OCS in the irradiation process as
well. Indeed, free S atoms readily react with CO to form OCS
\citep{ferrante08}.

Finally, we note that despite the correlation of OCS column densities
with the species mentioned above, the scatter in the correlations is
of the order of a factor of 2.  This implies that the OCS production
is quite strongly dependent on specific local conditions (radiation
fields, ice temperatures).  This is reminiscent of conclusions that
were drawn for the production of OCN$^-$ \citep{vanbroekhuizen05}.

\subsection{OCN$^-$}~\label{sec:dis_ocn}

The correlation of OCS with OCN$^-$ is as good as that with CH$_3$OH
(Fig.~\ref{f:correl}), and indeed OCN$^-$ and CH$_3$OH also correlate
well with eachother (Fig.~\ref{f:correlother}).  In a study of low
mass YSOs, \citet{vanbroekhuizen05} found that OCN$^-$ is present in
both CO-rich and polar environments, claiming that OCN$^-$ is formed
in a CO-rich environment (short-wavelength component) and survives in
a polar-CO environment (long-wavelength component; see also
\citealt{oberg11}). This is confirmed by a correlation of the
short-wavelength component with CO in an apolar environment, and of
the long-wavelength component with polar CO ices. For our MYSO sample,
the short wavelength component is detected in just a few cases. The
long-wavelength component indeed correlates much better with the polar
than with the apolar ices.  Thus, overall the idea of OCN$^-$ being
produced in a CO-rich environment is confirmed. For our MYSO sample,
much of the CO has already been converted to other species.

While it was shown that OCN$^-$ is easily produced at low temperatures
by acid base chemistry of HNCO and NH$_3$ mixtures \citep{raunier04},
the apolar ices are expected to be NH$_3$-poor, just like they are
H$_2$O-poor. Perhaps another base enables this reaction.  On the other
hand, it was shown that the presence of N in CO-rich ices leads
(non-energetically) to the formation of both HNCO and NH$_3$
\citep{fedoseev15} as a result of hydrogen addition and abstraction
reactions.

\subsection{Comparing YSOs, Dense Clouds, and Cometary Ices}~\label{sec:dis_comet}

In a statistical comparison of YSO and cometary ice abundances,
\citet{oberg11} concluded that comets are carbon and nitrogen-poor.
This could hint to the effects of ice processing in protoplanetary
disks, or that compared to the distribution of ice abundances of low
mass YSOs, the early Sun resembled the YSOs with low C and N ice
abundances. One possible cause of low C and N abundances could be the
presence of nearby massive stars, that sublimate or photodesorb the
most volatile (CO, N$_2$) ices. The Sun was indeed most likely formed
in a cluster \citep{adams10}.

Re-assessing the comparison between YSOs and comets, using the vastly
increased MYSO sample, we find that, in contrast to the lower
abundances in previous work, MYSOs have CO ice abundances that are
comparable to those of low mass YSOs and background stars
(Fig.~\ref{f:histo3}a-c). About 50\% of the comets have CO ice
abundances below those of MYSOs and low mass YSOs. The large
heterogeneity of the cometary CO ice abundances is also apparent,
probably reflecting their processing history. Comets are also
particularly poor in CH$_3$OH ices (Fig.~\ref{f:histo2})), especially
when compared to the new sample of MYSOs.

In contrast to carbon-bearing molecules, comets do not show a
deficiency in the OCS abundance (Fig.~\ref{f:histo1})). This is
surprising, considering that we find that OCS is mixed with
CH$_3$OH-rich ices. This could point to the additional production of
OCS in the protoplantary disk phase. Studies of OCS in low mass YSOs
will be enabled by JWST. A tentative detection by the AKARI space
telescope towards the edge-on disk IRC L1041-2 indicates a very high
OCS ice abundance of 1.1\% relative to H$_2$O \citep{aikawa12}, which
is 50\% more than the MYSO with the largest OCS abundance
(G010.8856+00.1221; Table~\ref{t:ocs})

As for the comparison between MYSOs, low mass YSOs and dense cloud
background stars, MYSOs are CH$_3$OH and OCN$^-$-rich. The total and
polar CO ice abundances are quite similar, but MYSOs do appear to have
significantly less apolar ices. This is consistent with a more
efficient conversion of CO to CH$_3$OH and OCN$^-$ in the ices.  One
should keep in mind, however, that the low mass YSO and dense cloud
background star samples are presently relatively small.

\section{Conclusions and Future Work}~\label{sec:concl}

Infrared 2-5 \mum\ spectra of a sample of twenty-three deeply embedded
MYSOs were studied with the primary goal to study the abundance of
solid state OCS via the C-O stretch mode at 4.90 \mum. The following
conclusions are drawn:

\begin{enumerate}

\item The OCS band at 4.90 \mum\ was detected in twenty MYSO lines of
  sight, increasing the sample by more than a factor of 5.

\item The absorption profile of the OCS band shows little variation
  from source to source and is consistent with mixtures of OCS
  embedded in CH$_3$OH-rich ices, or mildly heated proton-irradiated
  CO:H$_2$S ices.

\item All twenty-three MYSOs show CO and H$_2$O ice absorption, and
  most show absorption by OCN$^-$ and CH$_3$OH.

\item The OCS column density correlates well with that of CH$_3$OH and
  OCN$^-$, and polar CO ice, but not with H$_2$O and apolar CO ice.
  This firmly establishes that OCS ice is formed together with
  CH$_3$OH and OCN$^-$, deep inside dense clouds or protostellar
  envelopes.

\item The median ice abundance towards the MYSO sample, as a
  percentage of H$_2$O ice is CO:CH$_3$OH:OCN$^-$:OCS=24:20:1.53:0.15
    
\item It is deduced that CS is not the main precursor to OCS, because
  its abundance is too low by at least a factor of
  $\sim$40. Sulfurization of CO is likely needed. The source of this
  sulfur is unclear. Perhaps sulfur atoms or allotropes. The H$_2$S
  ice abundance is insufficiently constrained.

\item The OCS abundances relative to H$_2$O ice towards MYSOs and
  comets are remarkably similar. This contrasts with CH$_3$OH, which
  shows a deficit for comets. Perhaps additional OCS is formed in
  protoplanetary disk environments.
  
\item Compared to low mass YSOs, MYSOs show an excess of OCN$^-$ ice
  in a significant fraction of the targets. The CO and CH$_3$OH ice
  abundances are comparable to those observed towards low mass YSOs
  and dense cloud background stars. It should be noted that the low
  mass YSO and background star samples are fairly small, but this will
  be improved upon with JWST observations in the near
  future. Evidently, the current lack of detections of OCS ice towards
  low mass YSOs and background stars needs to be improved upon to form
  the trail of sulfur to comets. Indeed, the presented IRTF/SpeX 2-5
  \mum\ spectra of this sample of MYSO, most of which would saturate
  the JWST/NIRSpec and NIRCam WFSS detectors, will serve as a
  reference for the future JWST observations.

\end{enumerate}


\begin{acknowledgments}
This paper made use of information from the Red MSX Source survey
database at
\url{http://rms.leeds.ac.uk/cgi-bin/public/RMS_DATABASE.cgi} which was
constructed with support from the Science and Technology Facilities
Council of the UK. This research has made use of the SIMBAD database,
operated at CDS, Strasbourg, France.
\end{acknowledgments}

\bibliography{texreferences}{}

\end{document}